\documentclass[runningheads]{svjour3} 

\usepackage{geometry}
\usepackage{mathptmx}      
\usepackage{amsfonts}
\usepackage{amsmath}
\usepackage{textcomp}
\usepackage{amssymb}

\usepackage{graphicx}
\usepackage{epstopdf}
\usepackage{subfig}
\usepackage{mhchem}

\usepackage{setspace}
\usepackage{colortbl}

\usepackage{color}
\usepackage{ulem}

\usepackage[numbers,sort&compress]{natbib}
\newcommand{\D}[2]{\ensuremath{{\frac{d\!#1}{d\!#2}}}}
\newcommand{\DD}[2]{\ensuremath{{\frac{d^{2}\!#1}{{d\!#2}^{2}}}}}

\newcommand{\Rp}{\ensuremath{r_{1}}}
\newcommand{\Rm}{\ensuremath{r_{2}}}
\newcommand{\Rpm}{\ensuremath{r_{1,2}}} 

\newcommand{\KONST}[1]{\ensuremath{{K_{#1}}}}
\newcommand{\KM}{\KONST{M}} 
\newcommand{\KE}{\KONST{E}} 
\newcommand{\Kk}{\ensuremath{{k_{3}}}} 
\newcommand{\Knk}{\ensuremath{{k_{-3}}}} 

\newcommand{\MFN}{\ensuremath{{\mathcal{M}}_{O}}}
\newcommand{\MFA}{\ensuremath{{\mathcal{M}}_{A}}}
\newcommand{\MFI}{\ensuremath{{\mathcal{M}}_{C}}}

\newcommand{\Order}[1]{\ensuremath{\mathcal{O}(#1)}}

\newcommand{\FP}{\ensuremath{F^{*}}}

\begin{document}

\title{The Quasi-Steady State Assumption in an Enzymatically Open System}
\author{$\text{Ed Reznik}^{1,3}$, $\text{Daniel Segr\`{e}}^{1,2,3}$, $\text{William Erik Sherwood}^{3,4,*}$}
\authorrunning{Reznik,  Segr\`{e}, Sherwood}

\institute{* Corresponding author.\\ \ \\
              $^{1}$Department of Biomedical Engineering, 
              $^{2}$Bioinformatics Program and Department of Biology, \\
	     $^{3}$Center for BioDynamics
	     and $^{4}$Department of Mathematics, 
              Boston University, 
              Boston, MA 02215\\ \ \\
              {\it Email addresses:} ereznik@bu.edu (E. Reznik), dsegre@bu.edu (D. Segr\`{e}), wesher@bu.edu (W. E. Sherwood)\\
             }

\titlerunning{QSSA with Enzyme Input}

\date{\today}

\maketitle

\abstract{The quasi-steady state assumption (QSSA) forms the basis for rigorous mathematical justification of the Michaelis-Menten formalism 
commonly used in modeling a broad range of intracellular phenomena. A critical supposition of QSSA-based analyses is that the underlying biochemical reaction is enzymatically ``closed,'' so that free enzyme is neither added to nor removed from the reaction over the relevant time period. Yet there are multiple circumstances in living cells under which this assumption may not hold, {\it e.g.} during translation of genetic elements or metabolic regulatory events. Here we consider a modified version of the most basic enzyme-catalyzed reaction which incorporates enzyme input and removal. We extend the QSSA to this enzymatically ``open'' system, computing inner approximations to its dynamics, and we compare the behavior of the full open system, our approximations, and the closed system under broad range of kinetic parameters. We also derive conditions under which our new approximations are provably valid; numerical simulations demonstrate that our approximations remain quite accurate even when these conditions are not satisfied. Finally, we investigate the possibility of damped oscillatory behavior in the enzymatically open reaction.}


\section{Introduction}
\label{sec:intro}

The Michaelis-Menten formalism \cite{Michaelis:1913,Briggs:1925} is perhaps the most commonly used framework for modeling the dynamics of biochemical processes. Derivation of the formalism proceeds from the assumption that in an enzymatic reaction, the level of enzyme bound with substrate to form enzyme-substrate complex rapidly equilibrates at the start of the reaction. Following this initial transient, the levels of free and bound enzyme maintain near-equilibrium, or ``quasi-steady-state,'' for the remainder of the reaction, during which time their concentrations may be accurately modeled as simple rational functions of the substrate concentration. 

The quasi-steady-state assumption (QSSA) has been a topic of mathematical interest for nearly a century \cite{Briggs:1925, Bowen:1963, Segel:1988, Segel:1989, Borghans:1996, Schnell:2000, Stoleriu:2004, Tzafriri:2004, Stoleriu:2005, Flach:2006, Dingee:2008}. Its modern justification is based upon singular perturbation analyses that exploit separations between the time-scales on which reaction variables evolve in order to derive analytic approximations to the reaction dynamics and conditions under which these approximations are valid \cite{Bowen:1963, Segel:1989}. Classical QSSA analyses, like the original Michaelis-Menten derivation, consider biochemical systems which are ``closed.''  That is, they admit neither addition nor removal of unbound enzyme or substrate from the reaction once it is underway. This leads to a conservation relation which greatly simplifies the mathematical analysis. While this scenario may reasonably approximate laboratory experiments, biochemical isolation is not so common {\it in vivo}, where neither enzyme nor substrate levels remain constant. Recent studies have extended the QSSA for enzyme-catalyzed reactions to which substrate is added at either at a constant rate \cite{Stoleriu:2004} or as a periodic input \cite{Stoleriu:2005}. The question of the effect of variable enzyme levels on the QSSA has remained unaddressed up to now, though enzyme concentrations in living cells continually fluctuate due to multiple mechanisms: Transcription and translation of genetic elements may add to the available store of enzyme, while degradation (by proteases) and deactivation (for example, by kinases via phosphorylation)  may subtract from it. Similarly, allosteric regulation by small-molecule metabolites can substantially shrink or expand the number of enzyme molecules available and able to catalyze a reaction.
In fact, the time-scales on which such alterations in available enzyme levels take place may vary widely (from a few microseconds for changes in enzyme activation state to minutes for eukaryotic enzyme transcription and translation \cite{Alon:2006, Fell:1997}). Under a variety of different circumstances they may overlap with the intrinsic time scales of the biochemical reactions in question: the values for enzymatic reaction rate constants implied by estimates of turnover and reaction velocity span at least six orders of magnitude, ranging from microseconds to nearly a second, and these rates themselves are affected significantly by pH, temperature, and other factors \cite{Berg:2002}. 

The widespread application of the QSSA and the Michaelis-Menten formalism to the kinetic modeling of metabolic events rests in no small part on the common understanding that changes in enzymatic levels occur much more slowly than metabolic effects. Relatively minor modifications to the classical QSSA approach may be made to address some of the above mentioned scenarios in which the time-scales for changes in free enzyme levels and other reaction rates clearly overlap \cite{Keener:1998}, and indeed, the general success and flexibility of the classical QSSA approach has understandably led certain of its premises to remain essentially unexamined. Yet a quantitative assessment of the assumption of unchanging enzyme levels is of theoretical and practical interest. How does variation in the amount of available enzyme affect the dynamics of biochemical reactions? Can the assumption of conservation of total enzyme be relaxed?

Here we consider an enzymatically ``open'' biochemical reaction in which both enzyme input and removal are allowed. Our goals are threefold: First, we seek to extend the QSSA to this more biologically accurate situation. Second, we wish to assess the comparative accuracy with which the dynamics of the enzymatically open system are approximated by the enzymatically ``closed'' system on which classical QSSA analyses are based. Third, using the approximations derived in our extension of the QSSA, we aim to describe the range of possible dynamics the open system may exhibit in various parameter regimes. 

The remainder of this section briefly reviews the classical QSSA framework and its variants for the closed system, drawing primarily upon the exposition in \cite{Segel:1989}, and it introduces the mathematical description of the open system.

\subsection{The QSSA framework for enzymatically closed systems}
\label{subsec:QSSAframework}

The development of the Michaelis-Menten formalism, as well as the classical QSSA framework for its analysis, proceeds by first considering the ``canonical'' simplest enzyme-catalyzed biochemical reaction, to which we refer herein as the {\it closed system} or {\it closed reaction}:

\begin{equation}
\label{chem:iso}
\ce{S + E <=>[k_{1}][k_{-1}]  C ->[k_{2}] E + P}
\end{equation}

\noindent Substrate $S$ and free enzyme $E$ combine reversibly to form complex $SE = C$, which irreversibly yields free enzyme and product $P$ upon dissociation. Applying the law of mass action, the dynamics of species concentrations for this reaction are described by a set of coupled nonlinear ODEs:

\begin{subequations}
	\label{eq:iso}
	\begin{eqnarray}
		\D{E}{t} &=& -k_1ES + (k_{-1} + k_2)C\label{eq:isoE}\\
		\D{C}{t} &=& k_1ES - (k_{-1} + k_2)C\label{eq:isoC} \\
		\D{S}{t} &=& k_{-1}C - k_1ES\label{eq:isoS}\\
		\D{P}{t} &=& k_{2}C\label{eq:isoP}
	\end{eqnarray}
\end{subequations}

\noindent The concentration of complex and product is normally taken to be zero at the beginning of the reaction, {\it i.e.} $(E, C, S, P) = (E_0, 0, S_0,0)$ at time $t=0$. 

Two conservation relations obtain for the enzymatically closed system:

\begin{subequations}
	\begin{eqnarray}
		S_{0} &=& S(t) + C(t) + P(t)\label{eq:iso-conserve2}\\
		E_{0} &=& E(t) + C(t)\label{eq:iso-conserve1}
	\end{eqnarray}	
\end{subequations}

\noindent Both relations simply express the principle of conservation of mass. Equation (\ref{eq:iso-conserve2}) asserts that substrate is neither added to nor removed from the reaction once it has begun (except by conversion to product); Equation (\ref{eq:iso-conserve1}) is the analogous assertion for free enzyme, {\it i.e.}, that the reaction is enzymatically closed.

The first key step of any classical QSSA-based analysis of (\ref{eq:iso}) is to invoke (\ref{eq:iso-conserve1}) in order to reduce its dimensionality. Omitting the product formation equation (\ref{eq:isoP}), which decouples, the reduced system is 

\begin{subequations}
	\label{eq:iso-red}
	\begin{eqnarray}
		\D{C}{t} &=& k_1(E_0-C)S - (k_{-1} + k_2)C\label{eq:iso-redC} \\
		\D{S}{t} &=& k_{-1}C - k_1(E_0-C)S\label{eq:iso-redS}
	\end{eqnarray}
\end{subequations}

\noindent The goal of the analysis is then to derive approximate analytic expressions for the concentrations of complex and substrate over time.

The second essential step of the analysis is the recognition that the system (\ref{eq:iso-red}) possesses two distinct time-scales, so that its evolution proceeds in two phases. In the first phase, the concentrations of free enzyme and complex rapidly equilibrate, while the level of unbound substrate is assumed to remain nearly constant, {\it i.e.} $S \approx S_0$. During this initial transient (or pre-steady-state) period, the concentration of complex exponentially approaches its quasi-equilibrium value

\begin{equation}
\label{eq:iso-complex-equil}
C(S) = \frac{E_{0}S}{\KM + S}
\end{equation}

\noindent where $S$ is taken to be $S_{0}$ and $\KM = (k_{-1} + k_{2})/k_{1}$ is the Michaelis-Menten constant. Note that the quasi-equilibrium is a function of the substrate concentration; this is the basic functional form for Michaelis-Menten dynamics. Together with conservation relation (\ref{eq:iso-conserve2}), Equation (\ref{eq:iso-complex-equil}) also dictates the quasi-equilibrium level of free enzyme.

In the second phase, the quasi-steady-state period, the concentration of substrate falls as it is slowly converted to product according to the ODE

\begin{equation}
\label{eq:iso-substrate-equil}
\D{S}{t} = - \frac{k_{2}E_{0}S}{\KM + S}
\end{equation}




\noindent The duration of each phase is determined by the characteristic time scale for the species concentration undergoing the most rapid change within it. The time scale of the first phase is given by the time constant for the exponential relaxation of the complex concentration, holding substrate level constant,

\begin{equation}
t_c = \frac{1}{k_1(S_0+\KM)}
\label{eq:iso-tc}
\end{equation}

\noindent A useful estimate for the time scale of the second phase is given by the ratio of the maximum change in substrate concentration to the maximum rate of change in substrate concentration, which yields

\begin{equation}
t_s = \frac{\KM + S_0}{k_2E_0}
\label{eq:iso-ts}
\end{equation}

\noindent When the time scales for the two phases are well separated, {\it i.e.} $t_c \ll t_s$, and the change in substrate concentration during the initial transient period is small, the QSSA is valid, as are consequent analyses which invoke it. In particular, the accuracy of singular perturbation approximations to the transient (``inner solutions'' or ``inner approximations'') and quasi-steady-state dynamics (``outer solutions'' or ``outer approximations'') of the system is assured. The requisite separation of time scales was originally shown to hold when substrate concentration greatly exceeds free enzyme concentration, a condition which typically obtains for {\it in vitro} biochemical experiments \cite{Michaelis:1913, Briggs:1925, Bowen:1963, Segel:1989}. Subsequent analysis showed that the approximations which follow from the QSSA also obtain under opposite circumstances, when the level of free enzyme greatly exceeds that of substrate \cite{Frenzen:1989, Segel:1989, Schnell:2000}. QSSA which presumes $S_{0} \gg E_{0}$ is usually called {\it standard QSSA}, and QSSA which presumes $E_{0} \gg S_{0}$ is typically called {\it reverse QSSA}. In Section \ref{sec:numerical} of this paper, we investigate the behavior of an enzymatically open biochemical system in both the standard and reverse QSSA regimes. 

\subsection{The open reaction: Enzyme input and removal}
\label{subsec:non-isolated-reaction}

The biochemical system studied in this paper is a modification of the canonical closed reaction (\ref{chem:iso}) which admits the addition and removal of free enzyme from the system while the reaction is underway: 

\begin{equation}
\label{chem:noniso}
\ce{ \cmath{\emptyset} <=>[\Kk][\Knk] E + S <=>[k_{1}][k_{-1}]  C ->[k_{2}] E + P}
\end{equation}

\noindent  Free enzyme is introduced {\it de novo} to the closed reaction at a constant rate \Kk\ and removed or inactivated (but not bound to substrate) at a concentration-dependent rate with rate constant \Knk. This we refer to as the {\it open system} or {\it open reaction}. Its dynamics are described by ODEs which differ only slightly from (\ref{eq:iso})

\begin{subequations}
	\label{eq:input}
	\begin{eqnarray}
		\D{E}{t} &=& \Kk - \Knk E -k_1ES + (k_{-1} + k_2)C\label{eq:inputE}\\
		\D{C}{t} &=& k_1ES - (k_{-1} + k_2)C\label{eq:inputC} \\
		\D{S}{t} &=& k_{-1}C - k_1ES\label{eq:inputS}\\
		\D{P}{t} &=& k_{2}C\label{eq:inputP}
	\end{eqnarray}
\end{subequations}

\noindent The initial conditions are again taken to be $(E, C, S, P) = (E_0, 0, S_0,0)$. 

One possible scenario described by (\ref{chem:noniso}) and (\ref{eq:input}) is a simple enzyme-catalyzed reaction during which new enzyme is introduced due to transcription and translation of genetic elements, and also degraded in the cytosol.  In this case, we expect \Kk\ and \Knk\ to be several orders of magnitude smaller than the other rate constants for the reactions. For example, enzyme synthesis takes on the order of minutes in {\it E. coli}, while the time needed for diffusion-limited equilibrium binding of a small molecules to a protein, which corresponds to $k_{1}$ in (\ref{chem:noniso}), is on the order of microseconds to milliseconds \cite{Alon:2006}. However, as mentioned in the Introduction, the range of rate constants for enzymatic reactions is broad enough that for many biologically plausible circumstances reasonable values for $k_{1}$, $k_{-1}$, and  $k_{2}$ may approach the same orders of magnitude as \Kk\ and/or \Knk \cite{Berg:2002}. 
 
Though the previous biological example is worth bearing in mind, we emphasize that the biochemical (\ref{chem:noniso}) and mathematical (\ref{eq:input}) descriptions of the open system are generic and might usefully be applied to a range of biological situations. The rate of enzyme input, \Kk, may be taken to stand for the rate of increase in availability for biochemical participation of free, active enzyme by any of several means ranging from the level of allosteric effects to that of gene activity. And similarly, the interpretation of the rate of enzyme removal, \Knk, is not strictly tied to any specific biological mechanism. Furthermore, our analysis and numerical investigations in the following sections do not depend on \Kk\ or \Knk\ being substantially smaller than other rate constants in the system. Even if one discounts the possibility that the time-scales of the enzyme-substrate reaction and enzyme input/removal might intermingle {\it in vivo}, the framework of (\ref{chem:noniso}) and (\ref{eq:input}) may be considered as a means by which to examine in some generality the QSSA presumption that during biochemical reactions total enzyme levels remain essentially static, and in particular, its essentiality for QSSA-based analyses and the consequences for the QSSA were it not to hold.

The remainder of this paper is devoted to extending standard QSSA analysis to the open system, with particular emphasis on dynamics during the initial transient period. In Section \ref{sec:non-isolatedQSSA}, we apply the QSSA to derive ``inner'' approximations to the enzyme and complex dynamics of the open system. In Section \ref{sec:timescales}, we calculate time-scales for the initial transient and quasi-steady-state dynamics of the open system, and we derive conditions on the separation of these time-scales under which the approximations obtained in Section \ref{sec:non-isolatedQSSA} are valid. Section \ref{sec:numerical} presents the results of numerical simulations which we use to evaluate the accuracy of our inner approximations to the open system vis-\`{a}-vis the closed system across a broad range of parameter values. In Section \ref{sec:oscillation} we investigate the possibility of damped oscillatory behavior in the open system using two different approaches. We end with a brief discussion in Section \ref{sec:discussion}.


\section{QSSA for the enzymatically open system}
\label{sec:non-isolatedQSSA}

In attempting to recapitulate standard QSSA analysis in the case of the open system (\ref{eq:input}), we immediately face a significant obstacle. The usual first step is to exploit the conservation of total enzyme (free plus bound) to eliminate the differential equation for free enzyme concentration, but the inclusion of enzyme input and removal in the open system invalidates the balance relation (\ref{eq:iso-conserve1}). Instead, the rate of change of total enzyme concentration equals the difference between the rate of enzyme input and the rate of enzyme removal: 

\begin{equation}
\D{E}{t} + \D{C}{t} = k_3 - \Knk E
\label{eq:conserve1}
\end{equation} 

\noindent This modified conservation relation is more usefully recast in integral terms: 

\begin{equation}
E(t) + C(t) = E_0 + \int_0^t{(k_3-\Knk E)dt}
\label{eq:conserve2}
\end{equation}

The total concentration of enzyme in free and bound form is simply the initial enzyme concentration plus the sum over time of the net rate of enzyme accumulation. Though we are unable to reduce the dimensionality of the open system using this new balance relation, it does enable us to find closed form approximations to the enzyme and complex dynamics in the initial transient period. 

\subsection{Approximation to enzyme dynamics} 
\label{subsec:enzymeapprox}

Solving (\ref{eq:conserve2}) for $C$ and substituting into (\ref{eq:inputE}), we obtain an integro-differential equation:

\begin{equation}
\D{E}{t} = k_3 - \Knk E - k_1ES + (k_{-1} + k_2)\left(E_0 + \int_0^t{(k_3-\Knk E)d\!t} - E\right)
\label{eq:enew}
\end{equation}

\noindent Differentiating removes the integral and yields a second order differential equation

\begin{equation}
\DD{E}{t} + (\Knk  + k_1S + k_{-1} + k_2)\D{E}{t} + \left(\Knk (k_{-1} + k_2)\right)E = \Kk(k_{-1} + k_2)
\label{eq:2ndE}
\end{equation}

\noindent whose solution requires specification of initial conditions for $E$ and $\D{E}{t}$. 

Having $S$ multiplying the first derivative term of equation (\ref{eq:2ndE}) renders it nonlinear and would appear to stymie attempts to find a simple closed form solution. At this juncture, we invoke the quasi-steady-state assumption: we reason that if $S$ evolves on a much slower time scale than $E$, so that $\D{S}{E} \approx 0$, then we may treat $S \approx S_{0}$ as a quasi-static parameter. (We investigate conditions under which this assumption is justified in Section \ref{sec:timescales}.) 

Under this assumption, equation (\ref{eq:2ndE}) becomes a linear second order ODE, which we proceed to solve by considering its homogeneous and inhomogeneous components in turn. The characteristic equation for the homogeneous component

\begin{equation}
r^2 + (\Knk  + k_1S_{0} + k_{-1} + k_2)r + \left(\Knk (k_{-1} + k_2)\right) = 0
\label{eq:char1}
\end{equation}

\noindent has two roots:

\begin{eqnarray}
\Rpm &=& \frac{1}{2}\left(-\Knk  - k_1S_{0} - k_{-1} - k_2 \pm \sqrt{(\Knk +k_1S_{0} + k_{-1} + k_2)^2 - 4\Knk (k_{-1} + k_2)}\right) \nonumber \\
\ &=& \frac{1}{2}\left(-\Knk  - k_1S_{0} - k_{-1} - k_2 \pm (\Knk +k_1S_{0} + k_{-1} + k_2)\sqrt{1 - \frac{4\Knk (k_{-1} + k_2)}{(\Knk +k_1S_{0} + k_{-1} + k_2)^2}}\right) \label{eq:roots1}\\
\ &=&  \frac{\Knk +k_1(S_{0} + \KM)}{2}\left(-1 \pm \sqrt{1 - \frac{4\Knk \KM}{k_1(\Knk/k_{1}+S_{0} + \KM)^2}}\right) \label{eq:roots2}
\end{eqnarray}

\noindent The homogenous component of the solution is then $E_{h}(t) = Ae^{\Rp t} + Be^{\Rm t}$, where $A, B$ will be determined from boundary conditions. Note that both $\Rp$ and $\Rm$ necessarily have negative real components (we consider the possibility that the roots $\Rpm$ are complex in a subsequent section).

To find the inhomogeneous component of the solution, we make the ansatz $E_i(t) = Ct^2 + Dt + F$. Upon substitution into (\ref{eq:2ndE}), we find that $E_i(t) = \KE$, where $\KE = \Kk/\Knk $ is the {\it enzyme accumulation constant} for the reaction (\ref{chem:noniso}). Note that $\KE$ is not dimensionless, but is in fact in units of (enzyme) concentration. 

Thus, the full solution to (\ref{eq:2ndE}) is 
\begin{equation}
E(t) = E_{h}(t) + E_{i}(t) = Ae^{\Rp t} + Be^{\Rm t} + \KE
\label{eq:finale}
\end{equation}

\noindent For time $t=0$, the  initial conditions $E(0) = E_0$ and $\D{E}{t}\big|_{t=0} = k_3 - \Knk E_0 - k_1E_0S_0$  determine $A$ and $B$, 
\begin{subequations}
	\begin{eqnarray}
	A &=& \frac{-k_3 + E_0(\Knk  +k_1S_0) - \Rm(\KE - E_0)}{\Rm-\Rp} \label{eq:coefA}\\
	B &=& \frac{k_3 - E_0(\Knk  - k_1S_0) + \Rp(\KE - E_0)}{\Rm-\Rp} \label{eq:coefB}
	\end{eqnarray}
\end{subequations}

\noindent which completes the inner solution for enzyme dynamics during the initial transient period. 

\subsection{Approximation to complex dynamics} 
\label{subsec:complexapprox}

We find an approximate solution to the dynamics of the complex in a similar fashion. Substituting (\ref{eq:finale}) into (\ref{eq:inputC}), we obtain

\begin{equation}
\D{C}{t} + (k_{-1} + k_2)C = k_1S\left(\KE + Ae^{\Rp t} + Be^{\Rm t}\right)
\label{eq:deriveC}
\end{equation}
\noindent which is again linear if we treat $S$ as constant. The homogenous solution to (\ref{eq:deriveC}) is
\begin{equation}
C_h(t) = De^{-(k_{-1}+k_2)t} = De^{-k_{1}\KM t}
\label{eq:CHomog}
\end{equation}

\noindent where $D$ is a coefficient to be solved for. Our ansatz for the inhomogeneous solution is $C_i(t) = Xe^{\Rp t} + Ye^{\Rm t} + Z$, which we substitute into (\ref{eq:deriveC}) and solve for $X,Y,Z$: 
\begin{subequations}
	\begin{eqnarray}
	X &=& \frac{k_1A}{\Rp+k_{-1}+k_2}S_{0} = \frac{A}{\Rp/k_{1}+\KM}S_{0}\label{eq:cx}\\
	Y &=& \frac{k_1B}{r_2+k_{-1}+k_2}S_{0} = \frac{B}{\Rm/k_{1}+\KM}S_{0}\label{eq:cy}\\
	Z &=& \frac{k_3k_1}{\Knk (k_{-1}+k_2)}S_{0} = \frac{\KE}{\KM}S_{0}\label{eq:cz}
	\end{eqnarray}
\end{subequations}

\noindent The full solution is then

\begin{equation}
C(t) = C_h(t) + C_i(t) = S_{0}\left(\frac{Ae^{\Rp t}}{\Rp/k_{1}+\KM}+ \frac{Be^{\Rm t}}{\Rm/k_{1}+\KM} + \frac{\KE}{\KM}\right) + De^{-k_{1}\KM t}
\label{eq:complex}
\end{equation}

\noindent For the initial transient period, we use the $t=0$ initial condition $C(0) = 0$ to fix $D$:

\begin{equation}
D = -S_{0}\left(\frac{A}{\Rp/k_{1}+\KM}+ \frac{B}{\Rm/k_{1}+\KM} + \frac{\KE}{\KM}\right) 
\label{eq:complexD}
\end{equation}

\noindent The equation for $C(t)$ in the initial transient period becomes
\begin{equation}
	C(t) = S_{0}\left(\frac{A(e^{\Rp t} - e^{-k_{1}\KM t})}{\Rp/k_{1}+\KM}+ \frac{B(e^{\Rm t} - e^{-k_{1}\KM t})}{\Rm/k_{1}+\KM}\right) +
	\frac{\KE}{\KM}\left(1-e^{-k_{1}\KM t}\right)
\label{eq:tempc}
\end{equation}

\noindent Further simplification of (\ref{eq:tempc}) yields

\begin{equation}
C(t) = S_{0}\left(\frac{Ae^{\Rp t}}{\Rp/k_{1}+\KM}+ \frac{Be^{\Rm t}}{\Rm/k_{1}+\KM} + \frac{\KE}{\KM}\right)
\label{eq:finalc}
\end{equation}

The expressions we obtain for the approximate form of the enzyme and complex dynamics during the initial transient phase of the open reaction are more complicated than those for obtained by applying the QSSA for the closed system. Instead of a simple exponential relaxation to the quasi-steady-state, we find for the open system a double exponential modulation of the enzyme (respectively, complex) level set by the enzyme accumulation constant \KE. 

The substrate and product dynamics of the open system may be calculated by substituting (\ref{eq:finale}) and (\ref{eq:finalc}) into ODEs (\ref{eq:inputC}) and (\ref{eq:inputP}). 

\begin{subequations}
	\label{eq:approxSP}
	\begin{eqnarray}
		\D{S}{t} &=& k_{-1}C(t) - k_1E(t)S\label{eq:approxS}\\
		\D{P}{t} &=& k_{2}C(t)\label{eq:approxP}
	\end{eqnarray}
\end{subequations} 

For the remainder of the paper, we refer to the (non-autonomous differential-algebraic) system composed of Equations (\ref{eq:finale}), (\ref{eq:finalc}), (\ref{eq:approxS}), and ({\ref{eq:approxP}) as the {\it approximate open system}.  We computationally investigate solutions to the approximate open system in comparison to solutions to the `full' open system (\ref{eq:input}) and the closed system (\ref{eq:iso}) in Section \ref{sec:numerical}.

\subsection{Relation to classical Michaelis-Menten kinetics}
\label{subsec:michaelis-menten}

The closed system (\ref{eq:iso}) may be viewed as a special case of the open system (\ref{eq:input}) in which $\Kk = \Knk  = 0$, which we may investigate using the above approach. In this case, Equation (\ref{eq:2ndE}) simplifies to 

\begin{equation}
\DD{E}{t} + (k_1S + k_{-1} + k_2)\D{E}{t} = \DD{E}{t} + k_1(S + \KM)\D{E}{t}  = 0 
\label{eq:classic1}
\end{equation}

\noindent with initial conditions $E(0) = E_0$ and $\D{E}{t}\big|_{t=0} =  - k_1E_0S_0$. Again treating $S$ as a parameter, the solution to (\ref{eq:classic1}) is 

\begin{equation}
E(t) = E_0 - \frac{E_0S_0}{S_0+K_m} \left(1-e^{-k_1(S_0+K_m)t}\right)
\label{eq:classic2}
\end{equation}

\noindent Invoking the original conservation relation (\ref{eq:iso-conserve1}), we immediately obtain an expression for the complex:
\begin{equation}
C(t) = \frac{E_0S_0}{S_0+K_m} \left(1-e^{-k_1(S_0+K_m)t}\right)
\label{eq:classic3}
\end{equation}

\noindent Thus our approximations for the open system pass a basic consistency check: Equations (\ref{eq:classic2}) and (\ref{eq:classic3}) exactly match those obtained in classical derivations of Michaelis-Menten dynamics \cite{Michaelis:1913, Briggs:1925, Segel:1989}.


\section{Time scales and quasi-steady states}
\label{sec:timescales}

A key assumption underlying the derivations of the previous section is that the level of substrate does not decline appreciably from its initial value. This is obviously not true for all time, but as in other QSSA analyses, we expect the assumption to hold during an initial transient period as the levels of enzyme and complex evolve on a much faster time scale than the level of substrate. In the initial transient period, Equations (\ref{eq:finale}) and (\ref{eq:finalc}) should provide accurate inner approximations to the full dynamics of the reaction. The first aim of this section is to estimate the fast time scale of the open reaction (\ref{eq:input}), that is, to estimate the duration of the initial transient period. 
	
QSSA analyses of the closed reaction (\ref{eq:iso}) show that by the end of the transient period, the enzyme and complex reach a quasi-equilibrium with respect to one another. This quasi-equilibrium is maintained as the substrate level slowly declines and the reaction eventually extinguishes itself.  In the open system,  enzyme (and hence complex) levels are altered exogenously and at potentially rapid rates, so we do not necessarily expect the dynamics of the full system to be organized by slowly changing quasi-equilibria for enzyme and complex in that case. Nonetheless, we may derive expressions for the enzyme and complex quasi-equilibria that may be operative at the end of the transient period, and compare these expressions with those for the closed reaction. This is the second aim of this section.

The section's third aim is to establish conditions on the reaction parameters under which we may be confident that our inner approximations accurately reflect the dynamics of the full open system (\ref{eq:input}). In particular, we seek inequalities in non-dimensionalized form which, when satisfied, imply equations (\ref{eq:finale}) and (\ref{eq:finalc}) closely approximate the true solutions $E(t)$ and $C(t)$ for the open system during the initial transient period.

\subsection{Fast time scale (enzyme-complex)}
\label{subsec:fast-timescale} 

Initially the levels of enzyme and complex change quickly, evolving much more rapidly than the level of substrate. Thus we note first that the exponential form of equations (\ref{eq:finale}) and (\ref{eq:finalc}) present two immediate candidates for the fast time scale of the open reaction, namely the characteristic time scales $\Rp$ and $\Rm$. Since $|\Rm| \geq |\Rp|$, we expect the dominant fast time scale (and hence the time span of the initial transient period) to be

\begin{equation}
t_{f} = \frac{1}{| \Rm |} = \frac{1}{k_{1}(S_{0}+\KM) + \Knk }\label{eq:tfast}
\end{equation}

\noindent We observe that this expression differs only slightly from the usual estimate for the fast time scale in the closed system, which is
\begin{equation}
t_{f} = \frac{1}{k_{1}(S_{0}+\KM)}\label{eq:orig_tfast}
\end{equation}
\noindent In particular, the rate of enzyme removal \Knk\ reduces the duration of the transient period. The rate of enzyme input, by contrast, does not affect the estimate of $t_{f}$.  

We have neglected the contribution of $\Rp$ in setting the time scale of the initial transient. Indeed, we argue that under most biologically relevant parameter regimes, $|\Rm| \gg |\Rp|$. Let $\delta = (4\Knk \KM)/[k_1(\Knk /k_1+S + \KM)^2]$, and reexpress the discriminant of (\ref{eq:char1}) as 
$\sqrt{1-\delta} \approx 1 - \delta/2 - \Order{\delta^2}$ for $\delta \ll 1$ (using the binomial rule). When this inequality holds,  
\begin{eqnarray}
\Rp &\approx&  -\frac{(\Knk +k_1(S + \KM))\delta}{4}\\
\Rm &\approx& -\frac{\Knk +k_1(S + \KM)}{2}\Big(2+\frac{\delta}{2}\Big) \approx -(\Knk +k_1(S + \KM))\\
\Rm-\Rp &\approx&  \frac{\Knk +k_1(S + \KM)}{2}\Big(-1-1+\frac{\delta}{2}-\frac{\delta}{2}\Big) = -(\Knk +k_1(S + \KM)) \approx \Rm
\end{eqnarray}

\noindent Thus when $\delta \ll 1$, we have $\Rp \approx \Rm\delta/4$, so that $\Rp \ll \Rm$, and we are justified in neglecting the contribution of $\Rp$ in setting the fast transient time scale. 

Two biologically plausible ways of satisfying the condition $\delta \ll 1$ are immediately apparent. First, when the level of substrate is relatively high (as compared to the product of the Michaelis-Menten constant and the enzyme removal rate), the denominator $\Knk /k_1+S + \KM$ is large, and so $\delta$ is small. Relatively high initial substrate levels are assumed in standard QSSA analyses, and over the course of the initial transient phase, we expect little change from the starting level of substrate (an assumption whose validity we investigate in Subsection \ref{subsec:validity} below). Second, when $\Knk$ is small (relative to the Michaelis-Menten constant of the reaction), $\delta$ will be small regardless of the substrate level. Enzyme degradation rates are typically at least two orders of magnitude lower than reaction rates for complex and product formation; thus for biologically relevant parameter regimes we may reasonably assume $\Knk \ll \KM$ and hence $\delta \ll 1$. 

\subsection{Quasi-steady states}
\label{subsec:quasi-steady-states}

In the closed system, the relative levels of enzyme and complex quickly equilibrate with respect to the level of substrate. This rapid equilibration occurs within the initial transient period, and subsequently the enzyme and complex quasi-equilibria evolve on a slow time scale that tracks the slow depletion of substrate. Though we do not, in general, expect the enzyme and complex dynamics in the open system to be as strictly governed by quasi-steady states after the initial transient period, we may calculate the quasi-steady states just as for the closed reaction.
 
We begin by observing that the dominant characteristic decay time for equations (\ref{eq:finale}) and (\ref{eq:finalc}) is $t_{f}$. At the end of the initial transient period, when $\Order{t_f}$ time has elapsed,  those terms multiplied by $e^{\Rm t}$ approach zero, while (assuming $\delta \ll 1$ as above) $e^{\Rp t}$ remains near unity. Hence for $\bar{E}$, the quasi-steady state of the enzyme concentration, we find that at the end of the initial transient period
\begin{eqnarray}
\bar{E}  &\approx& A +\KE\nonumber \\ 
\ &\approx& \frac{-\Kk + \Knk E_0 + k_1E_0S - \Rm (\Kk/\Knk - E_0)}{\Rm} + \frac{\Kk}{\Knk}\\
&=&  \frac{-\Kk + \Knk E_0 + k_1E_0S}{-(k_{-1}+k_2 + k_1S + \Knk)} + E_0\\
&=& \frac{E_0\KM + \Kk/k_1}{S+\KM + \Knk/k_1}\label{eq:enzymebar}
\end{eqnarray}

\noindent The above expression bears close resemblance to the quasi-steady state one computes for the closed reaction, $\bar{E} = E_0\KM/(S_0 + \KM)$, with two intuitive differences. First, the putative quasi-steady state for the open reaction is elevated by $\Kk/k_1$, the ratio of the enzyme influx to the enzyme-substrate binding rate, a quantity representing the net accumulation of free enzyme at equilibrium. Second, the putative quasi-steady state is reduced by $\Knk/k_1$, the ratio of enzyme removal to the enzyme-substrate binding rate, which reflects the net subtraction of free enzyme at equilibrium. Whether the putative quasi-steady state for the open system is higher or lower than that of the closed system depends on the relative magnitudes of \Kk\ and \Knk.

Our derivation of $\bar{C}$, the quasi-equilibrium for complex concentration at the end of the initial transient period, proceeds similarly: 
\begin{align}
\bar{C} & \approx k_1S \left( \frac{A}{\Rp + k_{-1} + k_2} + \frac{\Kk/\Knk}{k_{-1} + k_2} \right) \\ 
%
%
&= k_1S \left(\frac{-\Kk + \Knk E_0 + k_1E_0 S}{-(k_{-1}+k_2+k_1S)(k_{-1}+k_{2})} + \frac{E_0 - \Kk/\Knk}{\frac{(k_{-1}+k_2+k_1S)(k_{-1}+k_{2})}{k_{-1}+k_2 + k_1S + \Knk}} + \frac{\Kk/\Knk}{k_{-1}+k_2} \right) \\
%
%
&= k_1S \left(\frac{(E_0 - \Kk/\Knk)(k_{-1} + k_2) - k_1\Kk S/\Knk}{(k_{-1}+k_2 + k_1S)(k_{-1}+k_2)} + \frac{\Kk/\Knk}{k_{-1}+k_2} \right) \\
 &= k_1S \left(\frac{E_0}{k_{-1}+k_2+k_1S} \right) = \frac{E_0S}{S+\KM}\label{eq:complexbar}
\end{align}

\noindent Rather remarkably, this final expression is identical to the quasi-steady state for the closed reaction (\ref{eq:iso-complex-equil}). 

Our computation of the quasi-steady states relies on two assumptions which may not necessarily hold true for the open reaction, depending upon parameter choices. First, we explicitly assumed that free enzyme and complex levels quickly equilibrate relative to one another. If free enzyme is added to or removed from the open system at a rate comparable to the usual enzyme-complex equilibration rate, our assumption does not necessarily hold, in which case the quasi-steady states calculated above are not meaningful for any substantial period. Second, we implicitly assumed that the amount of total enzyme (free and bound) remains essentially constant in the post-transient period. For relatively small values of $\Kk$ and $\Knk$, this assumption may hold nearly true for an extended period, during which time expressions (\ref{eq:enzymebar}) and (\ref{eq:complexbar}) may provide reasonably accurate approximations to the enzyme and complex concentrations of the system. However, even for relatively small values of $\Kk$ and $\Knk$, the accumulating deviation from constant total enzyme level will not remain negligible indefinitely. Thus (\ref{eq:enzymebar}) and (\ref{eq:complexbar}) may not accurately estimate enzyme and complex levels in the long term. Over shorter periods, however, and in the transient period in particular, we may be assured that the approximations from this and prior sections are valid, providing certain conditions are met, as discussed below.

\subsection{Validity conditions}
\label{subsec:validity}

To ensure the accuracy of the inner solutions (\ref{eq:finale}) and (\ref{eq:finalc}), we require foremost that the substrate concentration not deviate significantly from its initial level during the initial transient period. That is, for time $t \leq \Order{t_{f}}$, we require $S(t) \approx S_{0}$, or equivalently, $|\Delta S / S_{0}| \ll 1$. 

During the initial transient period, the total change in substrate concentration $\Delta S$ is coarsely bounded by $\left|\dot{S}_{max}\right| \cdot t_{f}$, the product of the maximum rate of change in substrate and the length of the transient period. Thus using the formula 
\begin{equation}
\left|\frac{\Delta S}{S_0}\right| \approx \frac{1}{S_{0}} \cdot {\left|\frac{dS}{{d\!t}}\right|}_{max} \cdot t_{f}
\label{eq:change}
\end{equation}
\noindent we certainly avoid underestimating the relative change in substrate. In the QSSA for the closed system, it can be shown that $\dot{S} = k_{-1}C(S,t) - k_{1}E(S,t)S$ achieves its maximum at $(S, t) \approx (S_{0}, 0)$. For small values of \Kk\ and \Knk, this is nearly true in the open system as well (our numerical investigations in Section \ref{sec:numerical} demonstrate that this is not always the case, {\it e.g.} when \Knk\ is small and \Kk\ is relatively large), and using this approximation we then find

\begin{equation}
\left|\frac{\Delta S}{S_0}\right| \approx \frac{k_1E_0S_0}{S_0 (k_{-1}+k_2+k_1S+\Knk )} = \frac{E_0}{S+\KM + \Knk /k_1}
\label{eq:change2}
\end{equation}

At this point we define several dimensionless parameters:  
\begin{equation}
\sigma = \frac{S_0}{K_m}, \quad\eta  = \frac{E_0}{K_m}, \quad \kappa = \frac{k_{-1}}{k_2}, \quad \alpha = \frac{\Knk }{k_1K_m}, \quad \rho = \frac{k_3}{k_1K_m^2}
\label{eq:nondimpars}
\end{equation}
\noindent Here $\alpha$ and $\rho$ augment the parameter set ($\sigma$, $\eta$, $\kappa$) familiar from the canonical nondimensionalization of the closed system reaction \cite{Segel:1989}. These dimensionless parameters may be used in conjunction with the rescaled variables $e = E/\bar{E}$, $c = C/\bar{C}$, $s = S/S_{0}$, $\tau = t/t_{f}$ to nondimensionalize the equations for the open system (\ref{eq:input}). 

Rewriting (\ref{eq:change2}) in dimensionless form, the condition $|\Delta S / S_{0}| \ll 1$ becomes
\begin{equation}
\eta \ll 1+ \sigma + \alpha
\label{eq:def_eps}
\end{equation}

\noindent When this inequality holds, we may be confident of the accuracy of our inner solutions during the initial transient period. We note that (\ref{eq:def_eps}) is weaker than the corresponding inequality for the closed system, namely $\eta \ll 1+ \sigma$ \cite{Segel:1989}. (Furthermore, the two inequalities coincide as $\Knk$ goes to zero, that is, for negligible enzyme removal.) Thus in any parameter regime for which the standard QSSA is valid, equations (\ref{eq:finale}) and (\ref{eq:finalc}) hold (with similar approximation errors) for $t \leq \Order{t_{f}}$. Numerical simulations support this conclusion, and show additionally that within a broad range of parameter values the approximations remain accurate well beyond time $t_{f}$.

\section{Numerical results}
\label{sec:numerical}
Subject to some constraints on rate constants, we expect the approximations derived in Sections \ref{subsec:enzymeapprox} and \ref{subsec:complexapprox} to provide good estimates of the enzyme and complex levels of the open system during the transient phase. It is not entirely clear from the derivation just how long the key assumption, namely that $S \approx S_{0}$, remains valid, nor how the domain of validity is affected by the rates of enzyme addition and removal, especially when one or both of $\Kk, \Knk $ approach the magnitude of the other reaction parameters. To gauge the efficacy of our approximations for different rates of enzyme input and removal, we numerically computed solutions for the full set of ODEs (\ref{eq:input}) of the open system and for the approximate open system (\ref{eq:finale}), (\ref{eq:finalc}), (\ref{eq:approxSP}) for broad ranges of values for the rates of enzyme addition and removal (parameters $\Kk$, $\Knk$, respectively). For the approximate open system, a differential-algebraic system of equations, we used the analytic expressions derived for $E(t)$ and $C(t)$ in the transient phase (equations (\ref{eq:finale}) and (\ref{eq:finalc}), respectively), and we numerically solved the (nonautonomous) differential equations (\ref{eq:approxS}) and (\ref{eq:approxP}), in which  $E(t)$ is given by equation  (\ref{eq:finale}) and $C(t)$ is given by equation (\ref{eq:finalc}). We also computed solutions to the closed system (\ref{eq:iso}) in order to better assess the utility of our approximation. Numerical integration was done in MATLAB using the $\mathtt{ode15s}$ stiff integrator with relative and absolute error tolerances of $10^{-10}$. We consider two regimes of initial conditions, the {\it standard QSSA regime} with $S_{0} \gg E_{0}$, and the {\it reverse QSSA regime} with $E_{0} \gg S_{0}$. The simulated initial conditions were $(S_{0}, E_{0}) = (1,0.1)$ and $(S_{0}, E_{0}) = (0.1,1)$ for the standard and reverse QSSA regimes, respectively.


%


In each regime of initial conditions, we perform a set of one-dimensional parameter sweeps: For each value $K \in \{10^{-4}, 0.1, 1\}$ (a sampling of parameter values which is representative of the range of different scenarios of enzyme addition and removal), we fix a parameter $k_{i}, i \in \{-3, 3\}$ at $k_{i} = K$, and we simulate the open system (both the full set of ODEs and our approximation) for 1600 values of $k_{-f}$ logarithmically spaced in the interval $[0.0001, 5]$. We display the portion of each simulation run spanning the time interval $[0,1]$ ($t_{f}) \ll 1$ for all parameter combinations considered), and we note that $t_{f}$ (the fast complex time scale) is less than 1 for every pair of $\Kk$, $\Knk$ values used. For the results presented below, we fixed the other parameters (which determine \KM) at the same standard values in each simulation, namely $k_{1} = k_{-1} = k_{2} = 1$.  (We also performed the same computations using other values for $k_{1}, k_{-1}, k_{2}$ in order to sample a range of \KM\ values, and we obtained results very similar to those presented below.) We did not exhaustively search the parameter space, but the selected examples we show give a comprehensive picture of the behavior of the full open system and our approximation under a wide range of conditions, including biologically relevant parameter ranges. 

\paragraph{Solution manifolds} We present our numerical results in Figures \ref{fig:Q_SE_kn3_d}--\ref{fig:Q_ES_k3_a} below, with each figure displaying the results of one parameter sweep (1600 individual simulations). We first organize the figures according to initial conditions (standard followed by reverse QSSA regimes), and then for each regime of initial conditions, we group the figures according to which of $\Kk, \Knk$ is the `free' parameter being varied in the sweep (first varying \Knk, then varying \Kk). A solution to either initial value problem (\ref{eq:input}) or (\ref{eq:finale}), (\ref{eq:finalc}), (\ref{eq:approxSP}) for a fixed value of the free parameter traces a one-dimensional curve in the four-dimensional phase space $(E, C, S, P)$, and if we smoothly vary the free parameter, we obtain a two-dimensional manifold of solutions in the five-dimensional space $(E, C, S, P, k_{i})$, where $k_{i} \in \{\Kk, \Knk\}$ is the free parameter. The results of a parameter sweep provide an approximation to these manifolds, and each figure depicts the solution manifolds for (\ref{eq:input}) and (\ref{eq:finale}), (\ref{eq:finalc}), (\ref{eq:approxSP}) for a single parameter sweep. Each figure comprises five subplots, and each of the first three subplots shows the projections of the solution manifolds onto one of the subspaces $(E, S, k_{i})$, $(C, S, k_{i})$, and $(E, C, k_{i})$. These projections may be thought of as the union, taken over the set of values swept for the free parameter $k$, of the two-dimensional phase plane diagrams of the solutions obtained at fixed values of $k_{i}$, for the $(E,S)$, $(C,S)$, and $(E,C)$ planes, respectively. We omit projections involving the phase variable $P$, since the equations for product formation decouple in (\ref{eq:iso}), (\ref{eq:input}) and (\ref{eq:finale}), (\ref{eq:finalc}), (\ref{eq:approxSP}), and thus this variable provides essentially redundant information.

Each three-dimensional subplot depicts three surfaces: the solution manifold of the full open system (\ref{eq:input}), denoted \MFN, in red; the solution manifold for the approximate open system (\ref{eq:input}) and (\ref{eq:finale}, \ref{eq:finalc}, \ref{eq:approxSP}), denoted \MFA, in blue; and the manifold of the full solution to the closed system (\ref{eq:iso}), denoted \MFI, in green\footnote{Neither \Kk\ nor \Knk\ appear in (\ref{eq:iso}), so the solution manifold \MFI\ is simply $[0.01, 5] \times \mathcal{C}$, where $\mathcal{C}$ denotes the solution of the initial value problem (\ref{eq:iso}) over the time interval $[0,1]$ (with $k_{1} = k_{-1} = k_{2} = 1$, and with initial conditions corresponding to the either the standard or reverse QSSA regimes) projected onto the appropriate phase plane.}.

We may think of the closed system (\ref{eq:iso}) as an approximation to the open system (\ref{eq:input}) that presumes the effects of enzyme input and removal to be negligible. For example, if $\Kk, \Knk \ll k_1,k_2,k_{-1}$, input and removal occur on a time-scale much slower than the individual chemical reactions which constitute the conversion of substrate to product. Viewed in this light, the projections of the manifold \MFI\ in Figures \ref{fig:Q_SE_kn3_d}--\ref{fig:Q_ES_k3_a} illustrate the degree to which this assumption of negligibility may mislead. Note that there are only two versions of \MFI, one for the standard QSSA regime and one for the reverse QSSA regime, so that the \MFI\ manifolds shown in Figures \ref{fig:Q_SE_kn3_d}--\ref{fig:Q_SE_k3_a} are identical to one another, as are the \MFI\ manifolds shown in Figures \ref{fig:Q_ES_kn3_d}--\ref{fig:Q_ES_k3_a}. 

In each three-dimensional subplot of Figures \ref{fig:Q_SE_kn3_d}--\ref{fig:Q_ES_k3_a}, the value of the free parameter is measured on a logarithmic scale on the vertical axis, and two specific phase space variables are measured on the two horizontal axes. Time is not measured explicitly as a coordinate in these phase plots, but the temporal orientations of the solution manifolds may be inferred by noting the location of the manifolds' edges corresponding to the initial conditions. For the standard QSSA regime, the solution manifolds ``start'' at the edge corresponding to $(E, C, S, P, k_{i}) = (0.1, 0, 1, 0, k_{i})$, and for the reverse QSSA regime, the solution manifolds start at the edge corresponding to $(E, C, S, P, k_{i}) = (1, 0, 0.1, 0, k_{i})$. The end of the transient period, which occurs at time $t \approx t_{f} < 1$, is indicated by a line drawn on each of the solution manifolds. The standard estimate for $t_{f}$ in the closed system, given by (\ref{eq:orig_tfast}), is drawn as a dashed line, and the estimate for $t_{f}$ derived in this paper for the open system, given by (\ref{eq:tfast}), is drawn as a solid line. For visual ease, both the dashed and solid lines are drawn  on the \MFN\ (red) and \MFA\ (blue) manifolds in black, and on the \MFN\ (green) manifold in magenta.

\paragraph{Approximation error} We have mentioned repeatedly that QSSA-based analyses presume $S\approx S_{0}$ during the pre-steady-state transient. The fractional decrease in substrate during this period, $\delta_{S} = (S(t_{f}) - S_{0})/S_{0}$, provides one measure of the error for the QSSA approximation. As discussed in \cite{Segel:1989}, given a prescribed tolerance for $\delta_{S}$, one useful heuristic for gauging whether the QSSA can be used (for the closed system) is the condition 

\begin{equation}
\label{eq:iso-tolerance}
\frac{E_{0}}{\KM + S_{0}} = \frac{\eta}{1+\sigma} < \delta_{S}
\end{equation}

\noindent For the open reaction, this condition becomes
\begin{equation}
\label{eq:noniso-tolerance}
\frac{E_{0}}{\KM + S_{0} + \Knk/{k_{1}}} = \frac{\eta}{1+\sigma+\alpha} < \delta_{S}
\end{equation}

\noindent Condition (\ref{eq:noniso-tolerance}) is weaker for the open system than condition (\ref{eq:iso-tolerance}) is for the closed system; both conditions imply in particular that the approximations should be accurate in the standard QSSA regime. Furthermore, we may expect the inner approximation for the open system to be more accurate at higher rates of enzyme removal. 

For each parameter sweep we present, we also plot a different measure of the approximation error. The fourth (two-dimensional) subplot in each of Figures  \ref{fig:Q_SE_kn3_d}--\ref{fig:Q_ES_k3_a} records the {\it signed} difference between the actual solution to (\ref{eq:input}) and the approximate solution (\ref{eq:finale}, \ref{eq:finalc}, \ref{eq:approxSP}). The fifth (two-dimensional) subplot in each of Figures  \ref{fig:Q_SE_kn3_d}--\ref{fig:Q_ES_k3_a} records the signed difference between the solution to the full open system (\ref{eq:input}) and the  solution of the full closed system (\ref{eq:iso}). In both cases, the signed error is taken at the end of the transient period, time $t_{f}$, where this time is given by the new estimate (\ref{eq:tfast}):

\begin{equation}
\label{eq:signed-error}
E_{X}(t_{f}) = X(t_{f}) - \hat{X}(t_{f})
\end{equation}

\noindent Here $X$ is one of $E, C, S$; $X(t_{f})$ denotes the $X$ component of the actual solution to (\ref{eq:input}) at time $t_{f}$; and $\hat{X}(t_{f})$ denotes the $X$ component either of the approximation (\ref{eq:finale}), (\ref{eq:finalc}), (\ref{eq:approxSP}) or the full closed system at time $t_{f}$. The signed error is calculated for each simulation as the free parameter $k_{i} \in \{\Kk, \Knk\}$ is varied. We use the signed difference between the actual and approximate solutions, rather than the absolute value of their difference, in order to make clear when the approximation underestimates or overestimates the true values of $E, C, S$.

\newcommand{\gwidth}{2in}
\newcommand{\ewidth}{1.4in}


\subsection{Standard QSSA regime}
When  $S_{0} \gg E_{0}$, our validity condition for inner approximation (\ref{eq:finale}), (\ref{eq:finalc}) is satisfied by a substantial margin. As reflected in the proximity of the red and blue manifolds in Figures \ref{fig:Q_SE_kn3_d}--\ref{fig:Q_SE_k3_a}, for a broad range of parameters our inner approximation provides a highly accurate estimate of the behavior of the open system up to, and often well past, times $\Order{t_{f}}$. Furthermore, the closed system poorly approximates the behavior of the open system, especially for non-negligible levels of enzyme input. 

\paragraph{Varying enzyme removal rate} The limitations of the closed system for approximating the open system are quite apparent in simulations where \Knk\ varies, shown in Figures \ref{fig:Q_SE_kn3_d}--\ref{fig:Q_SE_kn3_a}. When the rates of enzyme input and enzyme removal are both very low, the closed system does match the behavior of the open system well: the manifolds \MFN\ (red) and \MFI\ (green), as well as \MFA\ (blue), all lie very close together for values of \Knk\ two orders of magnitude and more below the value of \KM, as seen in Figure \ref{fig:Q_SE_kn3_d}. Above that rate of enzyme removal, the behavior of the open and closed systems begin to deviate substantially, with the open system overestimating the level of free enzyme, as is to be expected. Total enzyme levels decline as free enzyme is removed from the system relatively rapidly, rather than remaining approximately constant. Specifically, the results recorded in Figure \ref{fig:SE_kn3_d_ErrC} indicate that the signed error between the open and closed systems at time $t_{f}$ is $\Order{10^{-4}}$ for $\Knk = 10^{-3}$ and increases to $\Order{10^{-2}} $ as $\Knk$ approaches 1, with the sign of the error in the $E$ and $C$ components always positive and that of the $S$ component always negative.

At more rapid rates of enzyme input, the shape of \MFN\ differs greatly from that of \MFI, indicating that the closed system poorly approximates the behavior of the closed system at almost any level of enzyme removal. This can be seen in Figure \ref{fig:Q_SE_kn3_b}, in which the the rate of enzyme input is  an order of magnitude lower than the other reaction rate constants.  Here \MFN\ and \MFI\ intersect for at a unique value of \Knk (from Figure \ref{fig:SE_kn3_b_ErrC}, at $\Knk \approx 1$); at this parameter combination, the total amount of bound and unbound enzyme in the open system remains roughly constant over the transient period, as in the closed system. Elsewhere, the ratio between \Kk\ and \Knk\ determines whether the closed system over- or under-estimates reactant levels for the open reaction. The position of \MFI\ vis-\`{a}-vis \MFN\ reflects underestimation of enzyme and complex levels and overestimation of substrate levels when enzyme input is high and enzyme removal is low. Enzyme accumulates in free and bound form, rather than remaining constant. When enzyme input is low and enzyme removal is high, the manifold configuration is reversed. In both cases, the numerical results confirm our intuition regarding the behavior of the closed system considered as an approximation to the open reaction. In Figure \ref{fig:Q_SE_kn3_a}, which depicts results when the rate of enzyme input matches the other reaction rates, we see that for all values of \Knk, \MFN\ and \MFI\ intersect only at $(E_{0}, C_{0}, S_{0})$. The closed system always underestimates enzyme and complex levels and overestimates substrate levels (Figure \ref{fig:SE_kn3_a_ErrC}). The range of absolute discrepancies between the full open and closed systems grows from $\Order{10^{-3}}$  to $\Order{10^{-1}}$ as \Knk\ increases from $0.0001$ to $1$ (Figures \ref{fig:SE_kn3_d_ErrC}--\ref{fig:SE_kn3_a_ErrC}).

In each of Figures \ref{fig:Q_SE_kn3_d}--\ref{fig:Q_SE_kn3_a}, the manifolds \MFA\ and \MFN\ lie very close to one another for all values of \Knk, indicating that the inner approximation to the open system is very accurate, even when the rate of enzyme input is very rapid. In particular, the magnitude of the signed error between the full system and the inner approximation at $t_{f}$ recorded in Figures \ref{fig:SE_kn3_d_Err}--\ref{fig:SE_kn3_a_Err} is small, ranging only from $\Order{10^{-4}}$ when $\Knk=0.0001$ to $\Order{10^{-3}}$ when $\Knk=1$. The position of the \MFA\ manifold relative to the \MFI\ manifold reflects consistent underestimation of the level of enzyme and overestimation of the levels of complex and substrate. Since the inner approximation was derived by assuming constant substrate, the magnitude of $\Rm$, the dominant, negative exponent for enzyme and complex dynamics in Equations (\ref{eq:finale}) and (\ref{eq:complex}), does not decline as the substrate level falls. The decaying exponential component of Equation (\ref{eq:finale}) is overestimated, so the predicted enzyme level is lower than the actual enzyme concentration obtained from the full set of ODEs (\ref{eq:input}) of the open system. Furthermore, the approximate complex level given in Equation (\ref{eq:complex}) is directly proportional to $S_{0}$, and hence is overestimated. The substrate level, obtained via Equation (\ref{eq:approxS}), declines at a rate proportional to the enzyme level, and so it is underestimated in tandem with the overestimation of enzyme concentration. We note that the magnitudes of the signed errors in $E, S, C$ decrease as \Knk\ grows, particularly the signed errors in  enzyme and complex levels. This accords with the behavior of the error condition (\ref{eq:noniso-tolerance}) as \Knk\ increases, discussed above.
 
The curvature of the \MFN\ and \MFA\ manifolds seen in each of Figures \ref{fig:Q_SE_kn3_d}--\ref{fig:Q_SE_kn3_a} at higher levels of \Knk\ also accords with our intuition. For a given rate of enzyme addition, \Kk, we expect higher rates of enzyme removal to lead to lower levels of free enzyme; hence the $E$-$C$ and $E$-$S$ projections of \MFN\ and \MFA\ fold in the direction of lower enzyme levels as \Knk\ increases. Furthermore, more rapid enzyme depletion reduces the amount of enzyme available to catalyze the conversion of substrate to product, and thus at larger values of \Knk\, the level of substrate remaining at $t_{f}$ is greater. 


\begin{figure}[h]
 	\centering
	\subfloat[][Substrate-Enzyme]{\label{fig:SE_kn3_d_SE}\includegraphics[width= \gwidth]{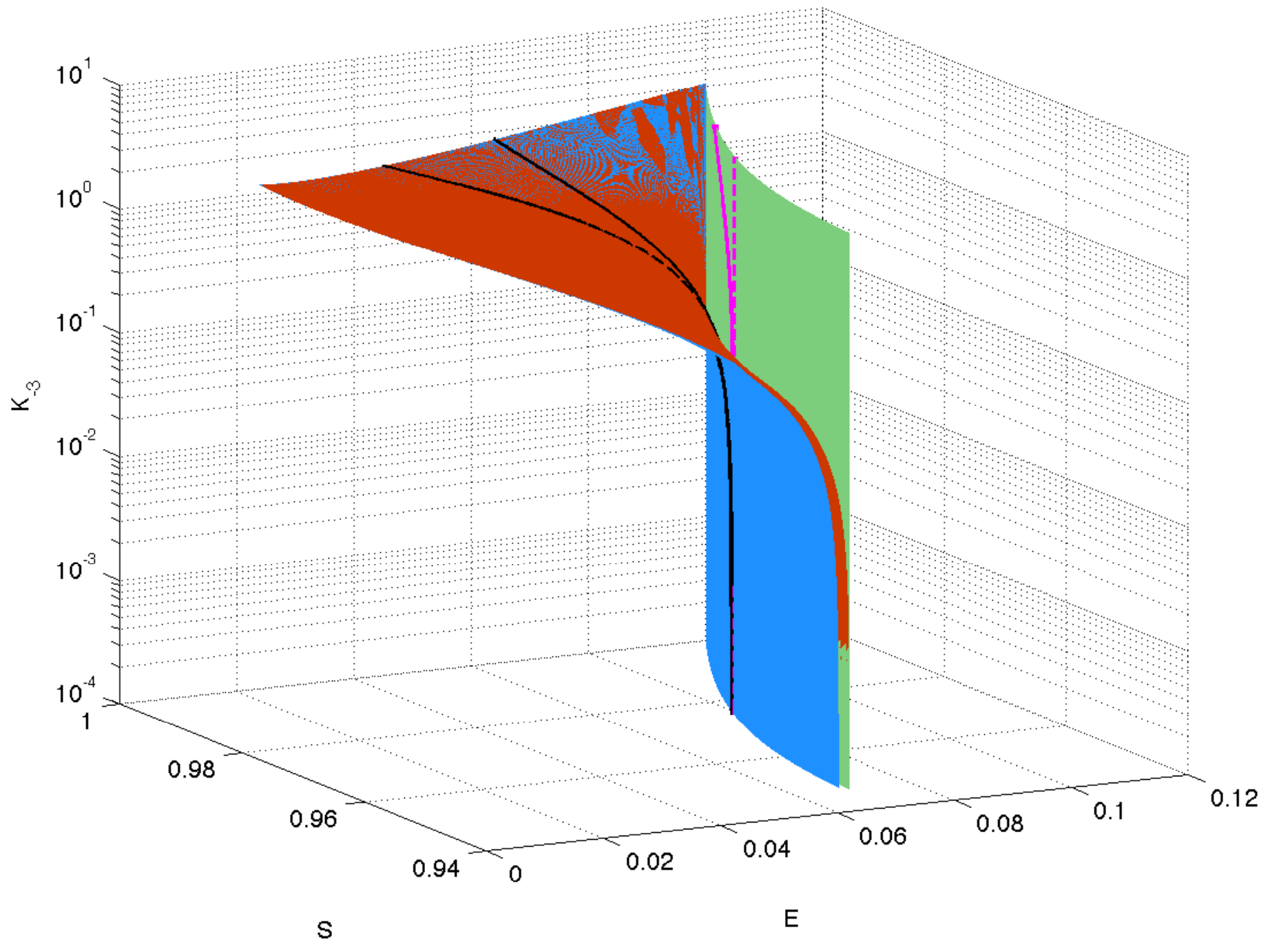}}
	\subfloat[][Complex-Enzyme]{\label{fig:SE_kn3_d_CE}\includegraphics[width= \gwidth]{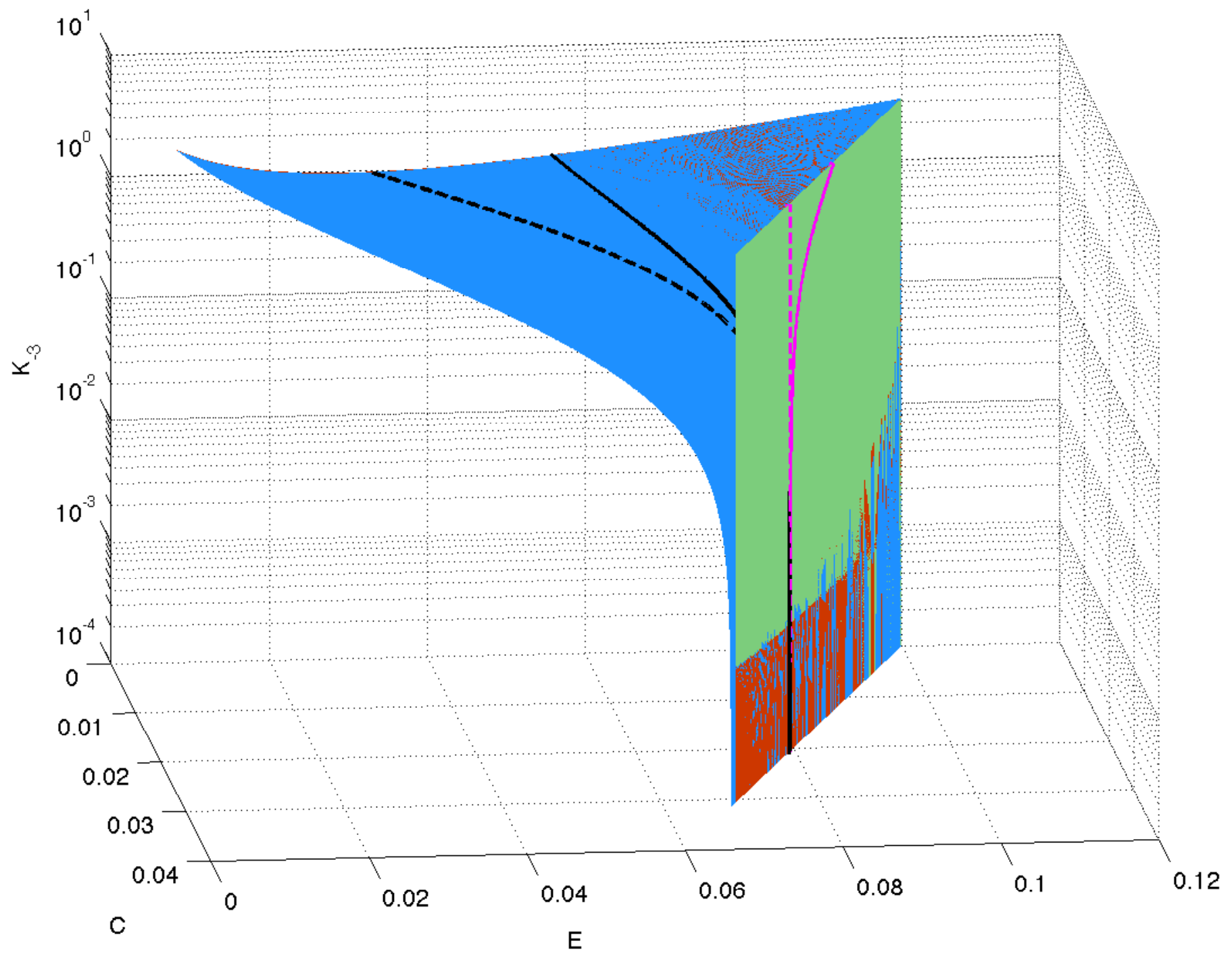}}
	\subfloat[][Complex-Substrate]{\label{fig:SE_kn3_d_CS}\includegraphics[width= \gwidth]{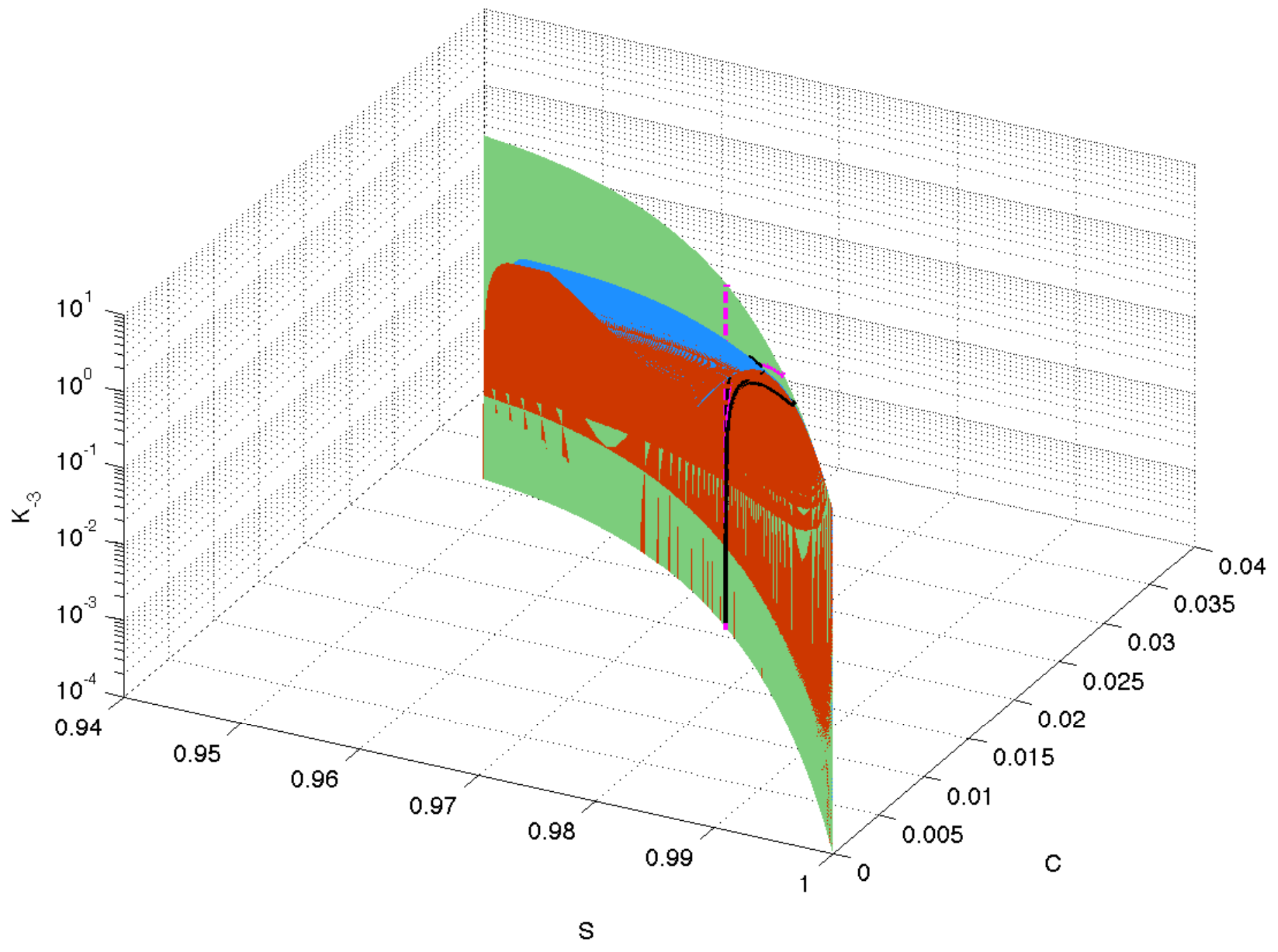}}\\
	\subfloat[][Signed error at $t_{f}$, full open \\system vs. open approximation]{\label{fig:SE_kn3_d_Err}\includegraphics[width = \ewidth]{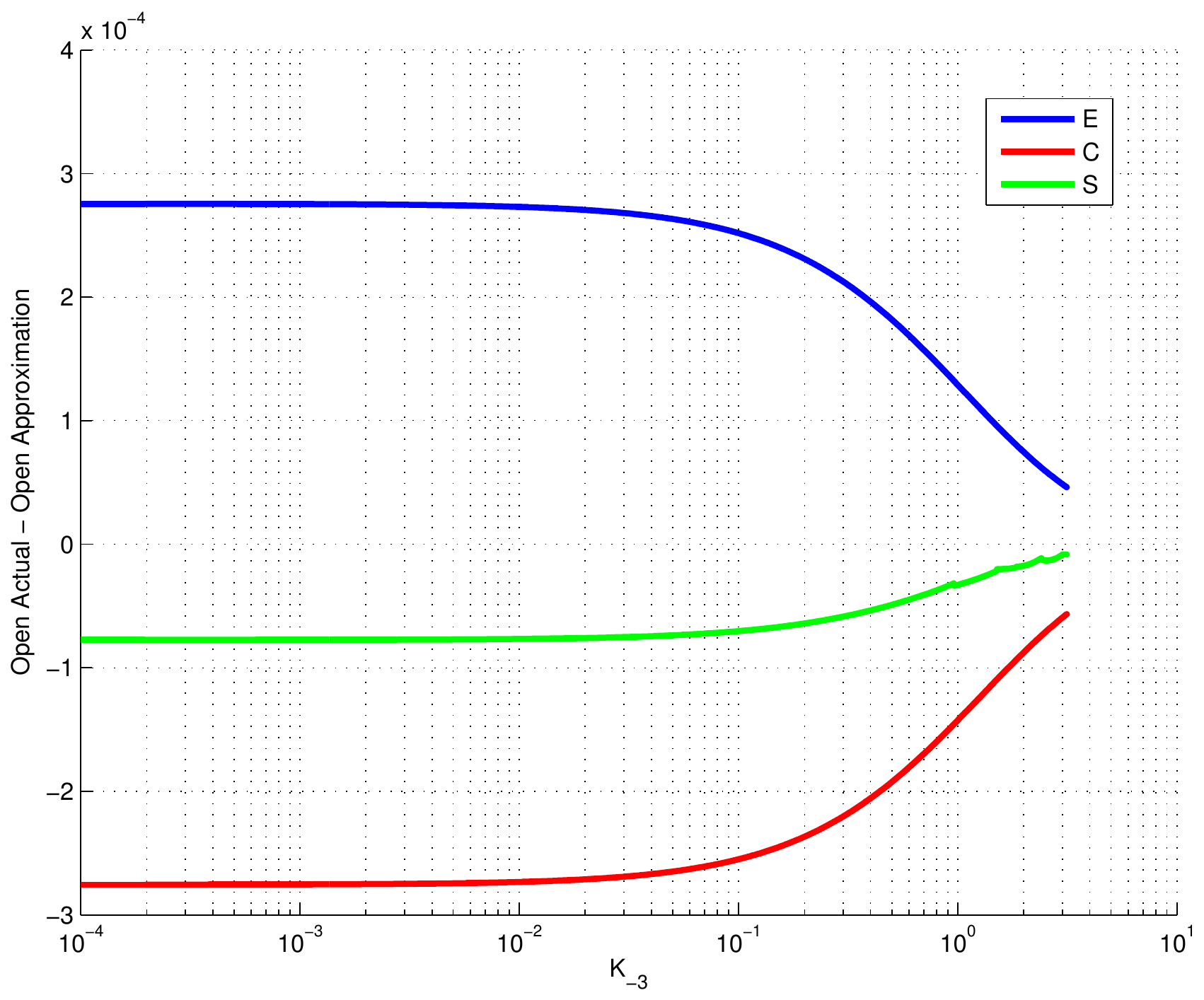}}\hspace{0.5cm}
	\subfloat[][Signed error at $t_{f}$, full open system vs. full closed system]{\label{fig:SE_kn3_d_ErrC}\includegraphics[width = \ewidth]{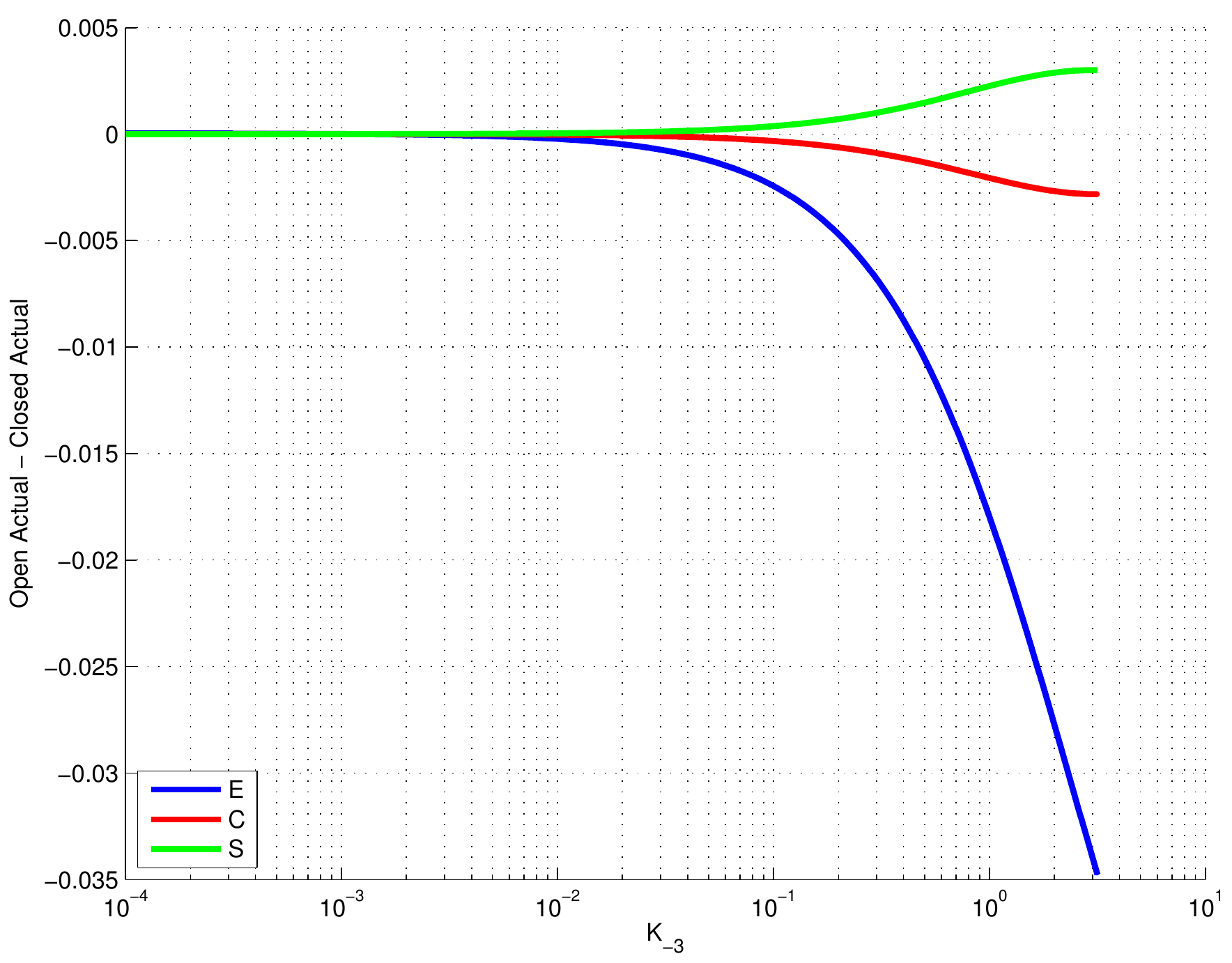}}\\
	\caption[Standard QSSA regime, varying $\Knk $. $\Kk = 0.0001$]{Phase plane portraits and signed errors in the standard QSSA regime as \Knk\ varies, with $\Kk = 0.0001$.}
	\label{fig:Q_SE_kn3_d}
\end{figure}

\begin{figure}[!ht]
 	\centering
	\subfloat[][Substrate-Enzyme]{\label{fig:SE_kn3_b_SE}\includegraphics[width= \gwidth]{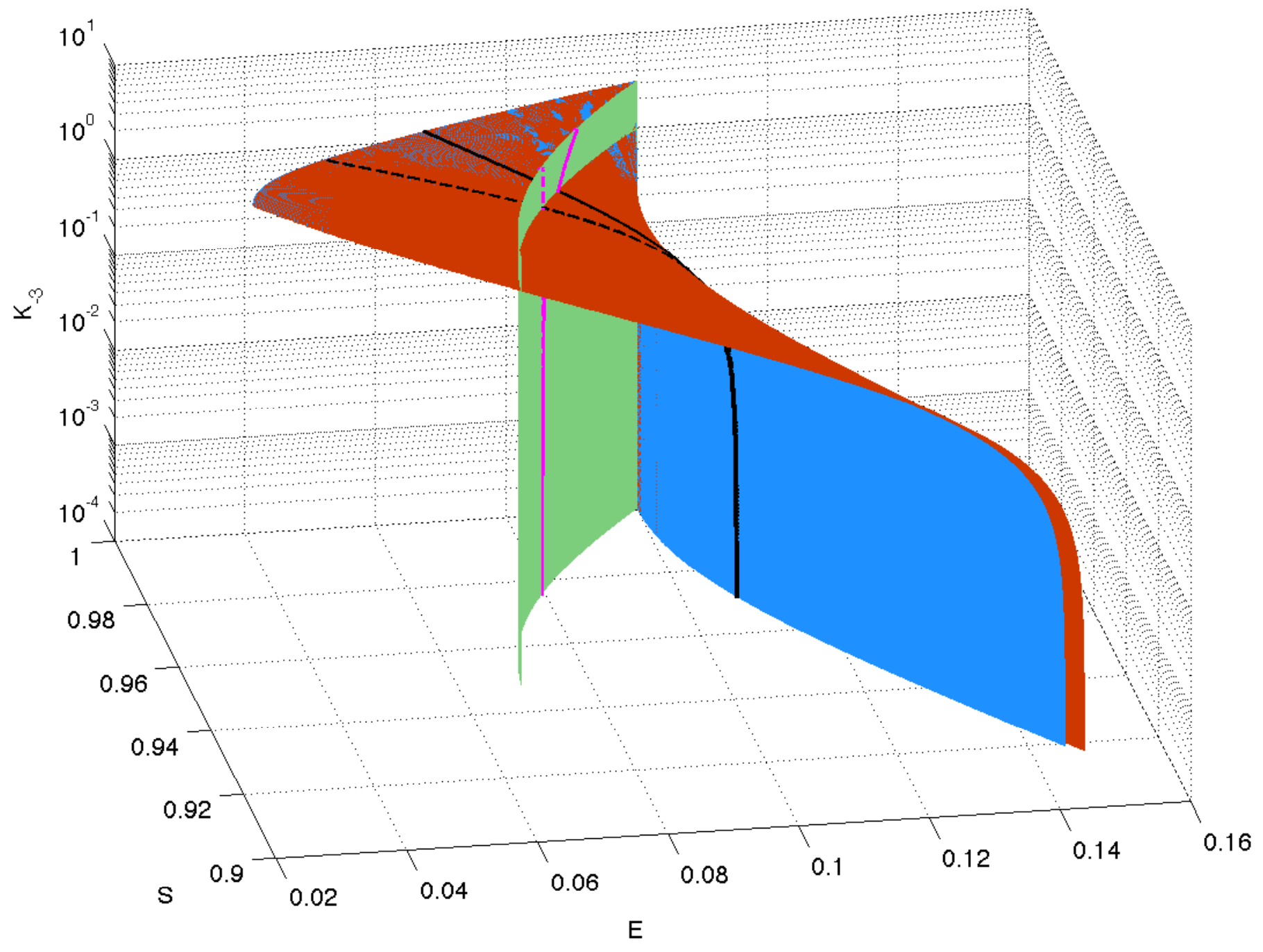}}
	\subfloat[][Complex-Enzyme]{\label{fig:SE_kn3_b_CE}\includegraphics[width= \gwidth]{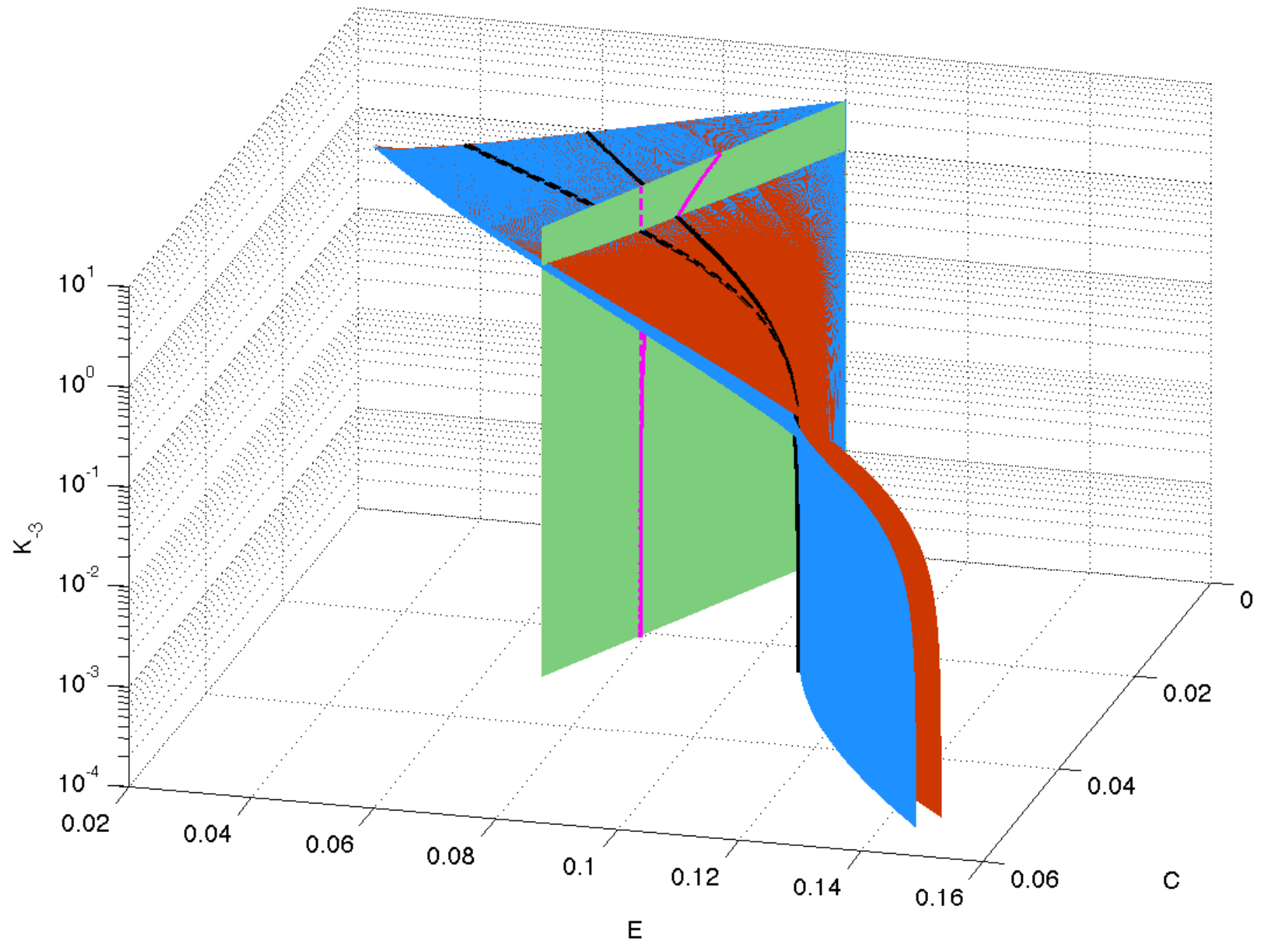}}
	\subfloat[][Complex-Substrate]{\label{fig:SE_kn3_b_CS}\includegraphics[width= \gwidth]{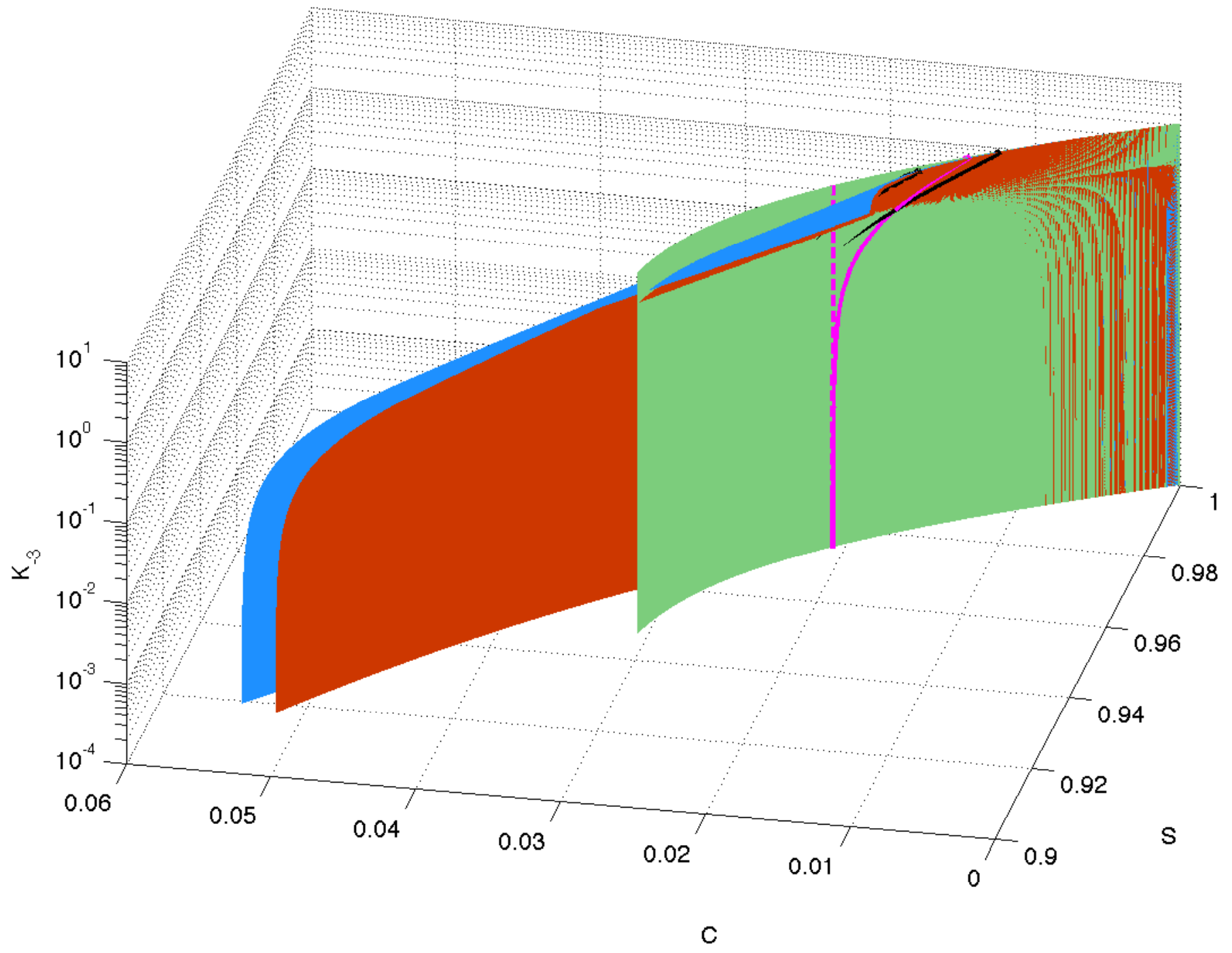}}\\
	\subfloat[][Signed error at $t_{f}$, full open \\system vs. open approximation]{\label{fig:SE_kn3_b_Err}\includegraphics[width = \ewidth]{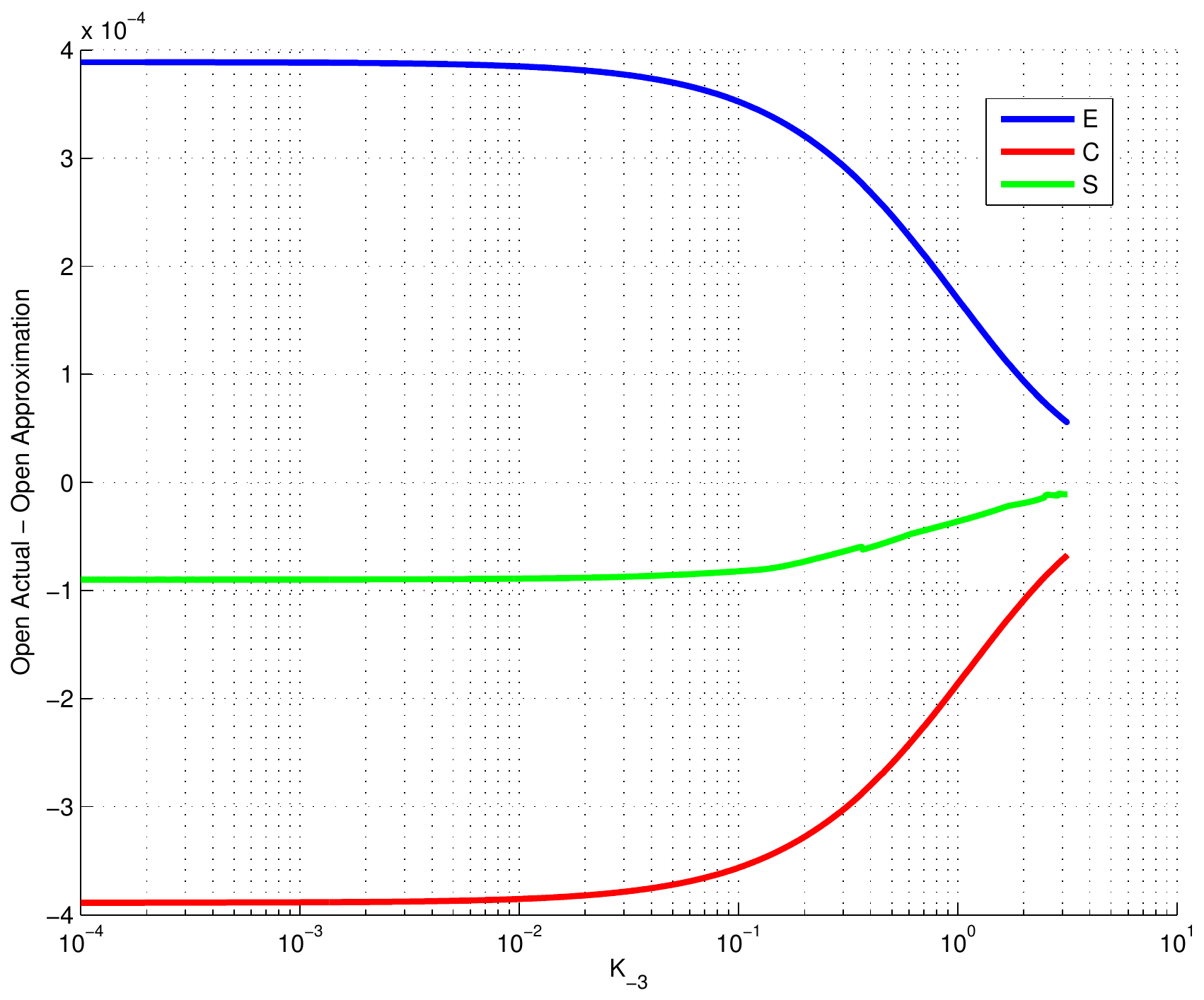}}\hspace{0.5cm}
	\subfloat[][Signed error at $t_{f}$, full open system vs. full closed system]{\label{fig:SE_kn3_b_ErrC}\includegraphics[width = \ewidth]{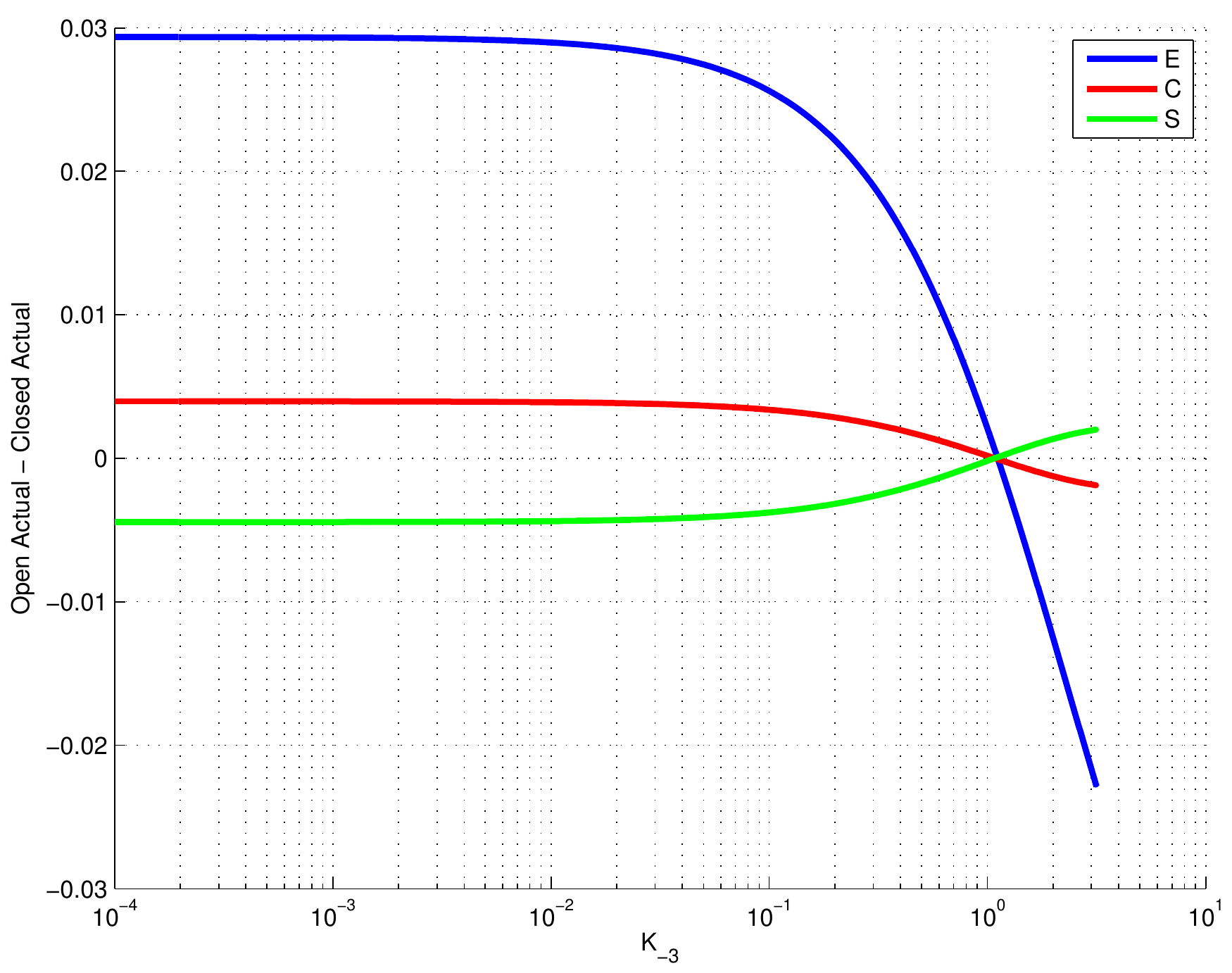}}\\
	\caption[Standard QSSA regime, varying $\Knk $. $\Kk = 0.1$]{Phase plane portraits and signed error in the standard QSSA regime as \Knk\ varies, with $\Kk = 0.1$.}
	\label{fig:Q_SE_kn3_b}
\end{figure}

\begin{figure}[!ht]
 	\centering
	\subfloat[][Substrate-Enzyme]{\label{fig:SE_kn3_a_SE}\includegraphics[width= \gwidth]{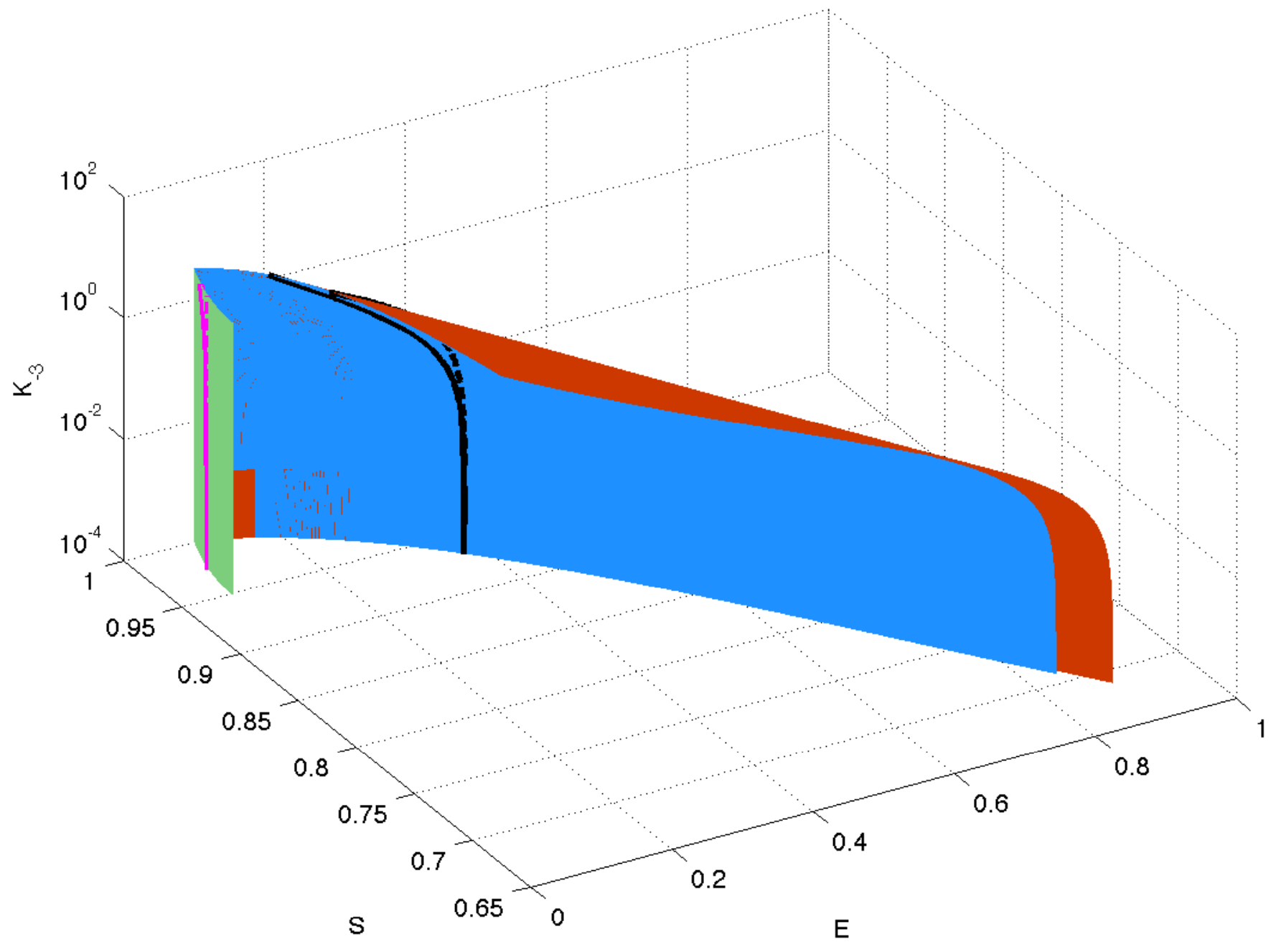}}
	\subfloat[][Complex-Enzyme]{\label{fig:SE_kn3_a_CE}\includegraphics[width= \gwidth]{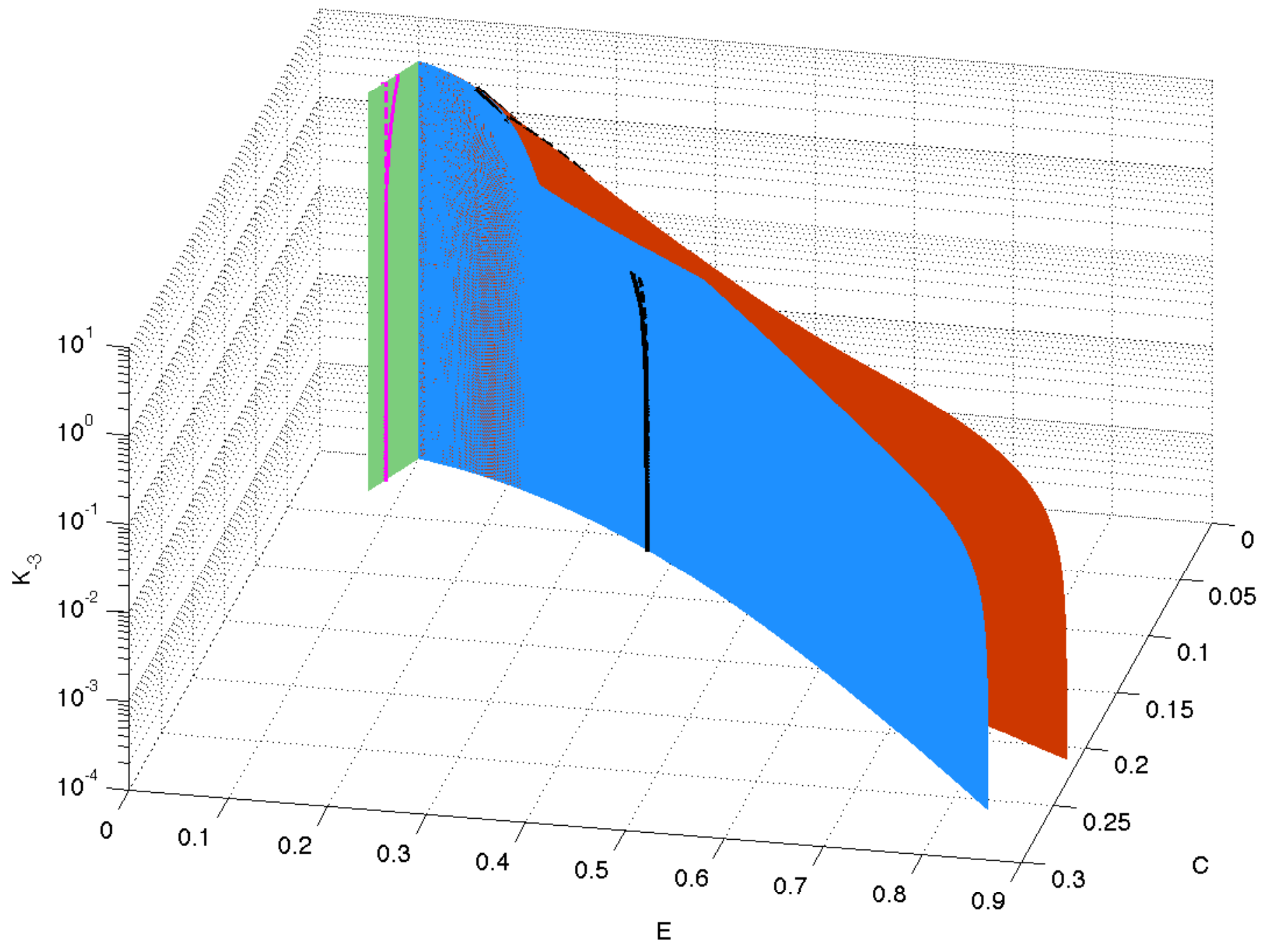}}
	\subfloat[][Complex-Substrate]{\label{fig:SE_kn3_a_CS}\includegraphics[width= \gwidth]{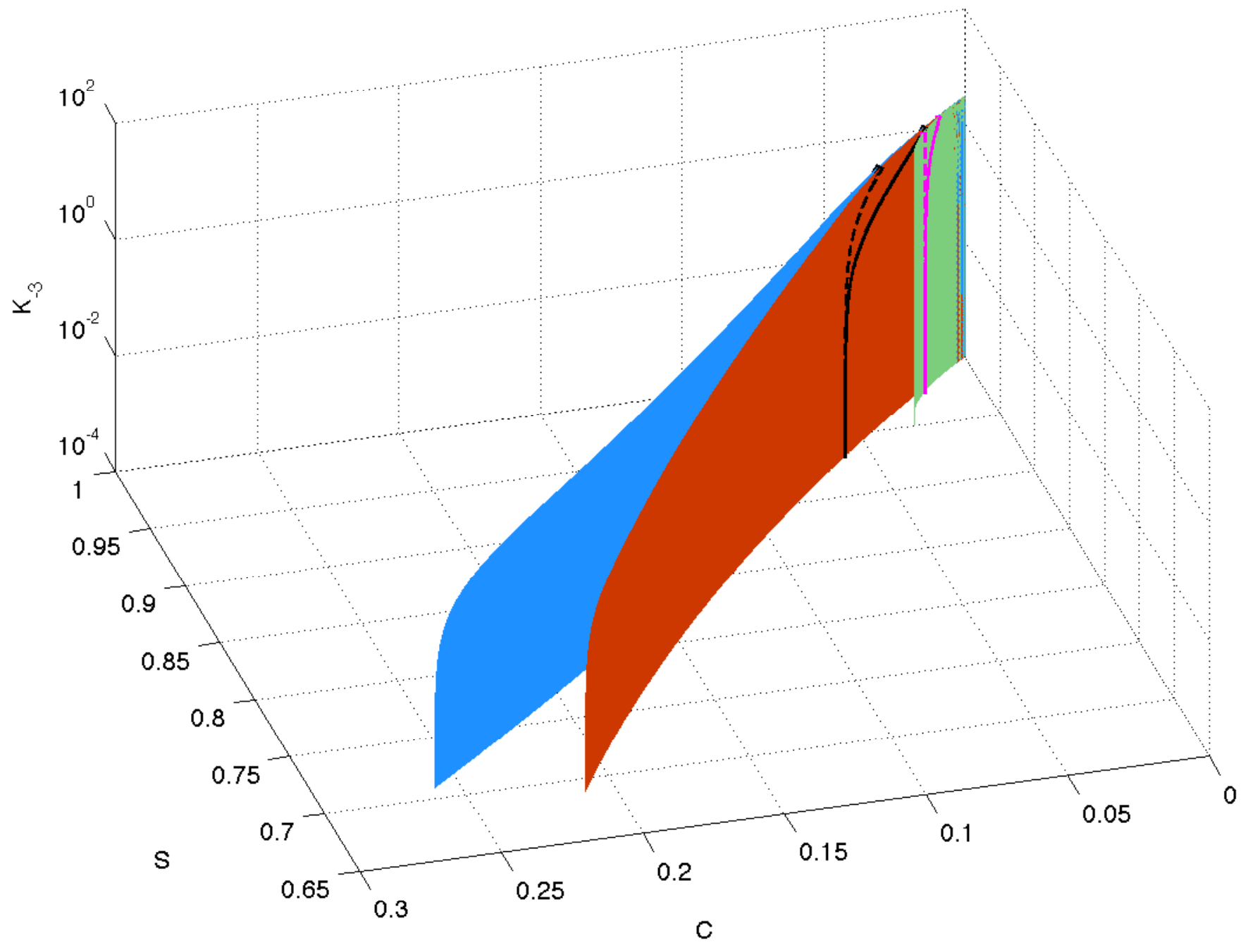}}\\
	\subfloat[][Signed error at $t_{f}$, full open \\system vs. open approximation]{\label{fig:SE_kn3_a_Err}\includegraphics[width = \ewidth]{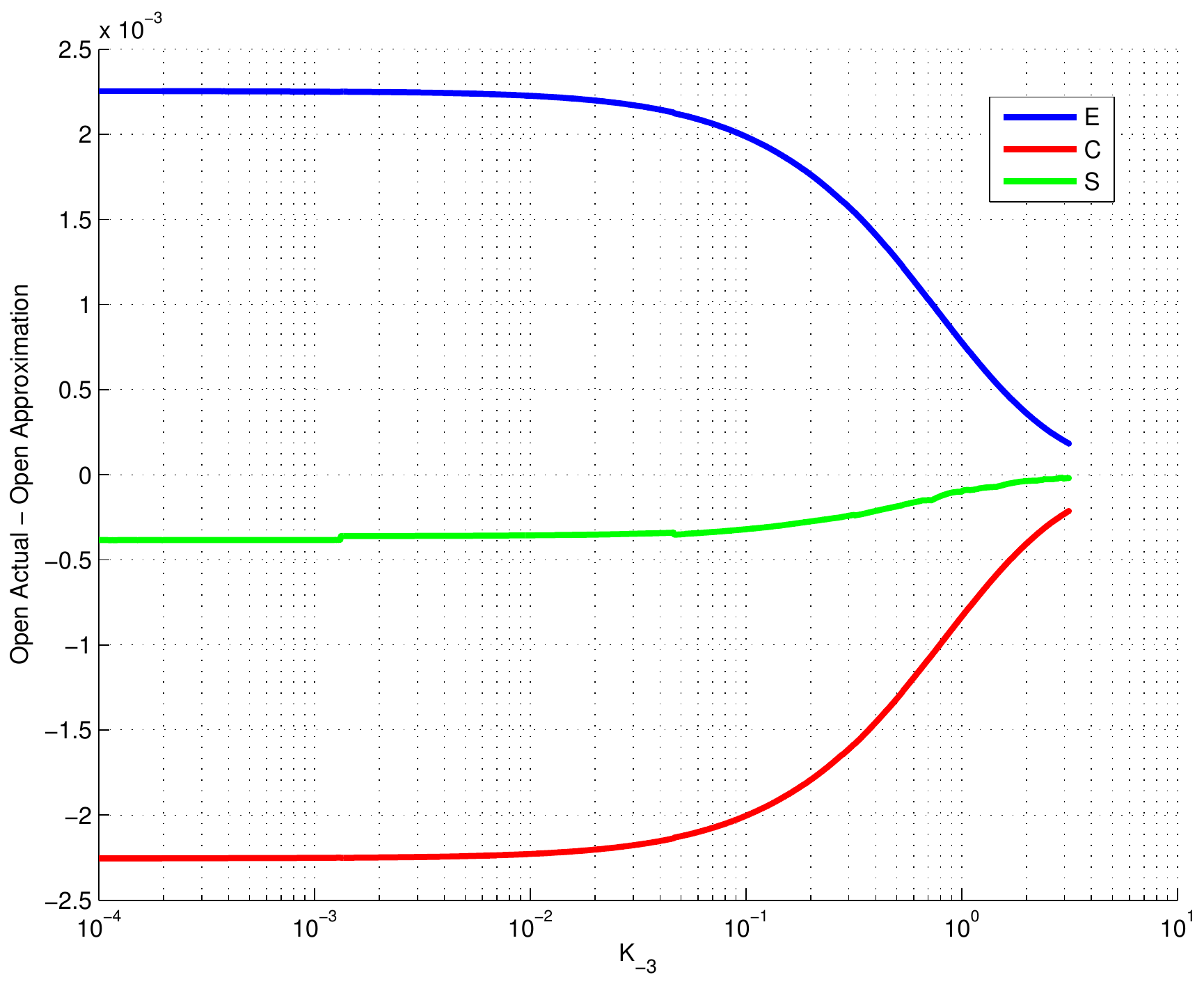}}\hspace{0.5cm}
	\subfloat[][Signed error at $t_{f}$, full open system vs. full closed system]{\label{fig:SE_kn3_a_ErrC}\includegraphics[width = \ewidth]{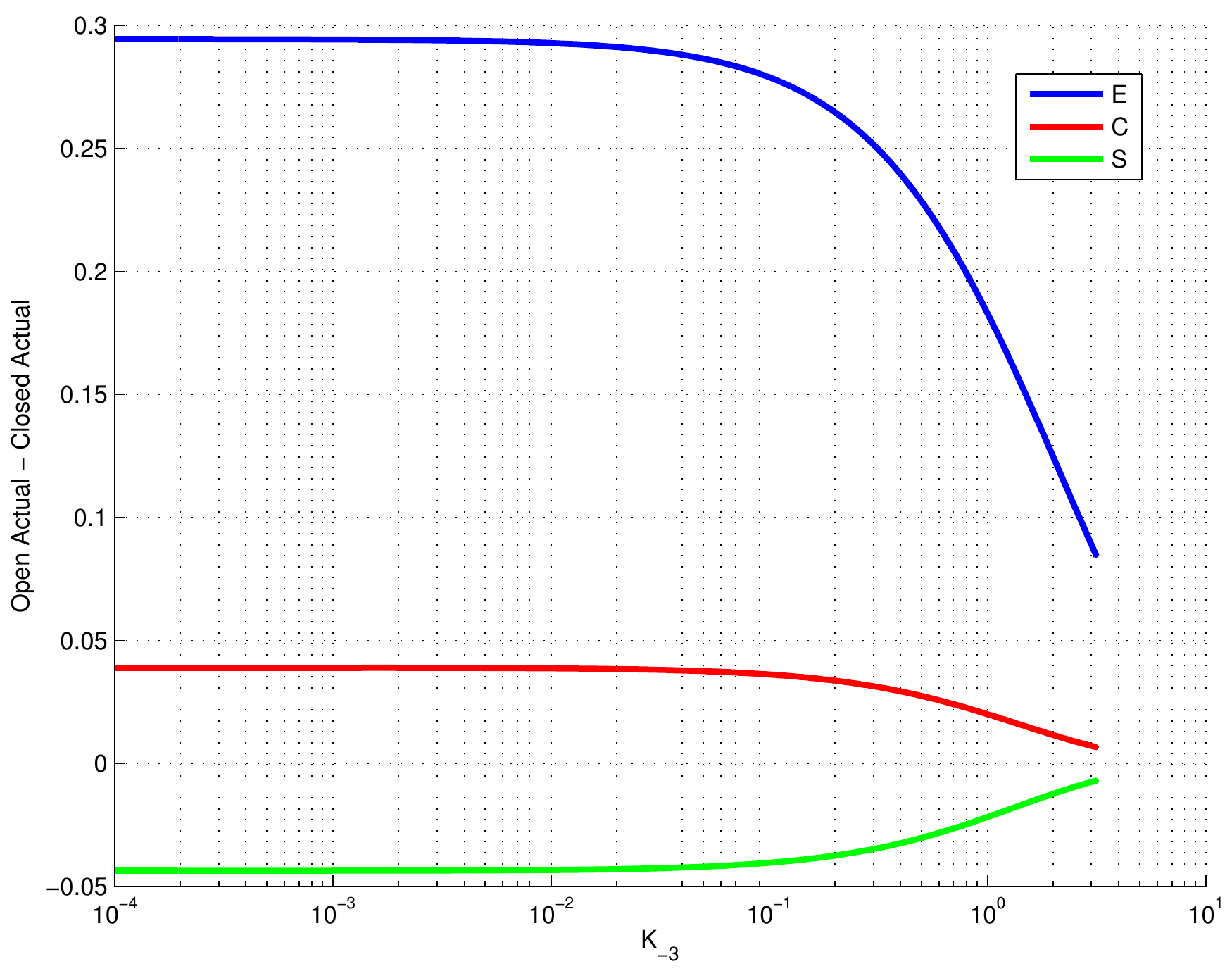}}\\	
	\caption[Standard QSSA regime, varying $\Knk $. $\Kk = 1$]{Phase plane portraits and signed error in the standard QSSA regime as \Knk\ varies, with $\Kk = 1$. }
	\label{fig:Q_SE_kn3_a}
\end{figure}

\paragraph{Varying enzyme input rate} Figures \ref{fig:Q_SE_k3_d}--\ref{fig:Q_SE_k3_a} show the solution manifolds \MFN, \MFA, and \MFI\ when \Kk\ is varied and \Knk\ is held fixed. Qualitatively, the respective solution manifolds at different \Knk\ values differ only little from one another;  the chief difference between them is the change in scale. As in the case above, where \Kk\ is held fixed and \Knk\ varied, we find that the open approximation outperforms the closed system as an estimator of the true behavior of the open reaction. The degree to which the open approximation is superior increases with higher \Knk\ values, as one would expect. 

For each fixed value of \Knk\, the  \MFN\ and  \MFA\ manifolds deviate substantially from the \MFI\ manifold once \Kk\ is above a certain value, which depends on \Knk. In each of Figures \ref{fig:Q_SE_k3_d}--\ref{fig:Q_SE_k3_a}, the \MFN\ and \MFA\ manifolds form large `flaps' that extend far away from the \MFI\ manifold in the directions of increased enzyme and complex and decreased substrate. These flaps correspond to parameter regimes in which the rate of enzyme input is relatively rapid and thus assuming that total enzyme remains constant is a clearly poor approximation to the behavior of the open reaction. Free enzyme accumulates, as does complex, and substrate is rapidly degraded. The deviation between the \MFN-\MFA\ manifolds and the \MFI\ manifold is obviously greatest at the uppermost rates of enzyme input, where \Kk\ significantly exceeds \Knk\ and begins to attain the same order of magnitude as \KM. We note, however, that even at these high values of \Kk, the open approximation (\ref{eq:finale}), (\ref{eq:finalc}) lies remarkably close to the solution of the full equations for the open system up to time $t_{f}$. As the value of \Kk\ decreases, the discrepancy between the open and closed systems decreases, as expected. As shown in Figures \ref{fig:SE_k3_d_ErrC}--\ref{fig:SE_k3_a_ErrC}, while for each \Knk\ value examined the signed error between the full open and closed systems rises to $\Order{1}$ as \Kk\ increases beyond approximately $10^{-1}$, within the three lowest decades of  \Kk\ values, the discrepancy between the full open and closed systems ranges from \Order{10^{-4}} when $\Knk = 0.0001$ to \Order{10^{-2}} when $\Knk =1$. As an estimate of the full open reaction, the open approximation is always at least as accurate as the closed system, even at low levels of enzyme input, and it is typically more accurate by more than an order of magnitude. For low \Kk\ values, this is difficult to discern in the three-dimensional subplots of Figures \ref{fig:Q_SE_k3_d}--\ref{fig:Q_SE_k3_a} due to the size of the flaps of the \MFN\ and \MFA\ manifolds. Figures \ref{fig:SE_k3_d_Err}--\ref{fig:SE_k3_a_Err}, however, show that for each \Kk\ value the signed error between the full open system and the open approximation is \Order{10^{-4}} over the three lowest decades of \Knk values and rises to at most \Order{10^{-2}} for $\Knk = 0.1$. Comparing with Figures \ref{fig:SE_k3_d_ErrC}--\ref{fig:SE_k3_a_ErrC}, we see that at identical parameter values, the open approximation is generally one to two orders of magnitude more accurate than the closed system as an estimate of the full open system.

The position of the \MFA\ manifold relative to that of the \MFN\ manifold, at all \Kk\ values, indicates that the open approximation consistently underestimates the level of free enzyme and overestimates the level of complex in the open reaction. This situation can also be seen in the subplots of the signed approximation error, \ref{fig:SE_k3_d_Err}--\ref{fig:SE_k3_a_Err}, which also show that the estimate for substrate levels is quite accurate at nearly all \Kk\ values.  The reasons for this arrangement of estimation errors are largely the same as those given above. As an addendum to the previous explanation, we note that the open approximation depends in part on the rate of addition of enzyme to the system being small relative to other reaction rates. In particular, we assume in Equation (\ref{eq:change2}) that the maximal rate of change in substrate concentration occurs at time $t = 0$, which is certainly the case in the closed reaction and is valid at lower values of \Kk. As \Kk\ increases, this assumption becomes less accurate; to consider a limiting case, the introduction of enzyme via an instantaneous pulse will produce an immediate, substantial increase in the rate of substrate degradation, and thus a greater deviation from the presumed value $S \approx S_{0}$ which appears in (\ref{eq:finale}), (\ref{eq:finalc}). We note also that the error condition (\ref{eq:noniso-tolerance}) remains unchanged as \Kk\ varies. Hence we would expect the increase in signed error with increasing \Kk\ shown in the Figures \ref{fig:SE_k3_d_Err}--\ref{fig:SE_k3_a_Err},  which is the reverse of what is seen when \Kk\ is held constant and \Knk\ is varied.

\

Before turning to the results for the reverse QSSA regime, we make two related observations which may aid in interpreting the figures in this subsection and the next. First we note that the intersection of the  \MFN\ and \MFI\ manifolds marks the location of parameter combinations for which  the total amount of enzyme remains approximately constant  in the open system  over the transient period. This occurs roughly where $\Kk \approx \Knk$, as is somewhat more easily seen in Figures \ref{fig:Q_SE_kn3_d}--\ref{fig:Q_SE_kn3_a} than in Figures \ref{fig:Q_SE_k3_d}--\ref{fig:Q_SE_k3_a}. To reiterate the conclusions detailed above, when $\Kk < \Knk$, the closed system overestimates the level of enzyme, and when $\Kk > \Knk$, the situation is reversed. 

Second, loosely speaking, each of Figures \ref{fig:Q_SE_kn3_d}--\ref{fig:Q_SE_kn3_a} is ``identical'' up to an extension (scaling) and rotation of the \MFI\ and \MFA\ manifolds. To see this relationship, observe first in Figure \ref{fig:SE_kn3_d_SE} how the \MFN\ and \MFA\ manifolds lie close to the \MFI\ manifold at low values of \Knk\ and peel away from it as \Knk\ increases, folding over in the direction of lower enzyme concentration. The same folding is seen in Figure  \ref{fig:SE_kn3_b_SE}, but here the fold intersects \MFI\ where $\Kk \approx \Knk$. The \MFN\ and \MFA\ manifolds extend substantially past the \MFI\ manifold in the directions of higher enzyme concentration and lower substrate concentration; this is most noticeable in the lower portions of \MFN\ and \MFA. This extension relative to the unchanging \MFI\ manifold is due to the input of free enzyme, which increases enzyme concentration and accelerates depletion of substrate. The increase in \Kk\ between Figure  \ref{fig:SE_kn3_d_SE} and Figure  \ref{fig:SE_kn3_b_SE} effectively rotates and stretches \MFN\ and \MFA\ relative to \MFI. At the highest level of \Kk, shown in Figure \ref{fig:SE_kn3_a_SE}, \MFN\ and \MFA\ retain their fold shape, but their elongation is much more pronounced and the rotation is even greater, so that \MFN\ and \MFA\ lie entirely on the other side of \MFI. 

Similar relations hold between the manifolds depicted in Figures \ref{fig:SE_kn3_d_CE}--\ref{fig:SE_kn3_a_CE} and Figures \ref{fig:SE_kn3_d_CS}--\ref{fig:SE_kn3_a_CS}. Furthermore, Figures \ref{fig:Q_SE_kn3_d}--\ref{fig:Q_SE_kn3_a} in this subsection are related to one another in an analogous fashion, as are Figures \ref{fig:Q_ES_k3_d}--\ref{fig:Q_ES_k3_a} and Figures \ref{fig:Q_ES_kn3_d}--\ref{fig:Q_ES_kn3_a} in the following subsection.  


\begin{figure}[!ht]
 	\centering
	\subfloat[][Substrate-Enzyme]{\label{fig:SE_k3_d_SE}\includegraphics[width= \gwidth]{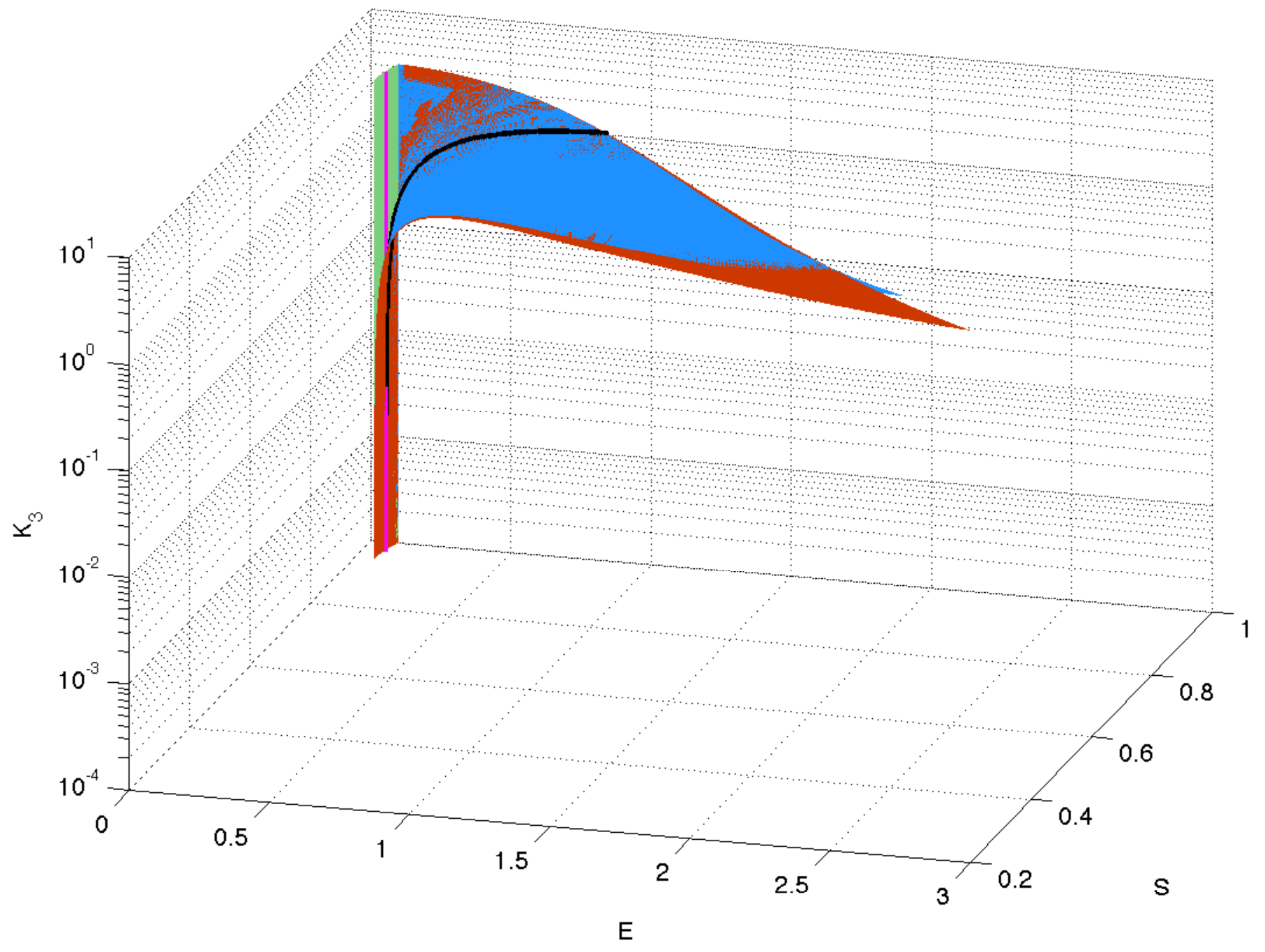}}
	\subfloat[][Complex-Enzyme]{\label{fig:SE_k3_d_CE}\includegraphics[width= \gwidth]{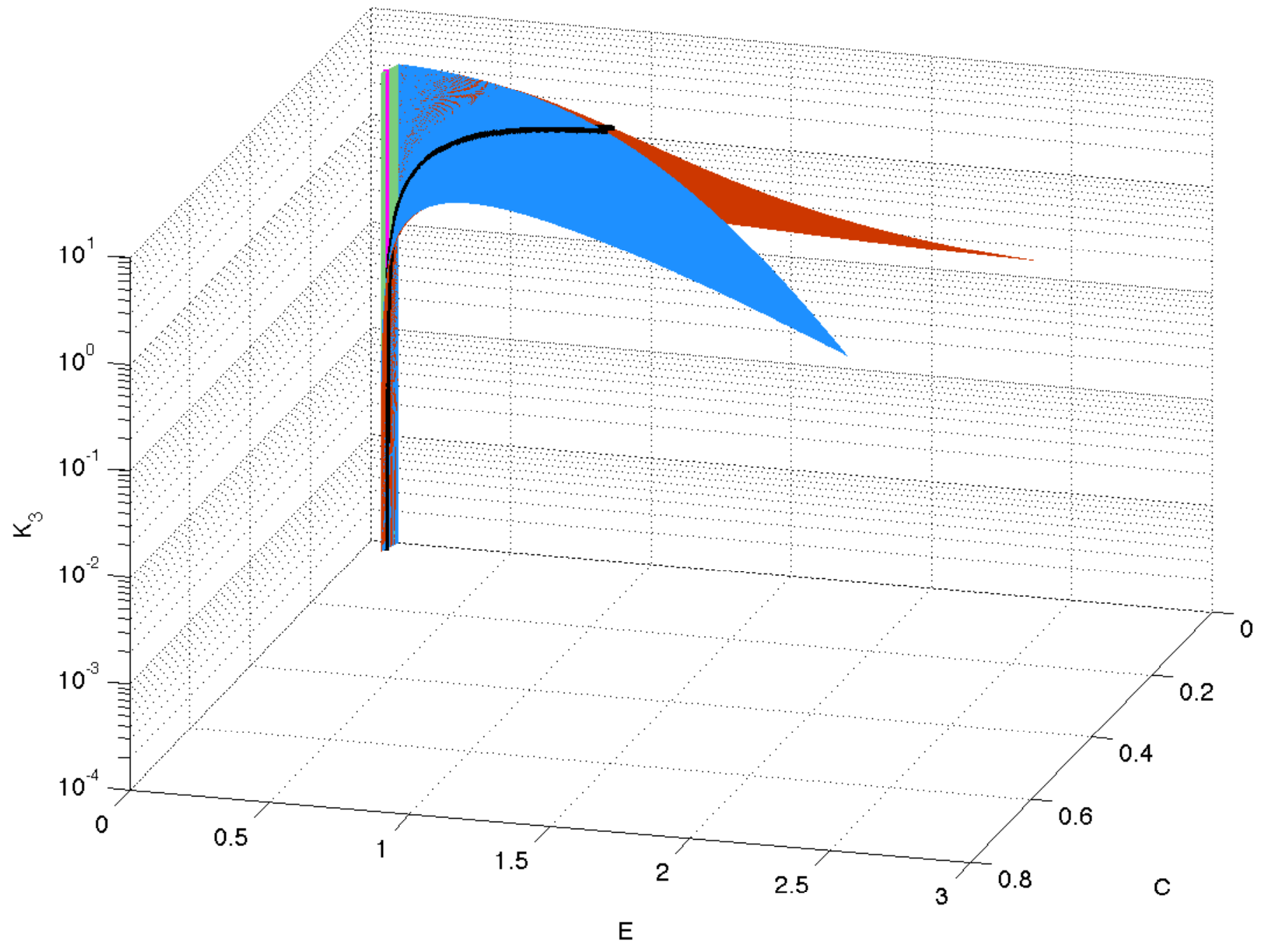}}
	\subfloat[][Complex-Substrate]{\label{fig:SE_k3_d_CS}\includegraphics[width= \gwidth]{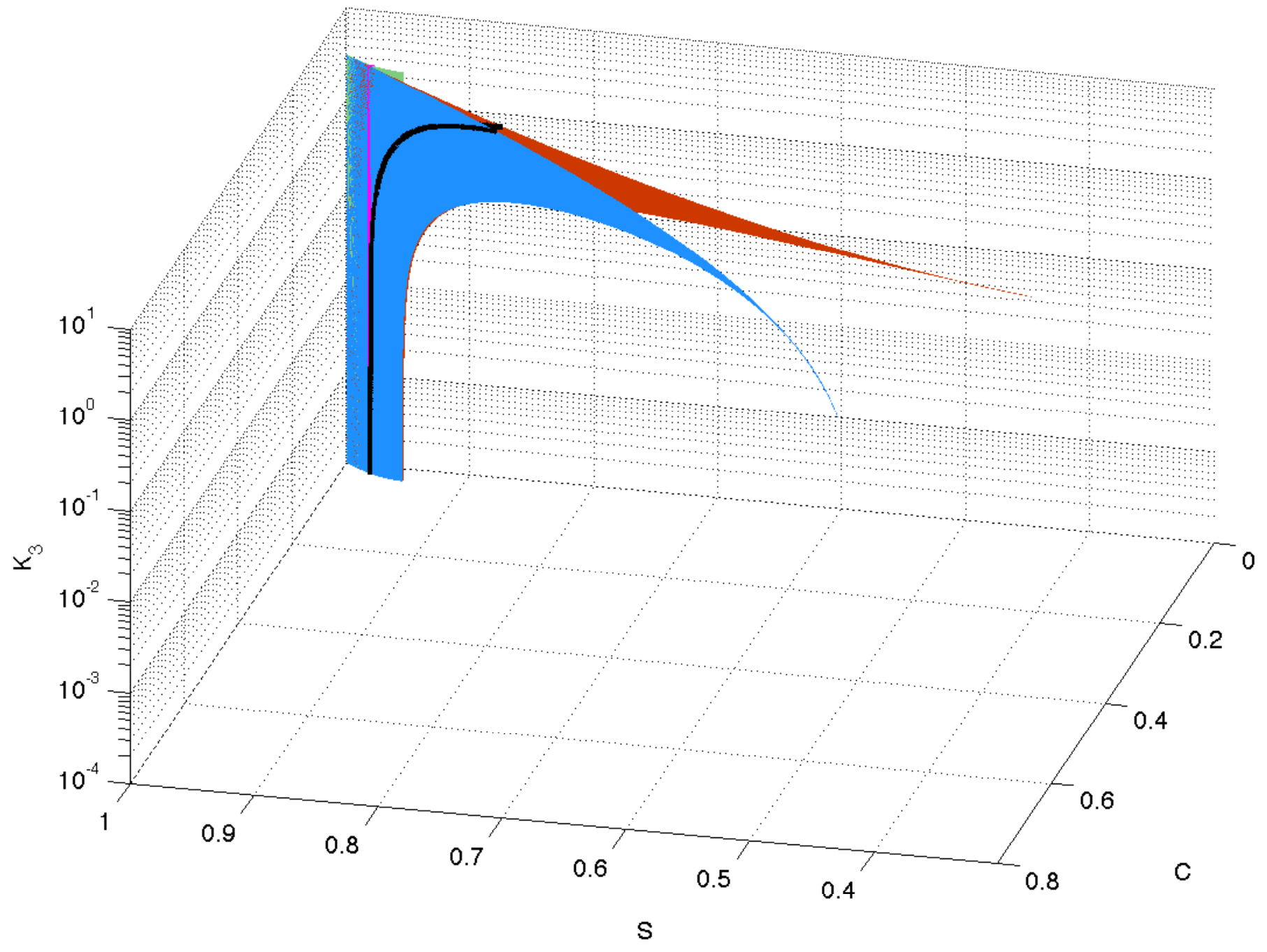}}\\
	\subfloat[][Signed error at $t_{f}$, full open \\system vs. open approximation]{\label{fig:SE_k3_d_Err}\includegraphics[width= \ewidth]{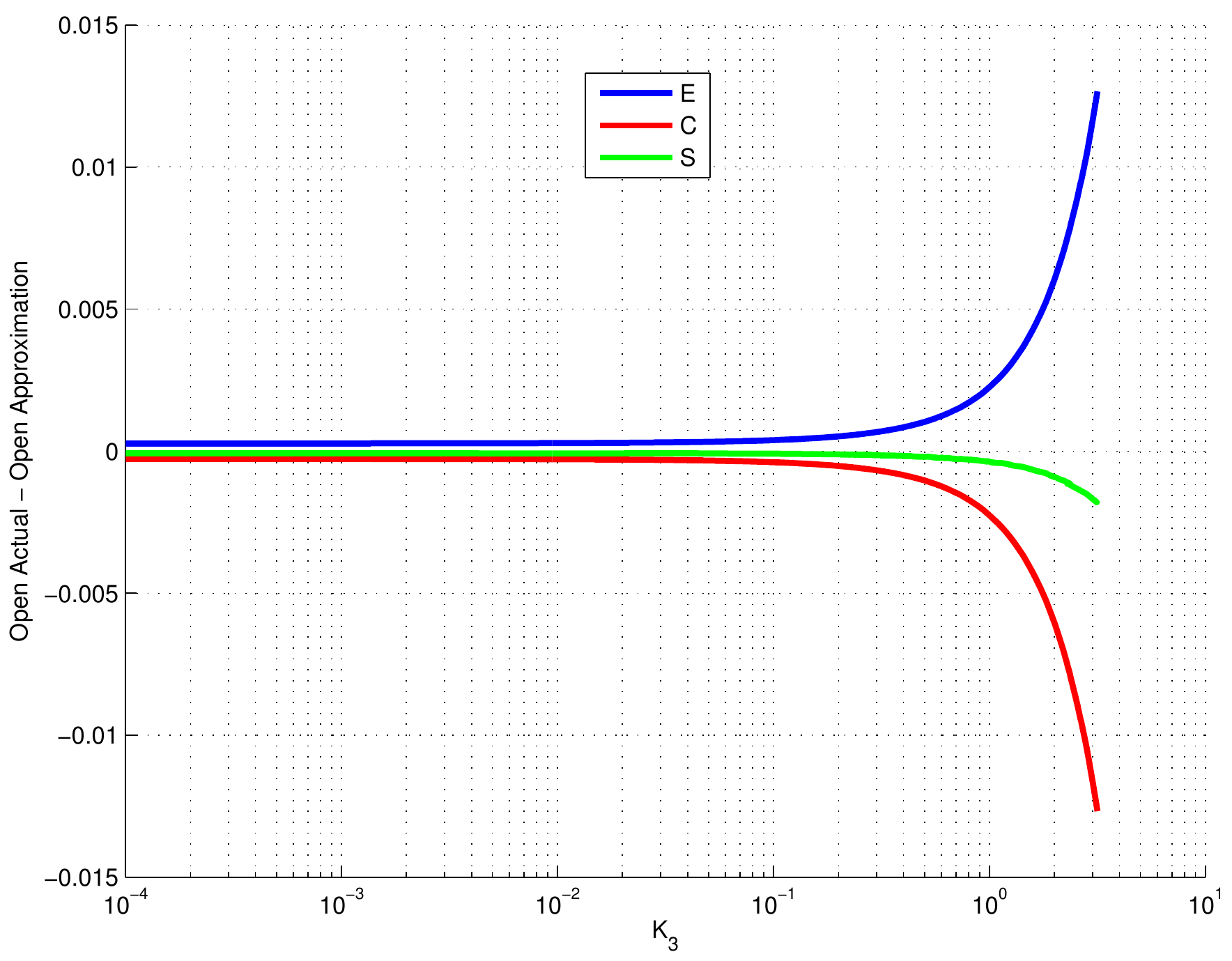}}\hspace{0.5cm}
	\subfloat[][Signed error at $t_{f}$, full open system vs. full closed system]{\label{fig:SE_k3_d_ErrC}\includegraphics[width= \ewidth]{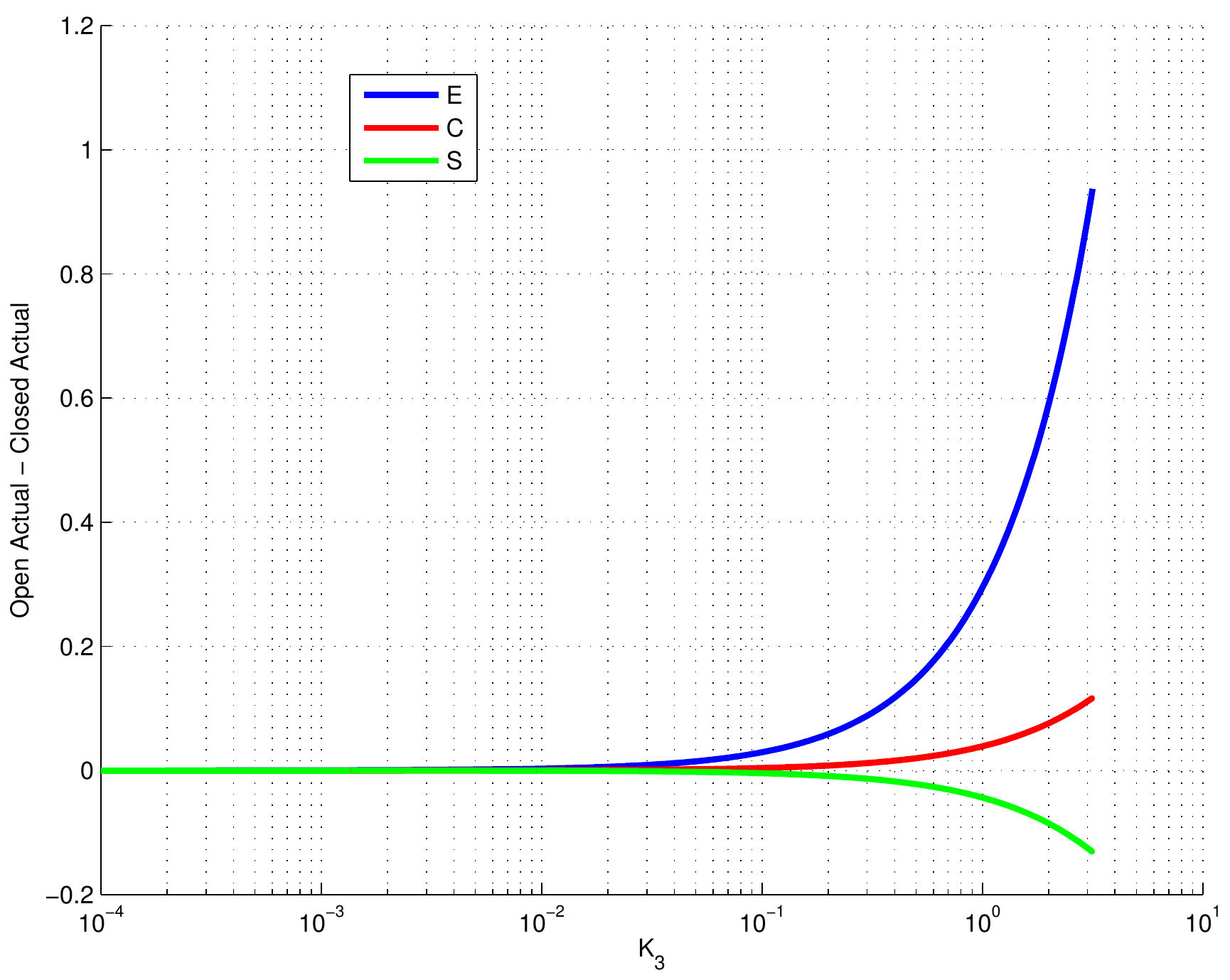}}\\
	\caption[Standard QSSA regime, varying $\Kk$. $\Knk  = 0.0001$]{Phase plane portraits and signed error in the standard QSSA regime as \Kk\ varies, with $\Knk = 0.0001$. }
	\label{fig:Q_SE_k3_d}
\end{figure}

\begin{figure}[!ht]
 	\centering
	\subfloat[][Substrate-Enzyme]{\label{fig:SE_k3_b_SE}\includegraphics[width= \gwidth]{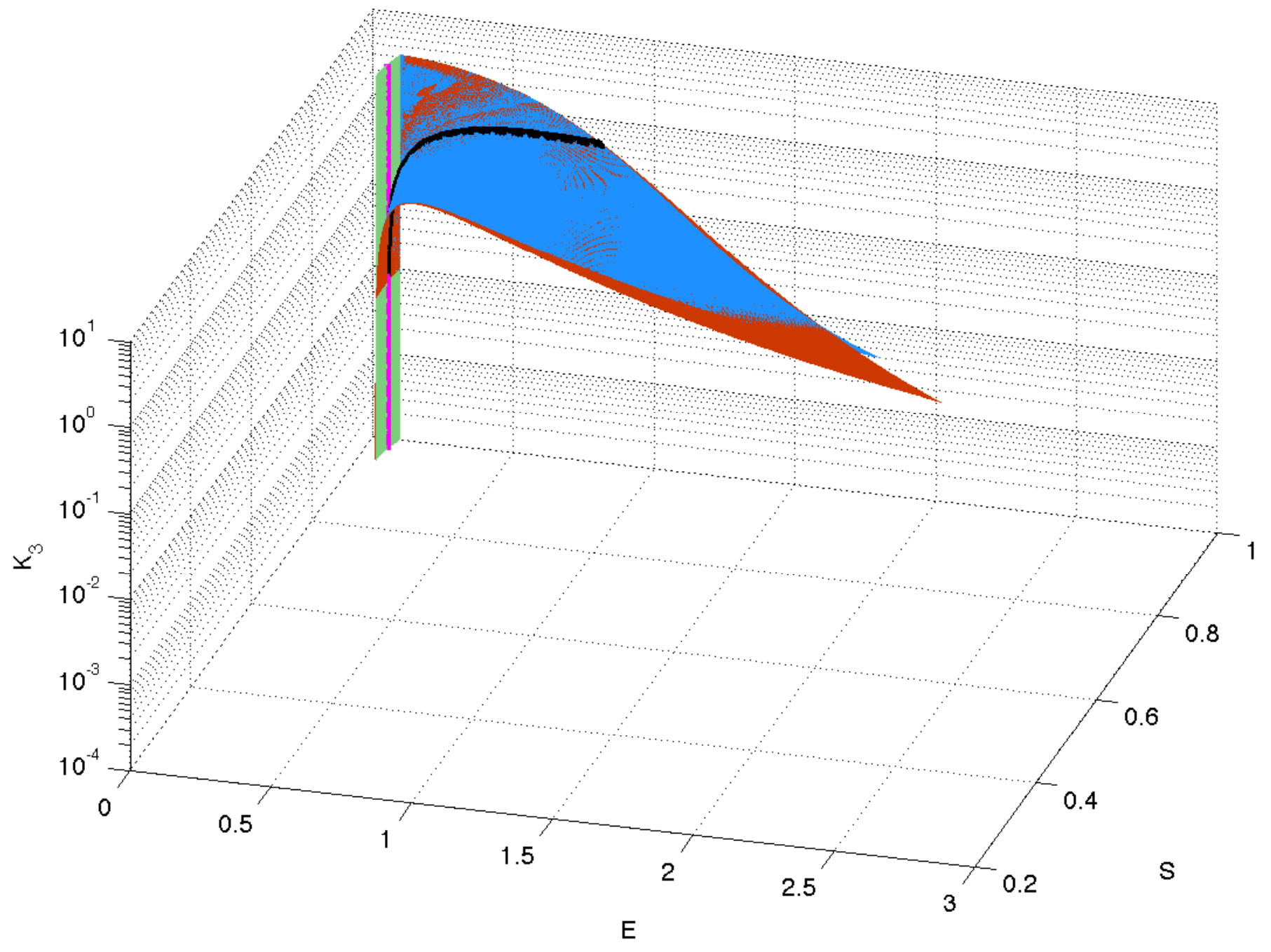}}
	\subfloat[][Complex-Enzyme]{\label{fig:SE_k3_b_CE}\includegraphics[width= \gwidth]{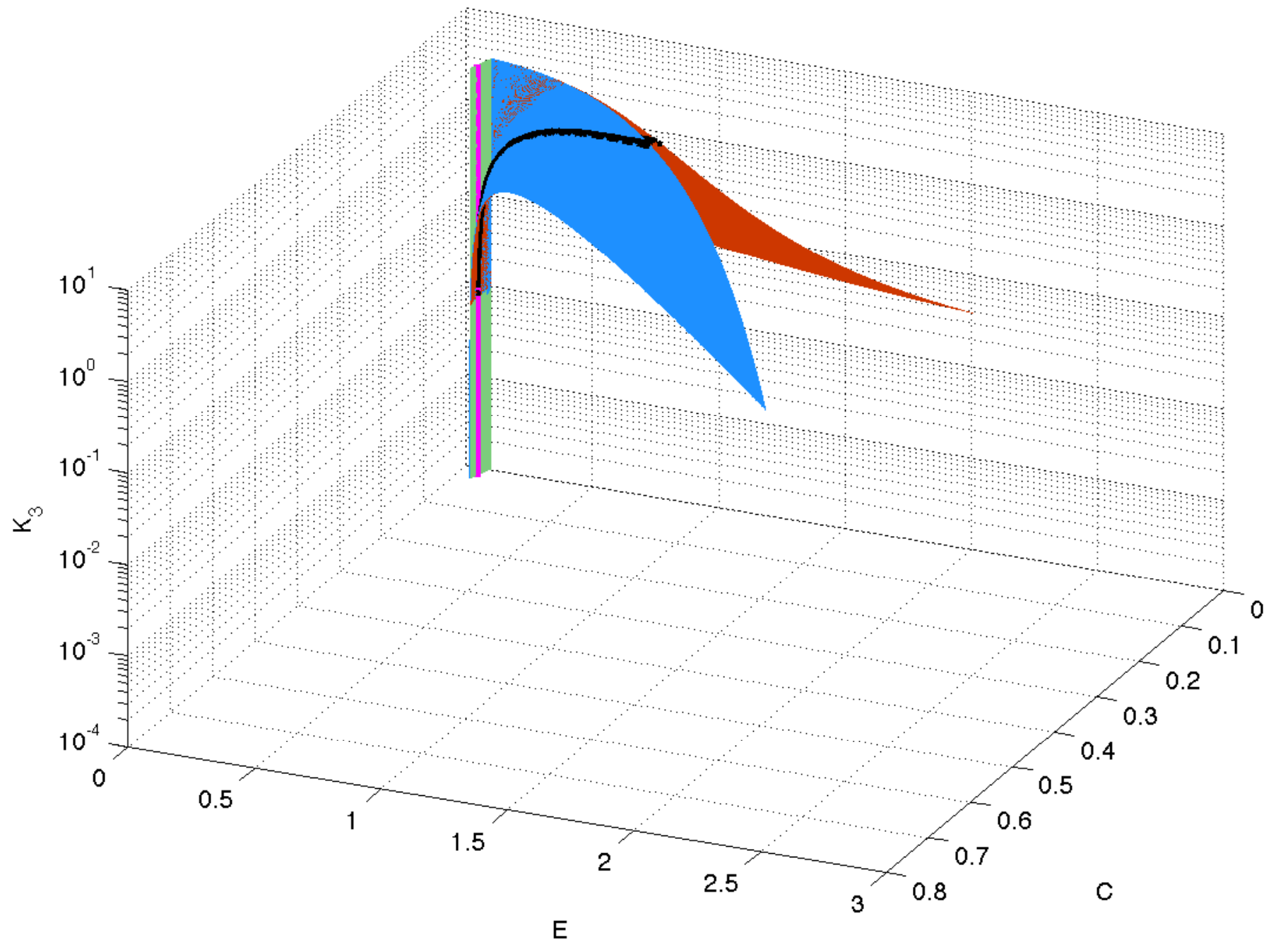}}
	\subfloat[][Complex-Substrate]{\label{fig:SE_k3_b_CS}\includegraphics[width= \gwidth]{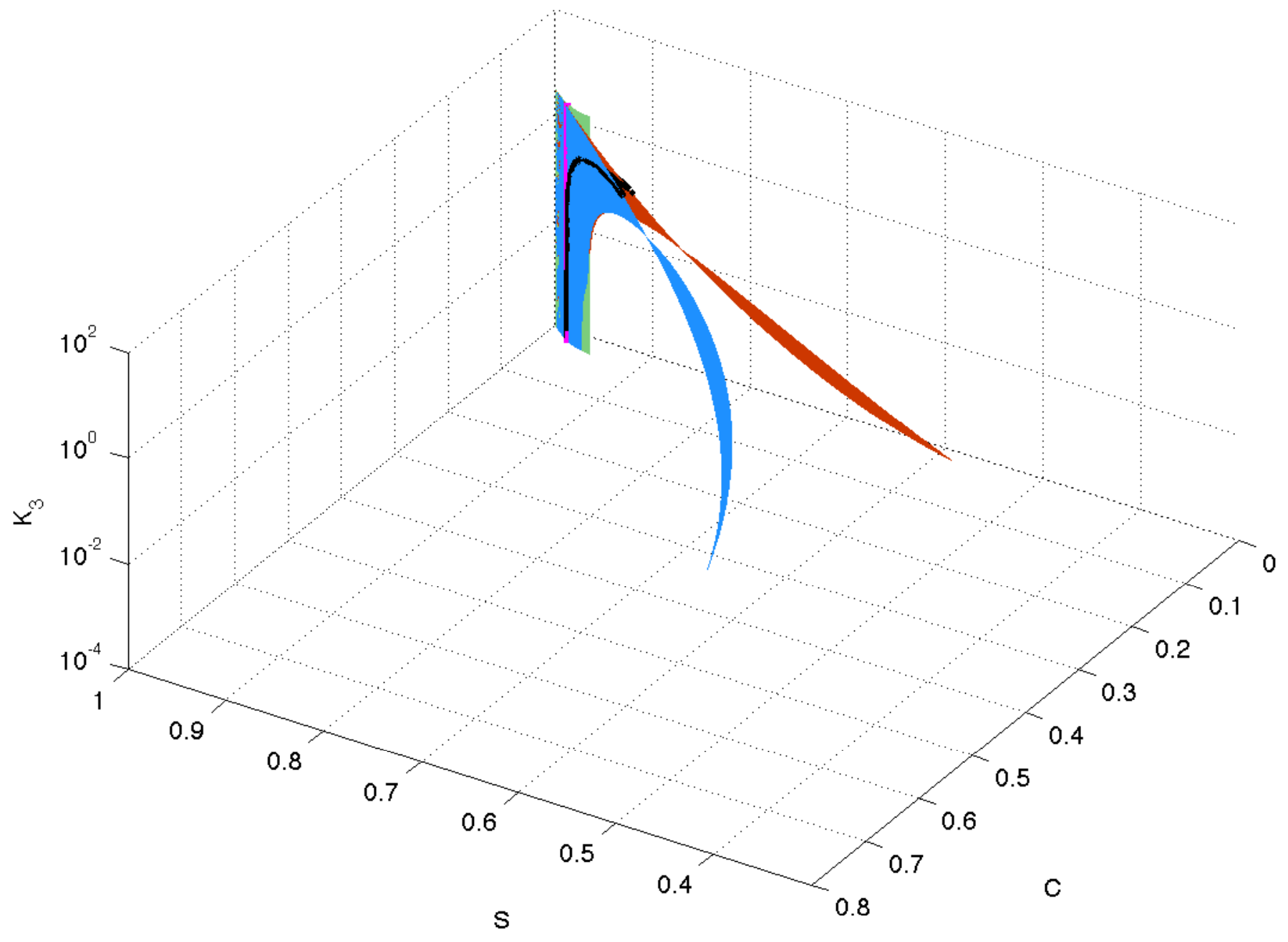}}\\
	\subfloat[][Signed error at $t_{f}$, full open \\system vs. open approximation]{\label{fig:SE_k3_b_Err}\includegraphics[width= \ewidth]{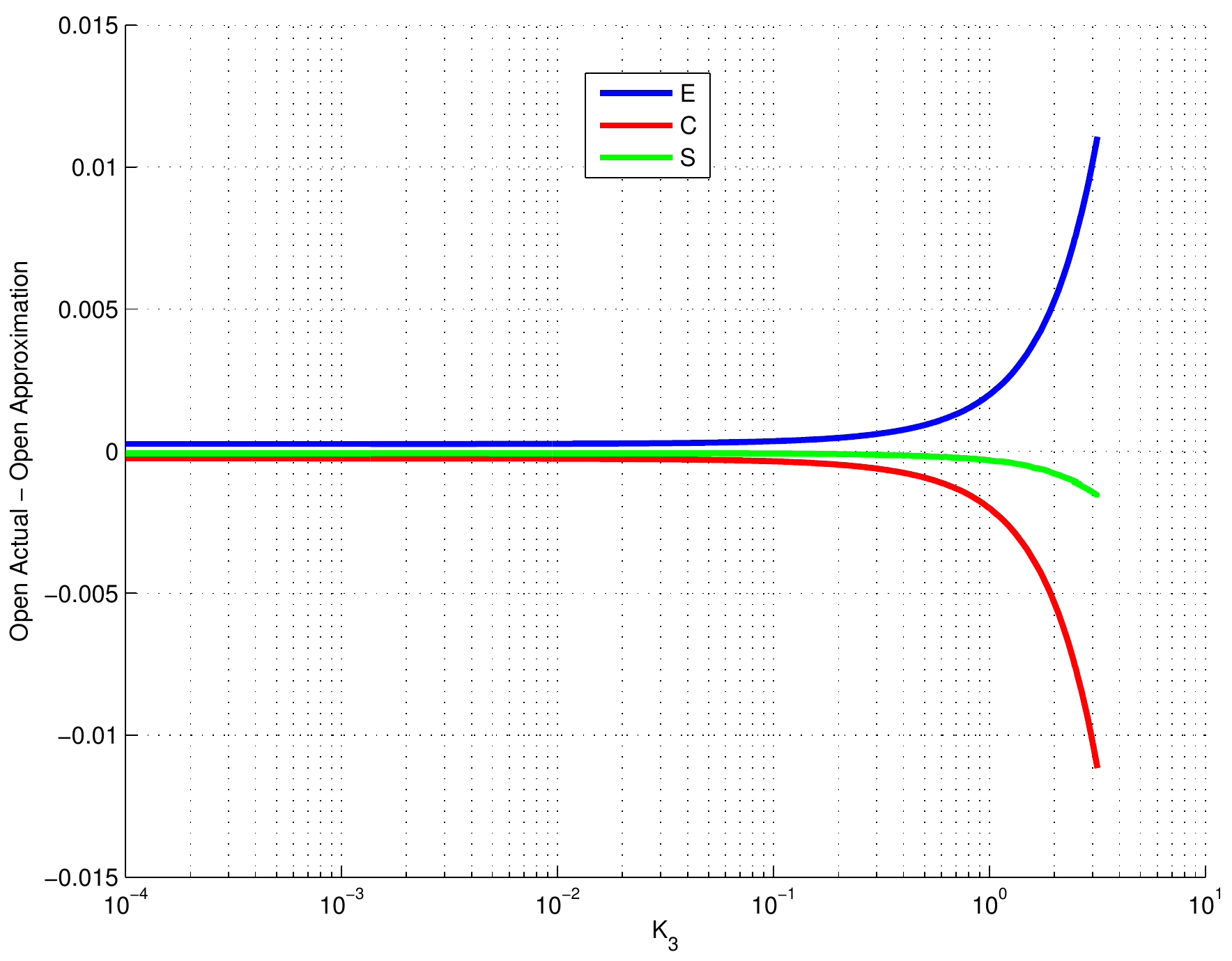}}\hspace{0.5cm}
	\subfloat[][Signed error at $t_{f}$, full open system vs. full closed system]{\label{fig:SE_k3_b_ErrC}\includegraphics[width= \ewidth]{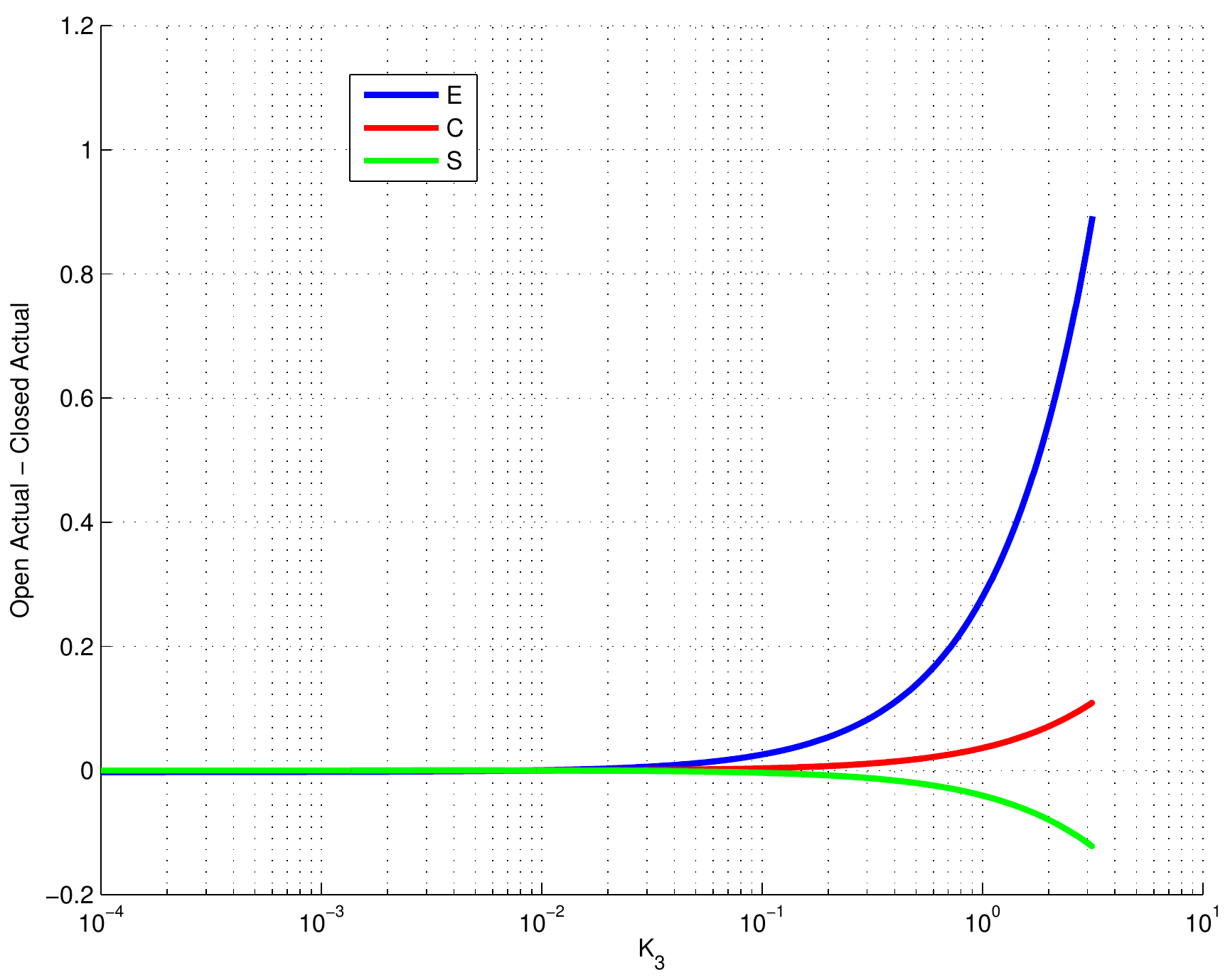}}\\
	\caption[Standard QSSA regime, varying $\Kk$. $\Knk  = 0.1$]{Phase plane portraits and signed error in the standard QSSA regime as \Kk\ varies, with $\Knk = 0.1$. }
	\label{fig:Q_SE_k3_b}
\end{figure}

\begin{figure}[!ht]
 	\centering
	\subfloat[][Substrate-Enzyme]{\label{fig:SE_k3_a_SE}\includegraphics[width= \gwidth]{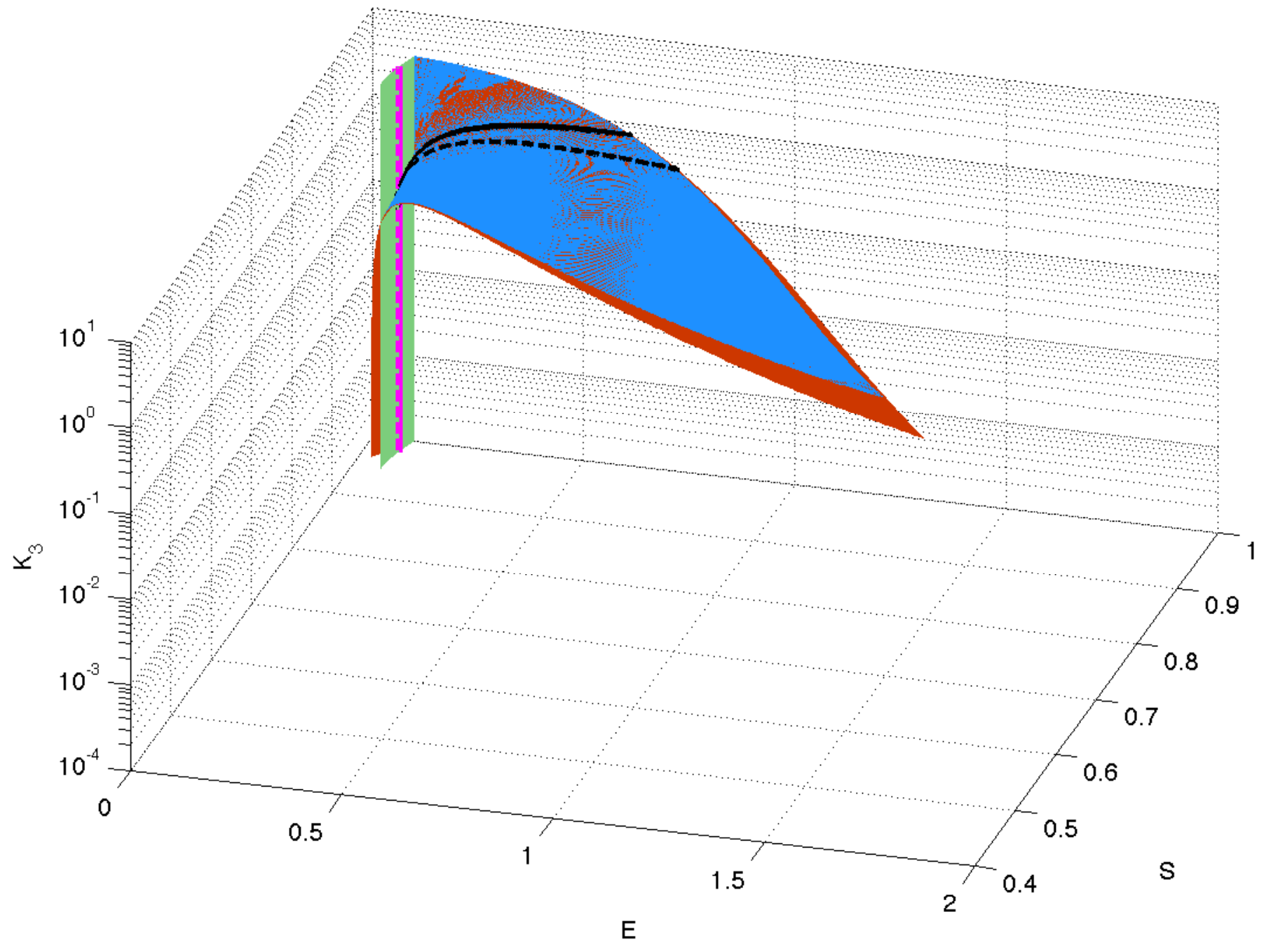}}
	\subfloat[][Complex-Enzyme]{\label{fig:SE_k3_a_CE}\includegraphics[width= \gwidth]{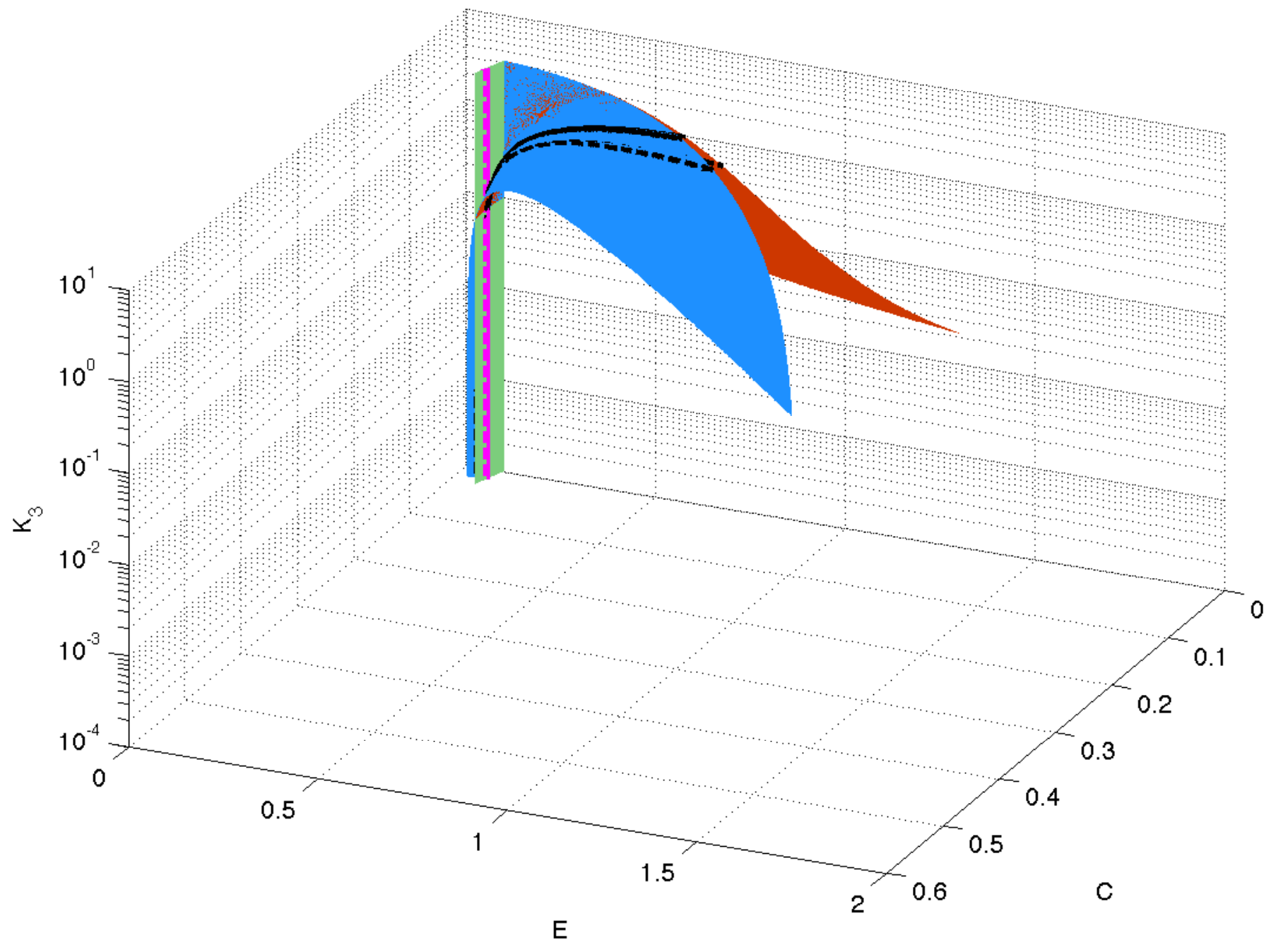}}
	\subfloat[][Complex-Substrate]{\label{fig:SE_k3_a_CS}\includegraphics[width= \gwidth]{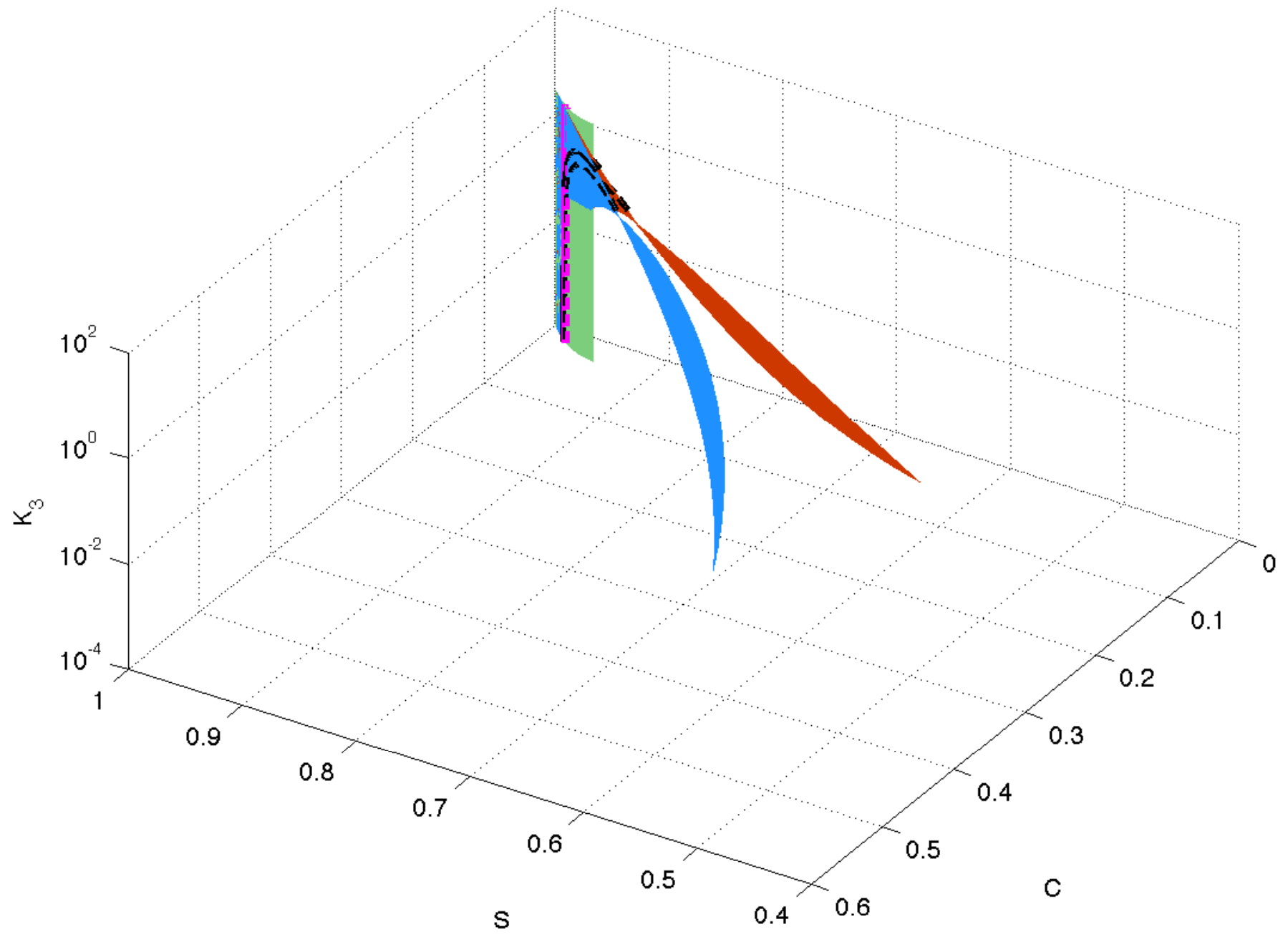}}\\
	\subfloat[][Signed error at $t_{f}$, full open \\system vs. open approximation]{\label{fig:SE_k3_a_Err}\includegraphics[width= \ewidth]{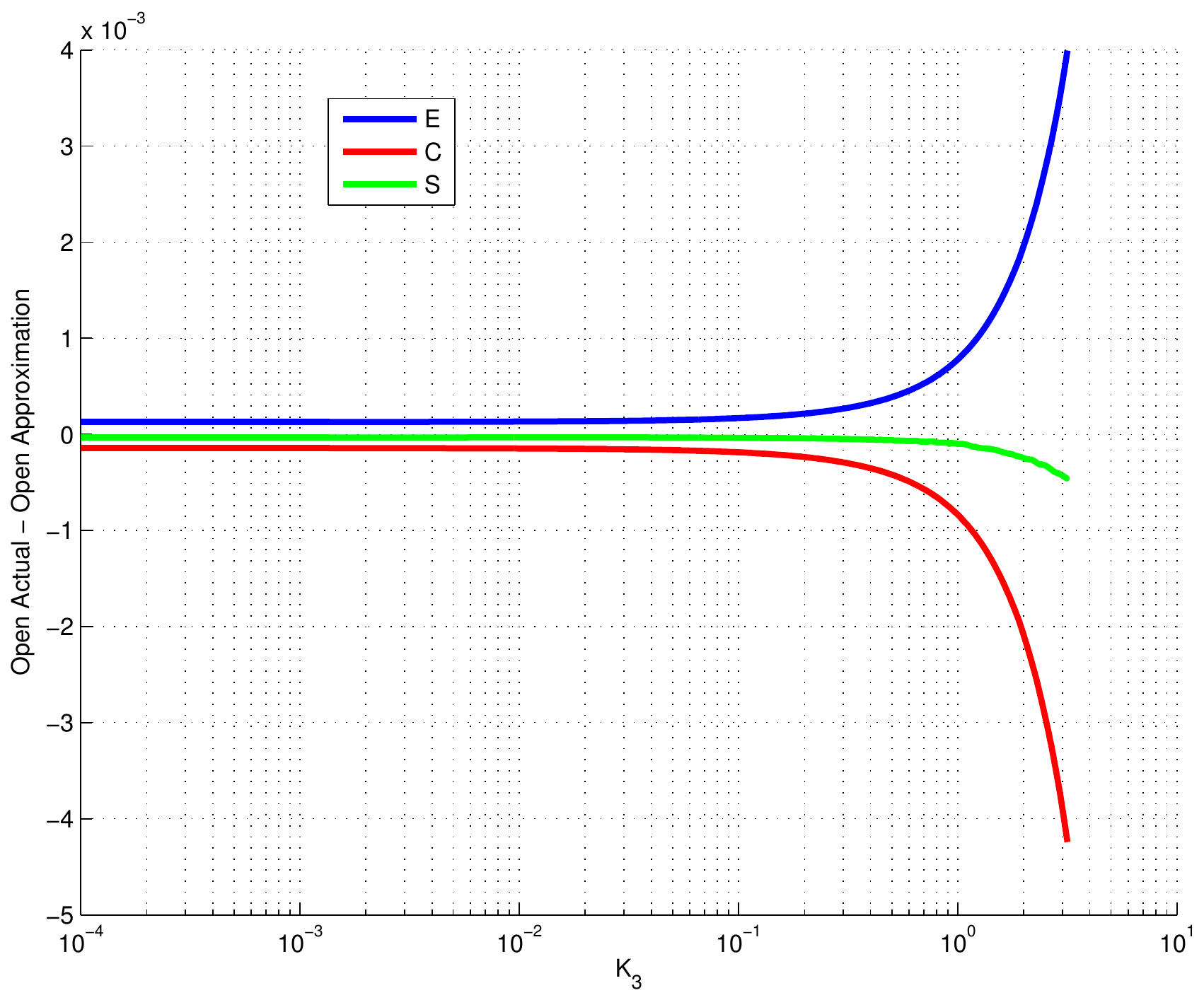}}\hspace{0.5cm}
	\subfloat[][Signed error at $t_{f}$, full open system vs. full closed system]{\label{fig:SE_k3_a_ErrC}\includegraphics[width= \ewidth]{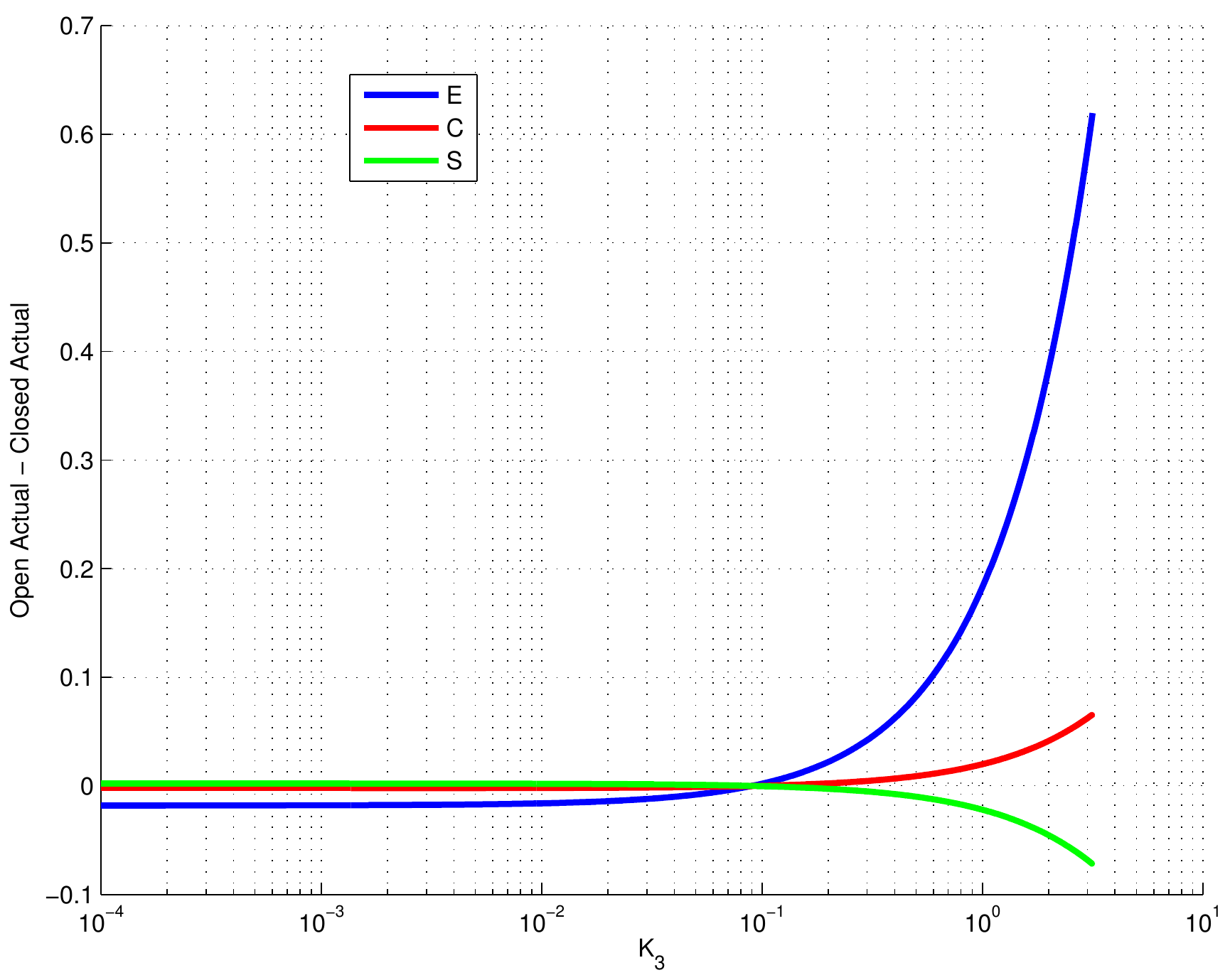}}\\
	\caption[Standard QSSA regime, varying $\Kk$. $\Knk  = 1$]{Phase plane portraits and signed error in the standard QSSA regime as \Kk\ varies, with $\Knk = 1$. }
	\label{fig:Q_SE_k3_a}
\end{figure}

\subsection{Reverse QSSA regime}

When $E_{0} \gg S_{0}$, the validity condition (\ref{eq:def_eps}) does not hold, {\it i.e.}  $\eta/(1+\alpha+\sigma) \nless 1$, and we cannot be confident that the open approximation (\ref{eq:finale}), (\ref{eq:finalc}) will accurately capture the behavior of the full open system. Also uncertain is the accuracy of the closed system as an approximation to the full open reaction. The results of numerical simulations in the reverse QSSA regime, however, bear striking overall resemblance to the corresponding results for the standard QSSA regime. In particular, our numerical results indicate that the open approximation provides a quite accurate approximation to the full open system. The closed system approximates the open reaction well only when the ratio of \Kk\ and \Knk\ implies that the change in enzyme level up to time $t_{f}$ is miniscule.

The reverse QSSA regime results shown in Figures \ref{fig:Q_ES_kn3_d}--\ref{fig:Q_ES_kn3_a}, in which \Knk\ varies, and in Figures \ref{fig:Q_ES_k3_d}--\ref{fig:Q_ES_k3_a}, in which \Kk\ varies, match up very closely, at least qualitatively, with the corresponding results for the standard QSSA regime. As in the standard QSSA regime, and for the same reasons, the open approximation consistently underestimates the concentration of free enzyme and overestimates the concentration of complex and substrate, at least up to time $t_{f}$, the estimated end of the transient period. The closed system over- or under-estimates enzyme, complex, and substrate levels, depending on the ratio of \Kk\ and \Knk. As can be seen in Figures  \ref{fig:ES_kn3_d_Err}--\ref{fig:ES_k3_a_Err}, the difference between the true quantities in the full open system and the estimates from (\ref{eq:finale}), (\ref{eq:finalc}) follows the same pattern in the reverse QSSA regime as in the standard QSSA regime when \Kk\ and \Knk\ are varied. We point out, however, that past time $t_{f}$ the open approximation may begin to underestimate the level of substrate. This phenomenon can be seen in Figures \ref{fig:ES_kn3_d_CS}--\ref{fig:ES_kn3_a_CS}, where the $C$-$S$ projection of the \MFA\ manifold intersects and folds behind the  $C$-$S$ projection of the \MFN\ manifold. This occurs in scenarios where \Knk\ significantly exceeds \Kk, such that the amount of total enzyme available, and hence the effective substrate degradation capacity, drops quickly and consequentially over the course of the reaction's early stages. We also note that in the reverse QSSA regime the closed system's estimate of complex and substrate levels roughly matches its accuracy in the standard QSSA regime, as can be seen by comparing Figures \ref{fig:ES_kn3_d_ErrC}--\ref{fig:ES_k3_a_ErrC} with Figures \ref{fig:SE_kn3_d_ErrC}--\ref{fig:SE_k3_a_ErrC}.  

The chief difference between corresponding sets of results from the standard and reverse QSSA regimes is quantitative. The signed error between the full open system and the open approximation in each of enzyme, complex, and susbstrate, measured at time $t_{f}$, is typically four to ten times greater in the reverse QSSA regime than in the standard QSSA regime, at least when the free parameter is smaller than \KM\  by an order of magnitude or more. To explain this phenomenon, we refer to condition (\ref{eq:def_eps}). In the reverse QSSA regime, $\eta > \sigma$, so for the condition to hold, $\alpha$, the non-dimensionalized surrogate for \Knk, must be quite large. At higher \Knk\ values, the accuracy of the open approximation should improve, and this is precisely what may be seen by comparing, for example, the error scale in Figure \ref{fig:ES_k3_d_Err} ($\Knk=0.0001$) with that of Figure \ref{fig:ES_k3_a_Err} ($\Knk=1$), or by examining any of Figures \ref{fig:ES_kn3_d_Err}--\ref{fig:ES_kn3_a_Err}.
In general, we would expect the magnitude of error between the full open system and the open approximation at identical values of \Knk and \Kk\ to be higher in the reverse QSSA regime than in the standard QSSA regime, which is indeed the outcome we find numerically.


\begin{figure}[!ht]
 	\centering
	\subfloat[][Substrate-Enzyme]{\label{fig:ES_kn3_d_SE}\includegraphics[width= \gwidth]{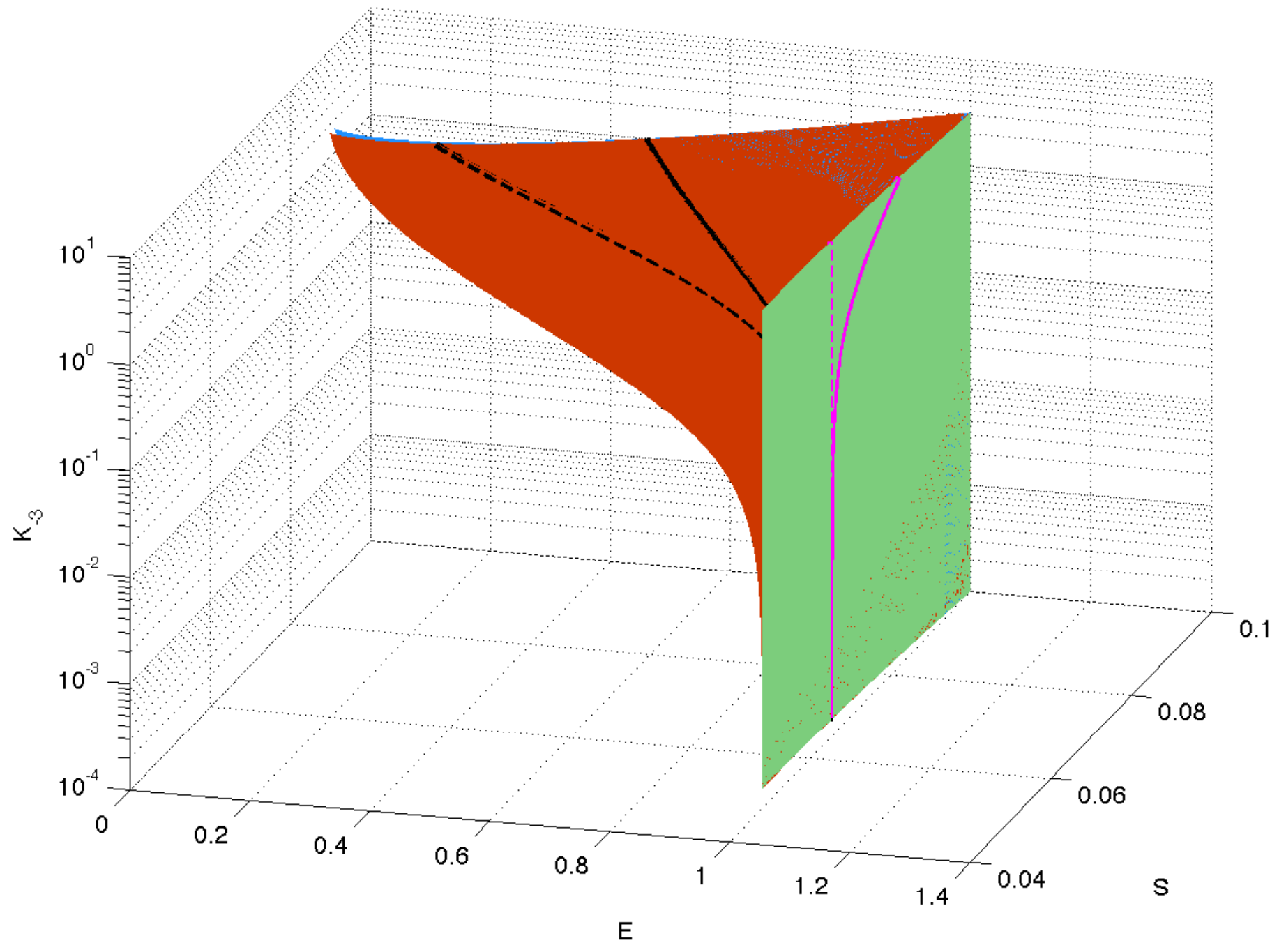}}
	\subfloat[][Complex-Enzyme]{\label{fig:ES_kn3_d_CE}\includegraphics[width= \gwidth]{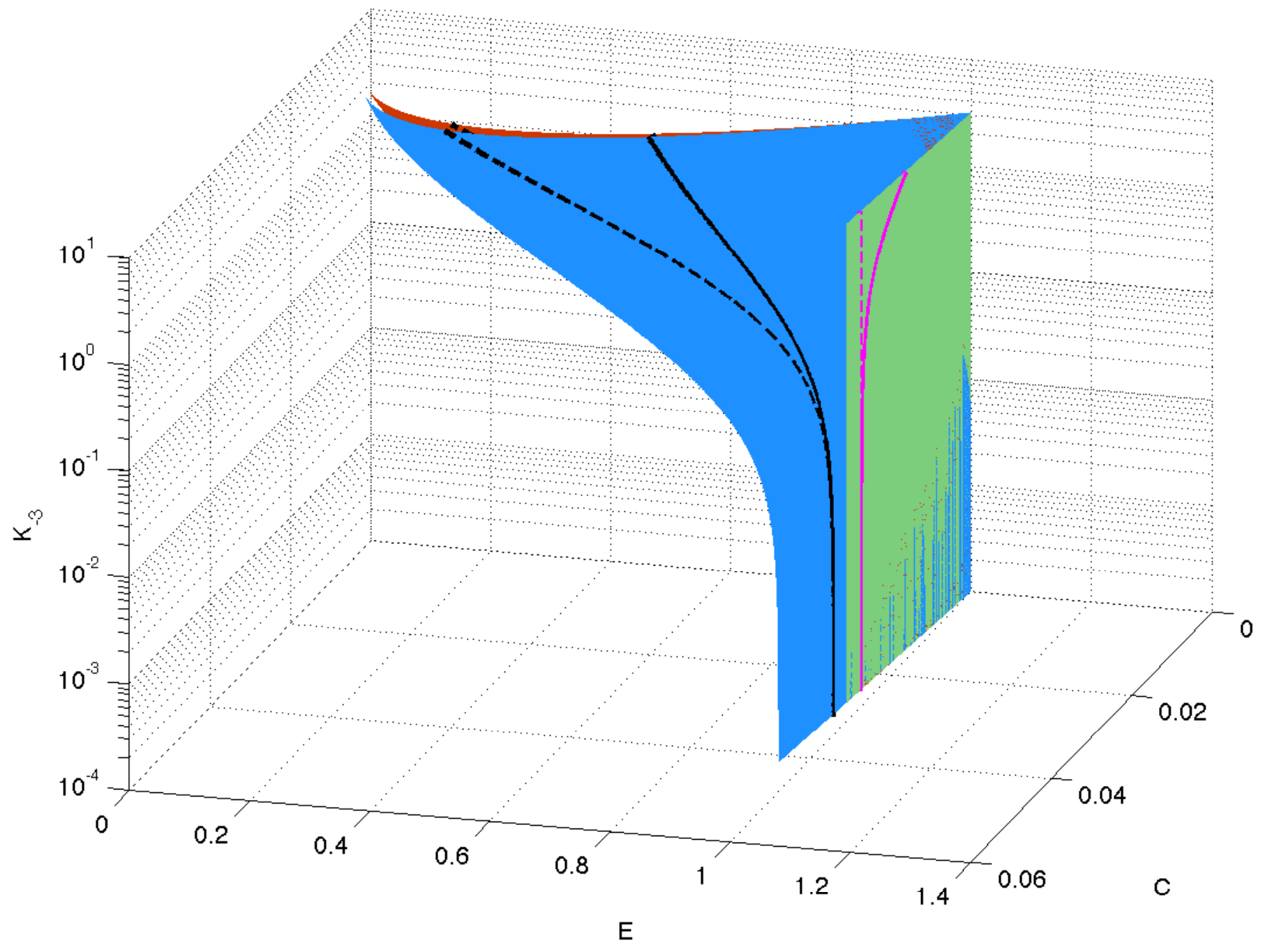}}
	\subfloat[][Complex-Substrate]{\label{fig:ES_kn3_d_CS}\includegraphics[width= \gwidth]{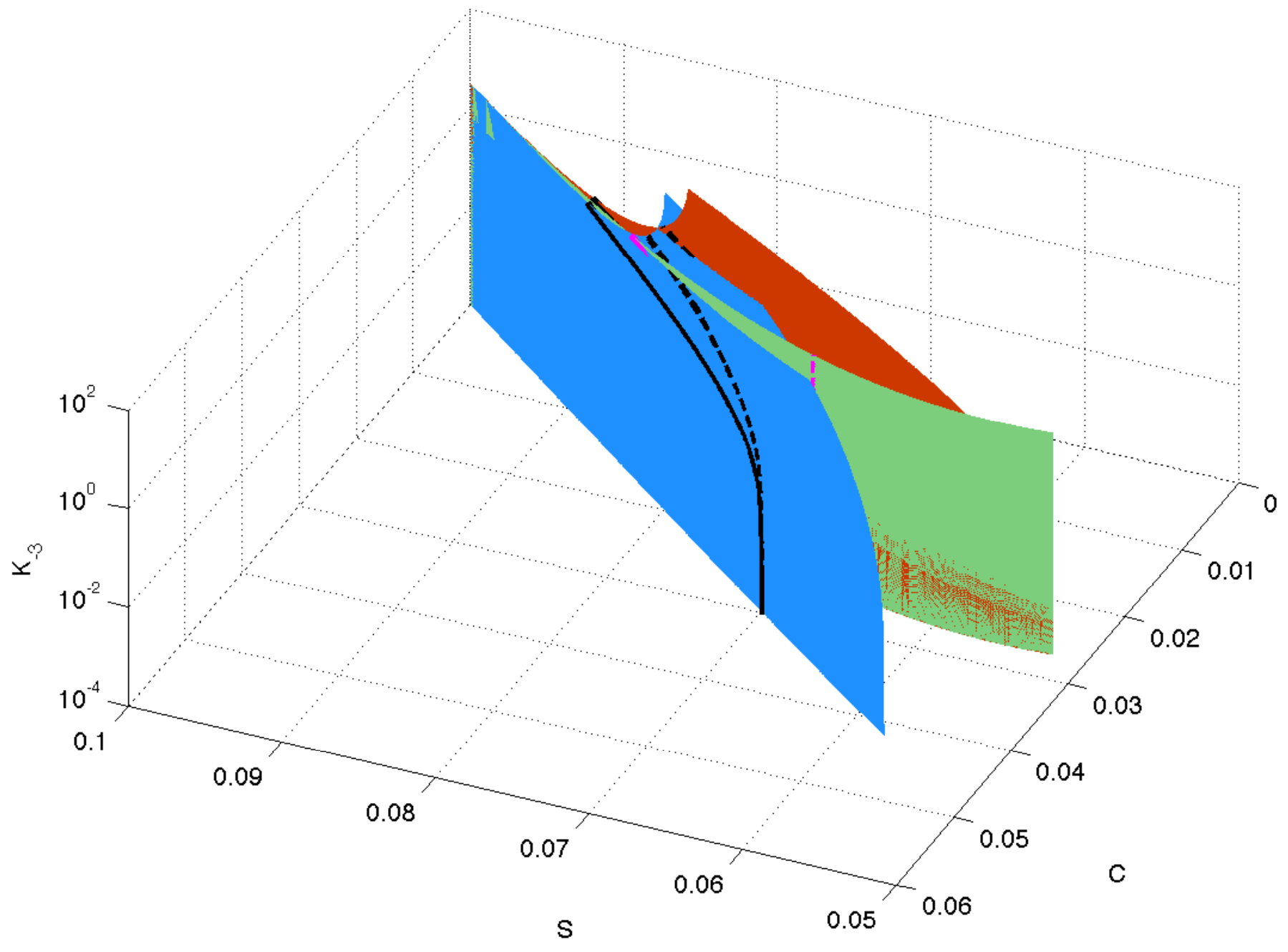}}\\
	\subfloat[][Signed error at $t_{f}$, full open \\system vs. open approximation]{\label{fig:ES_kn3_d_Err}\includegraphics[width= \ewidth]{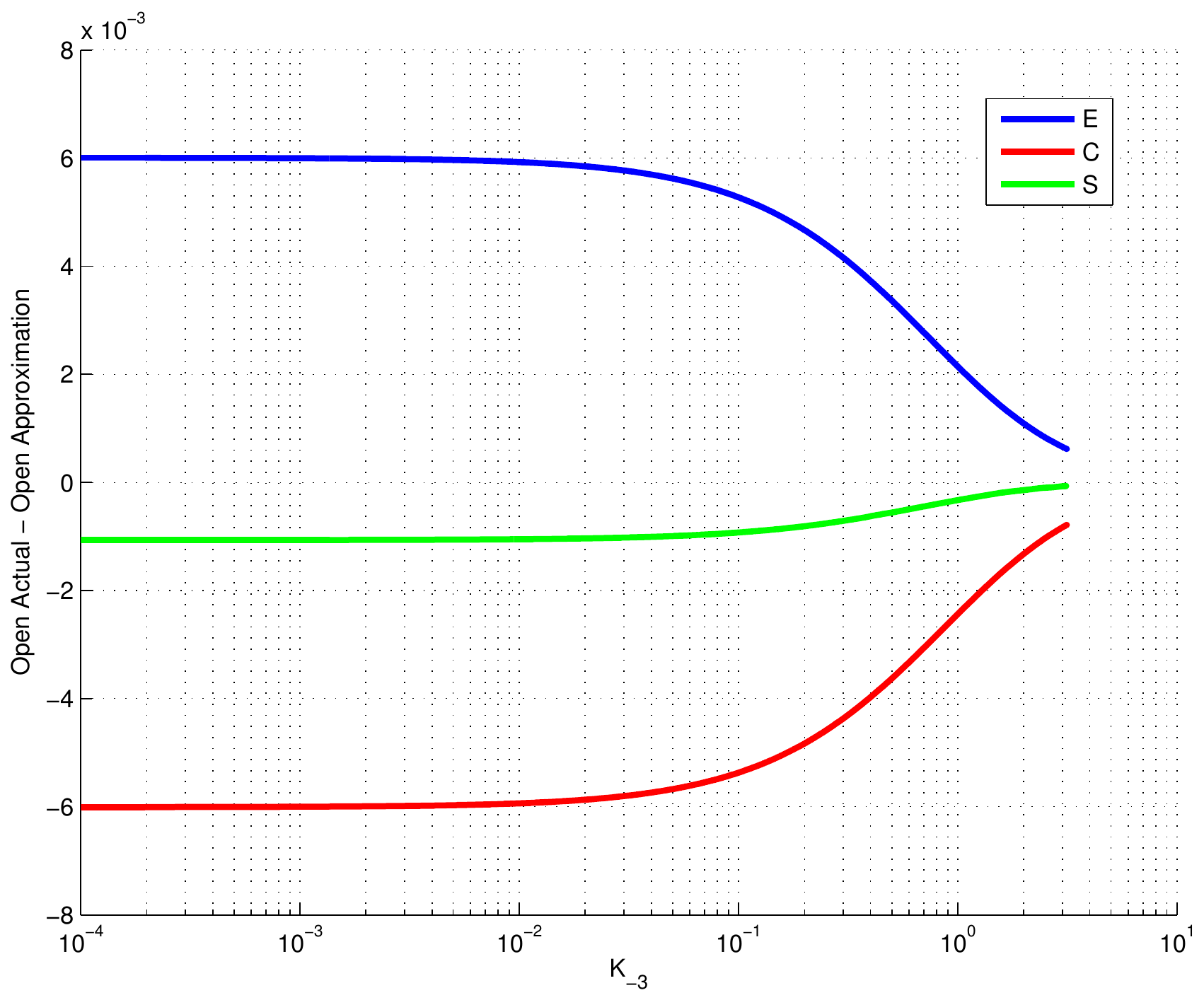}}\hspace{0.5cm}
	\subfloat[][Signed error at $t_{f}$, full open system vs. full closed system]{\label{fig:ES_kn3_d_ErrC}\includegraphics[width= \ewidth]{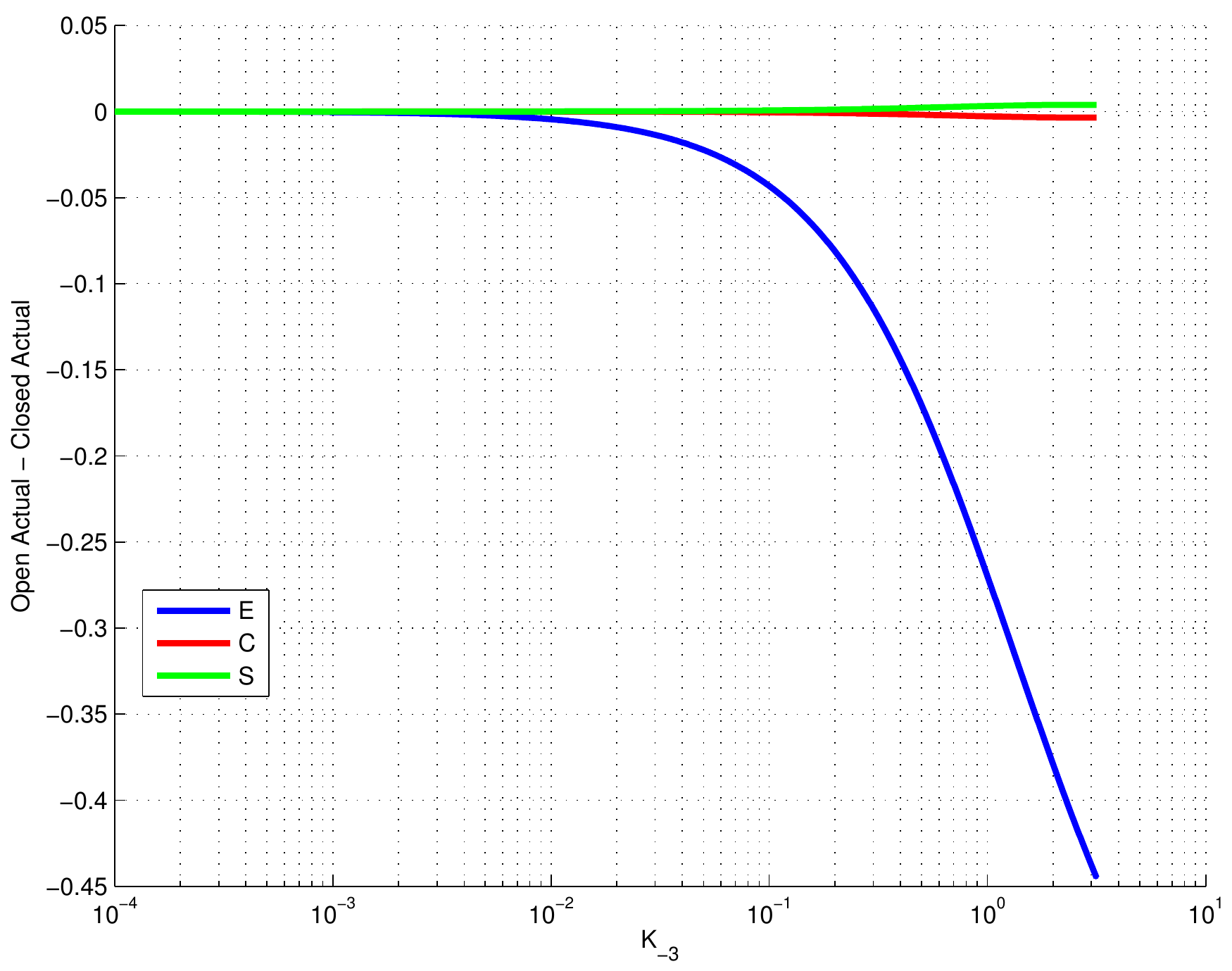}}\\
	\caption[Reverse QSSA regime, varying $\Knk $. $\Kk = 0.1$]{Phase plane portraits and signed error in the reverse QSSA regime as \Knk\ varies, with $\Kk = 0.0001$.  }
	\label{fig:Q_ES_kn3_d}
\end{figure}

\begin{figure}[!ht]
 	\centering
	\subfloat[][Substrate-Enzyme]{\label{fig:ES_kn3_b_SE}\includegraphics[width= \gwidth]{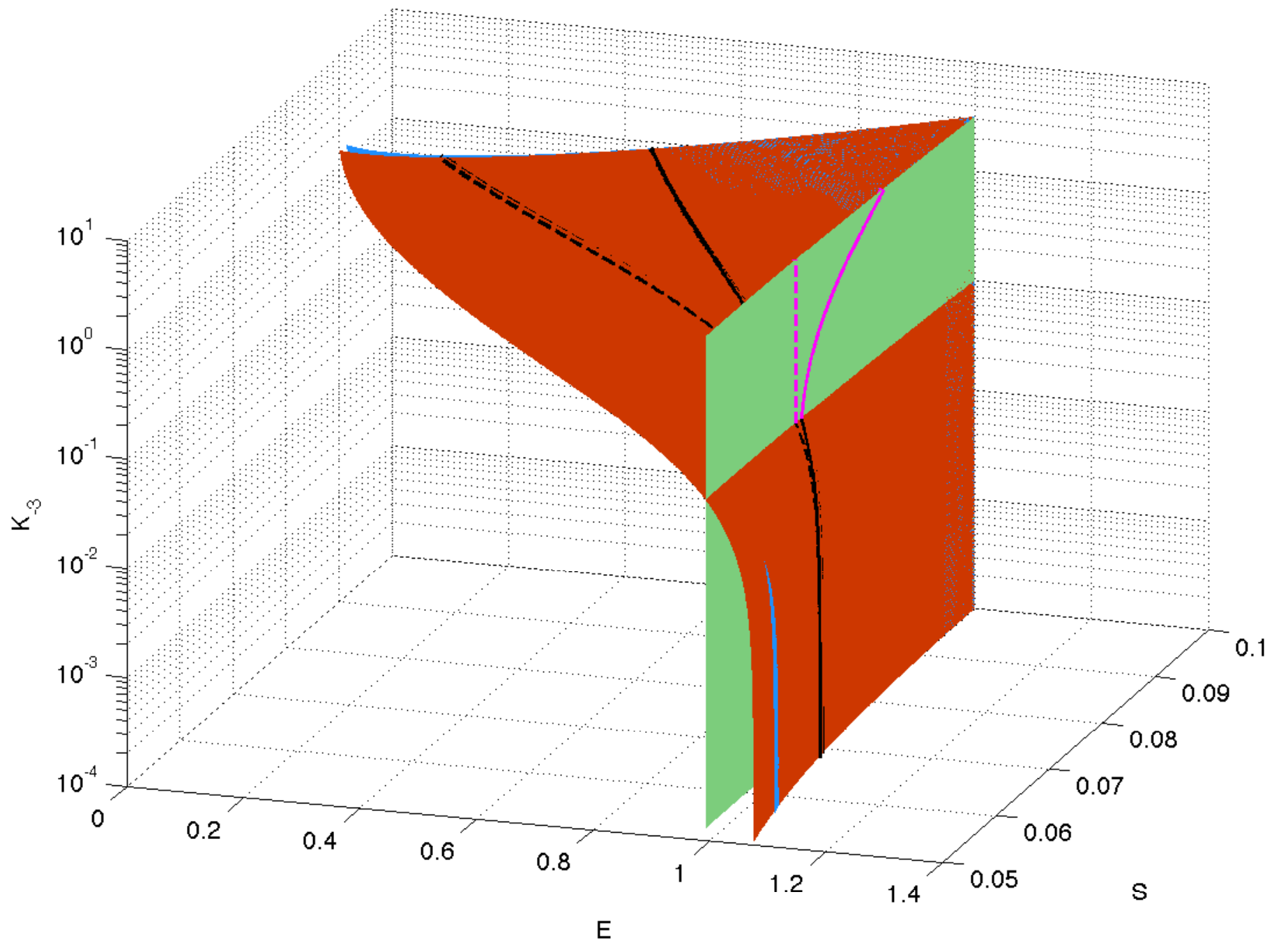}}
	\subfloat[][Complex-Enzyme]{\label{fig:ES_kn3_b_CE}\includegraphics[width= \gwidth]{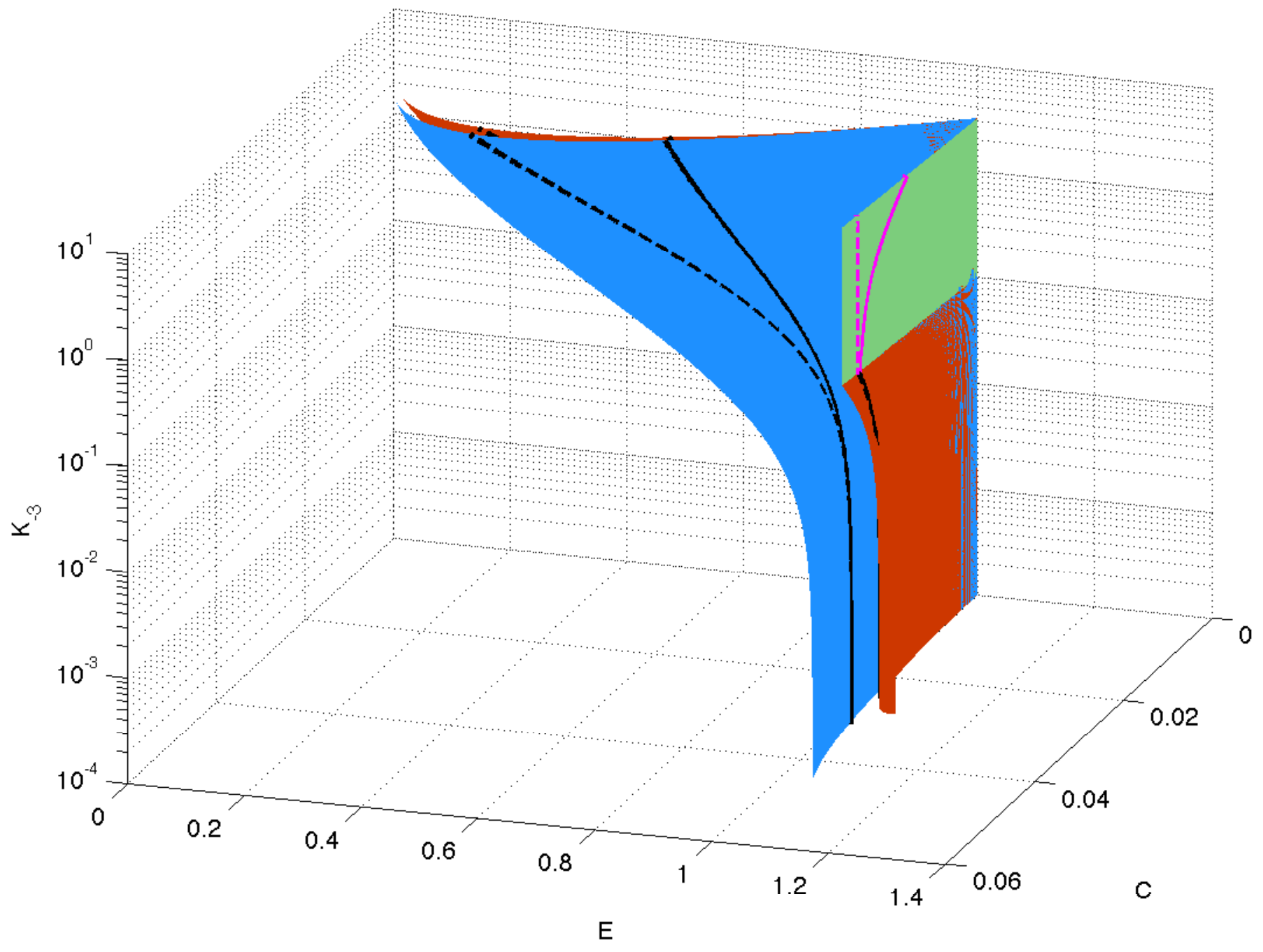}}
	\subfloat[][Complex-Substrate]{\label{fig:ES_kn3_b_CS}\includegraphics[width= \gwidth]{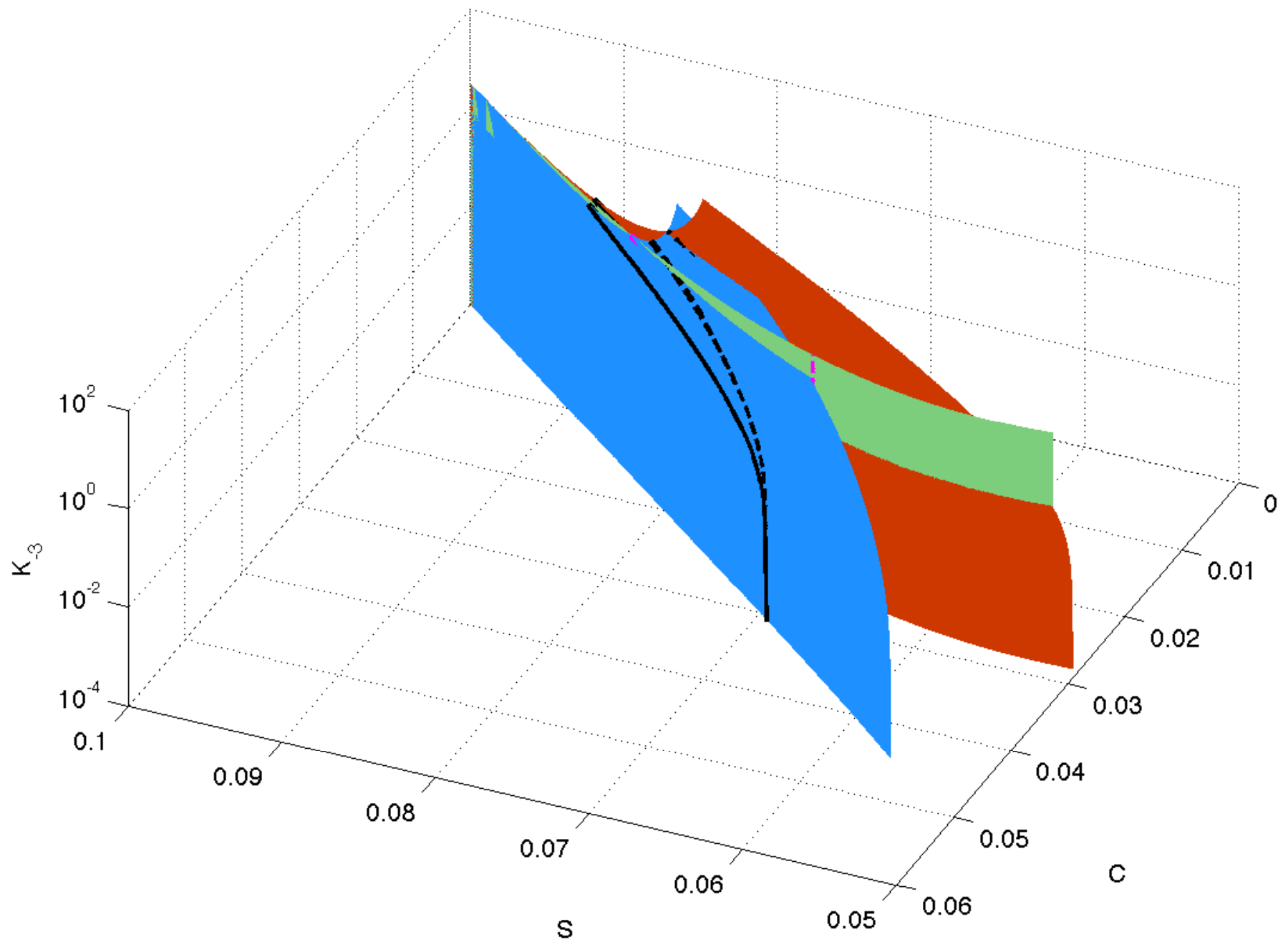}}\\
	\subfloat[][Signed error at $t_{f}$, full open \\system vs. open approximation]{\label{fig:ES_kn3_b_Err}\includegraphics[width= \ewidth]{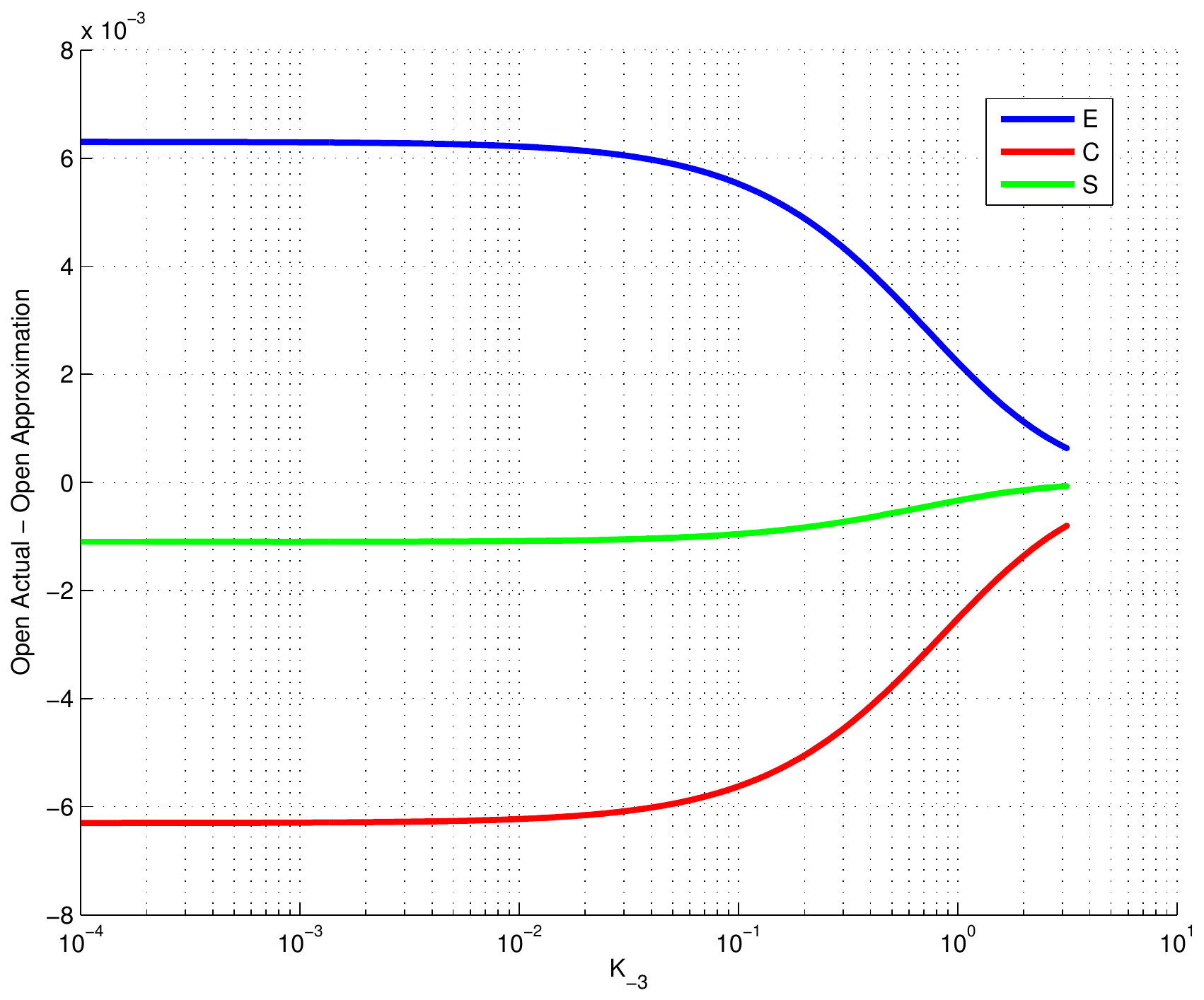}}\hspace{0.5cm}
	\subfloat[][Signed error at $t_{f}$, full open system vs. full closed system]{\label{fig:ES_kn3_b_ErrC}\includegraphics[width= \ewidth]{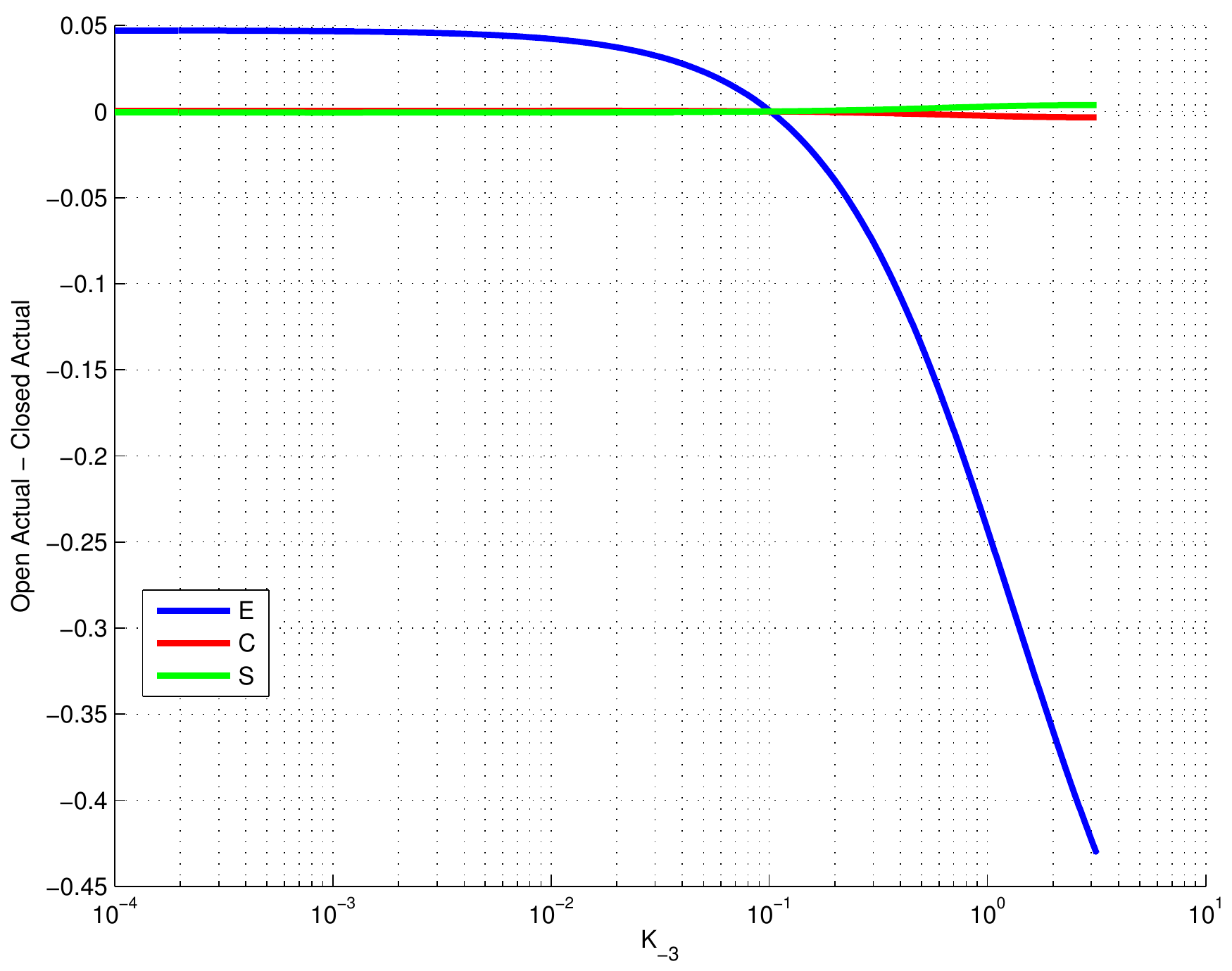}}\\
	\caption[Reverse QSSA regime, varying $\Knk $. $\Kk = 0.1$]{Phase plane portraits and signed error in the reverse QSSA regime as \Knk\ varies, with $\Kk = 0.1$. }
	\label{fig:Q_ES_kn3_b}
\end{figure}

\begin{figure}[!ht]
 	\centering
	\subfloat[][Substrate-Enzyme]{\label{fig:ES_kn3_a_SE}\includegraphics[width= \gwidth]{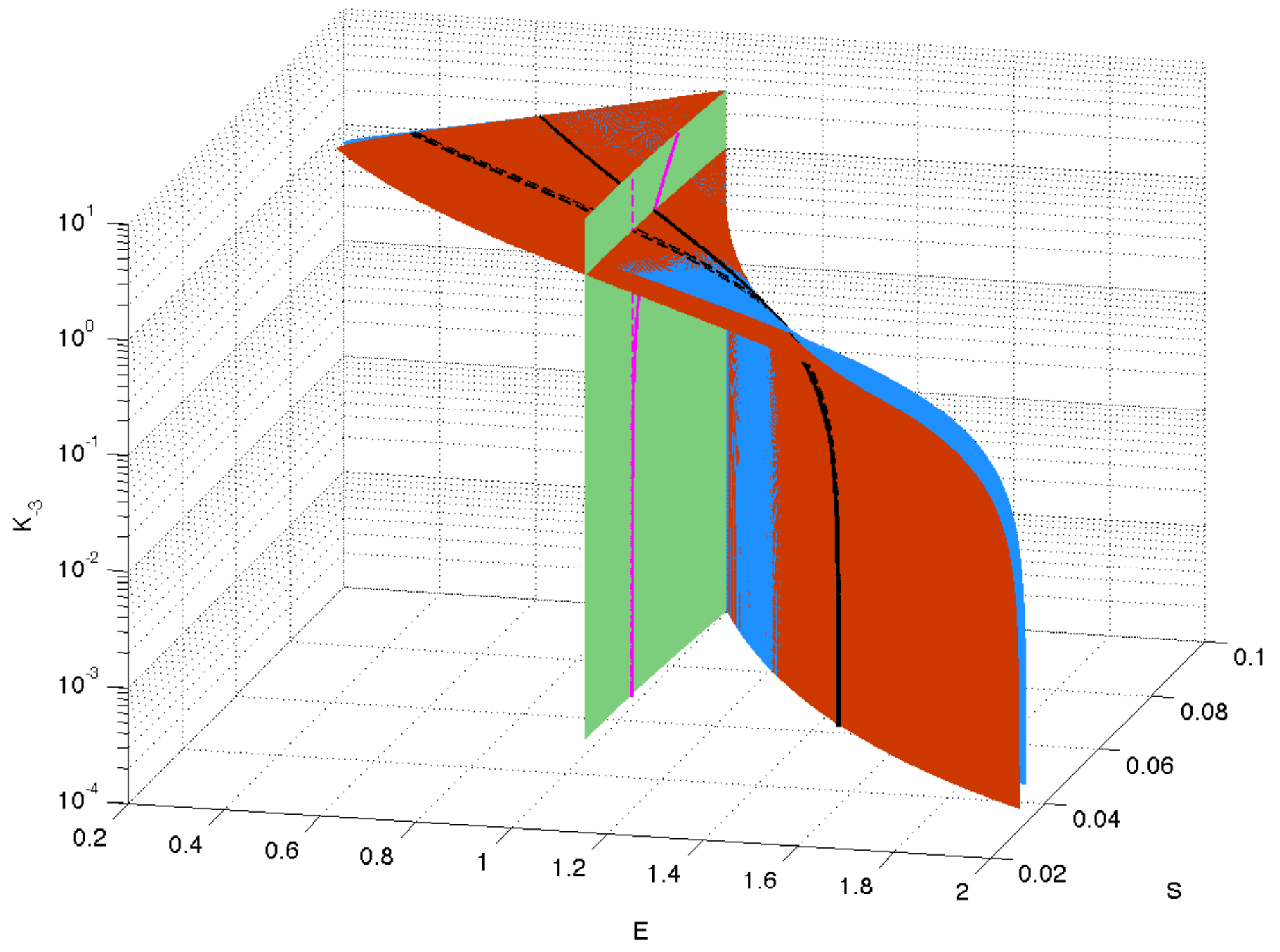}}
	\subfloat[][Complex-Enzyme]{\label{fig:ES_kn3_a_CE}\includegraphics[width= \gwidth]{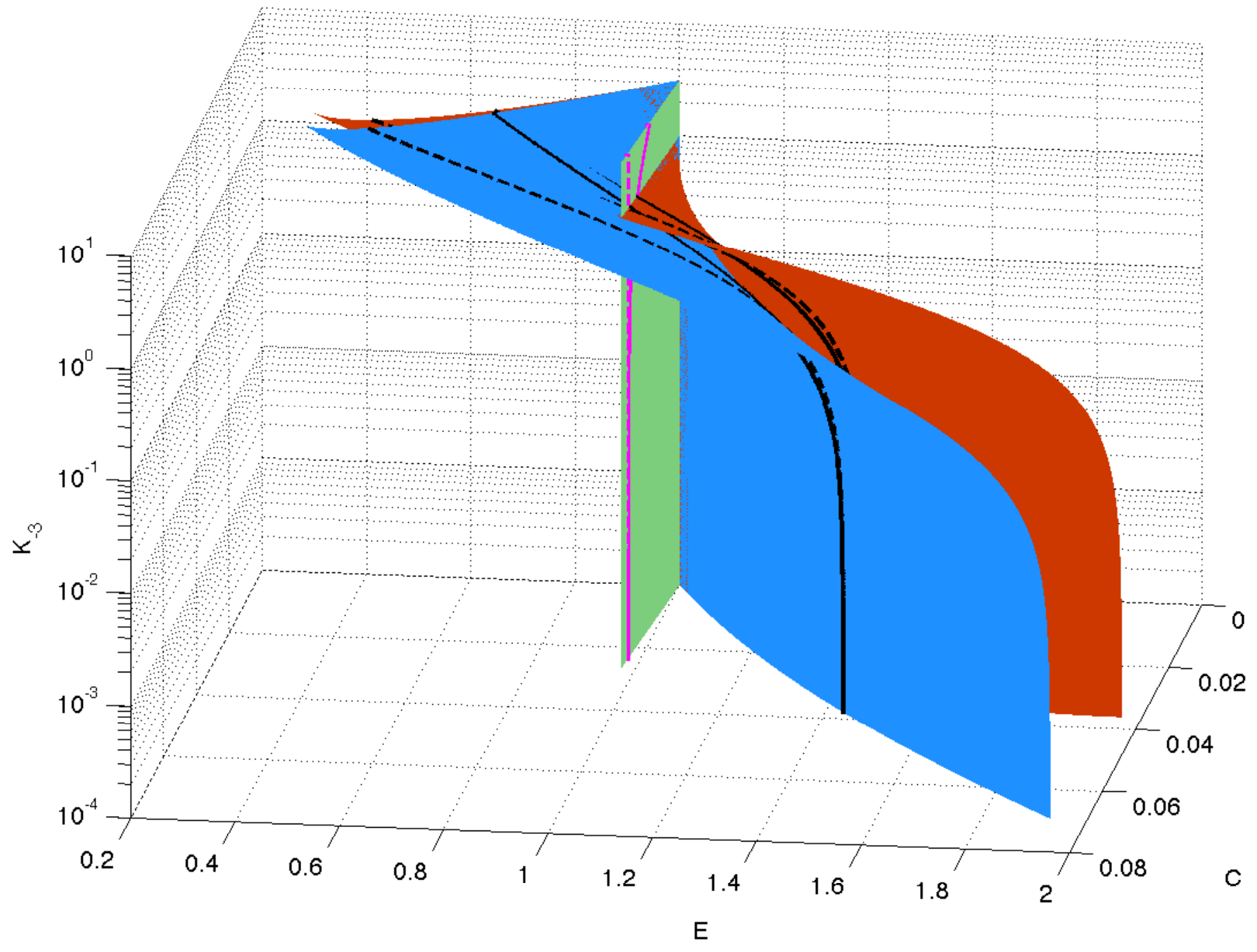}}
	\subfloat[][Complex-Substrate]{\label{fig:ES_kn3_a_CS}\includegraphics[width= \gwidth]{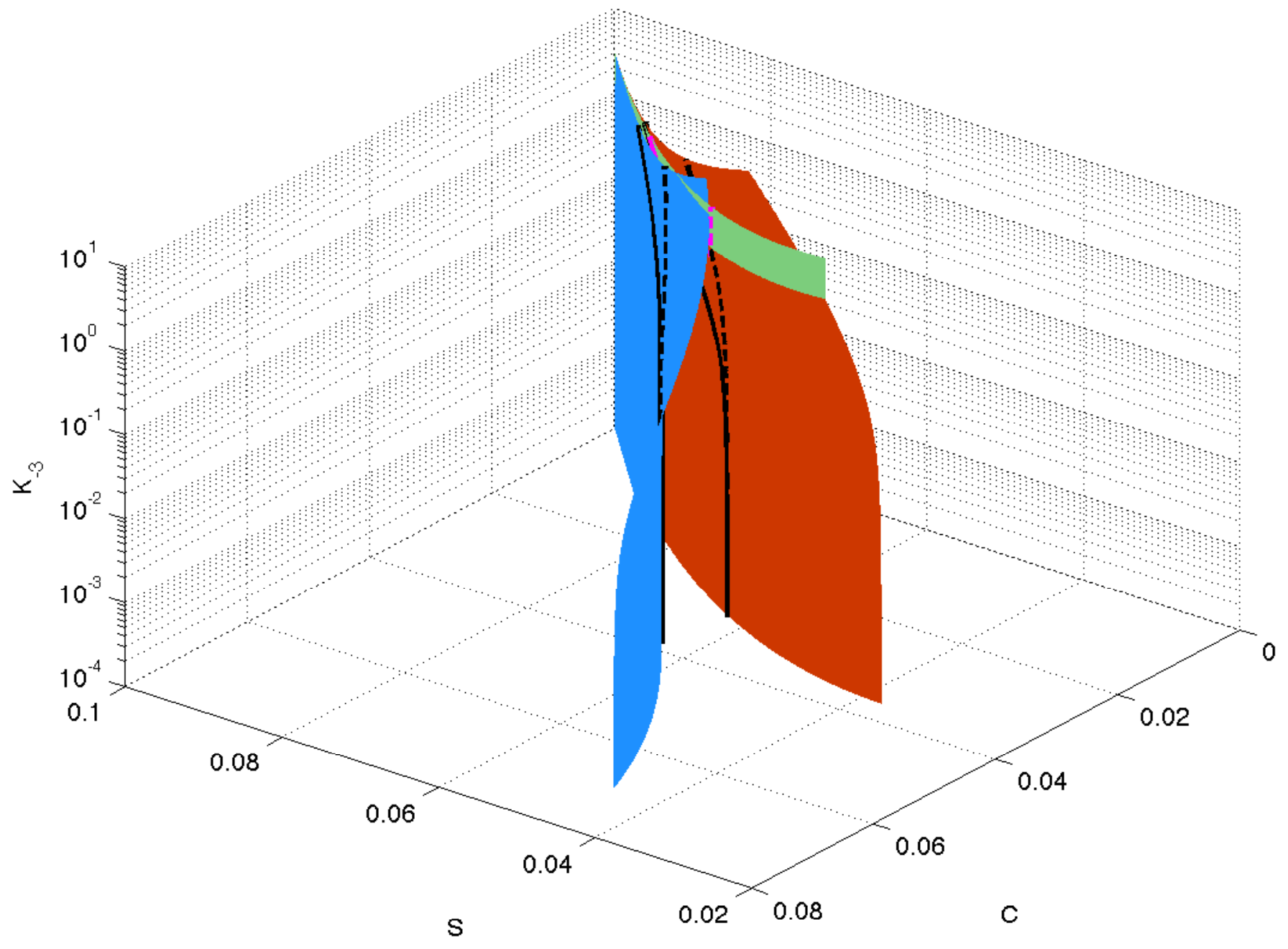}}\\
	\subfloat[][Signed error at $t_{f}$, full open \\system vs. open approximation]{\label{fig:ES_kn3_a_Err}\includegraphics[width= \ewidth]{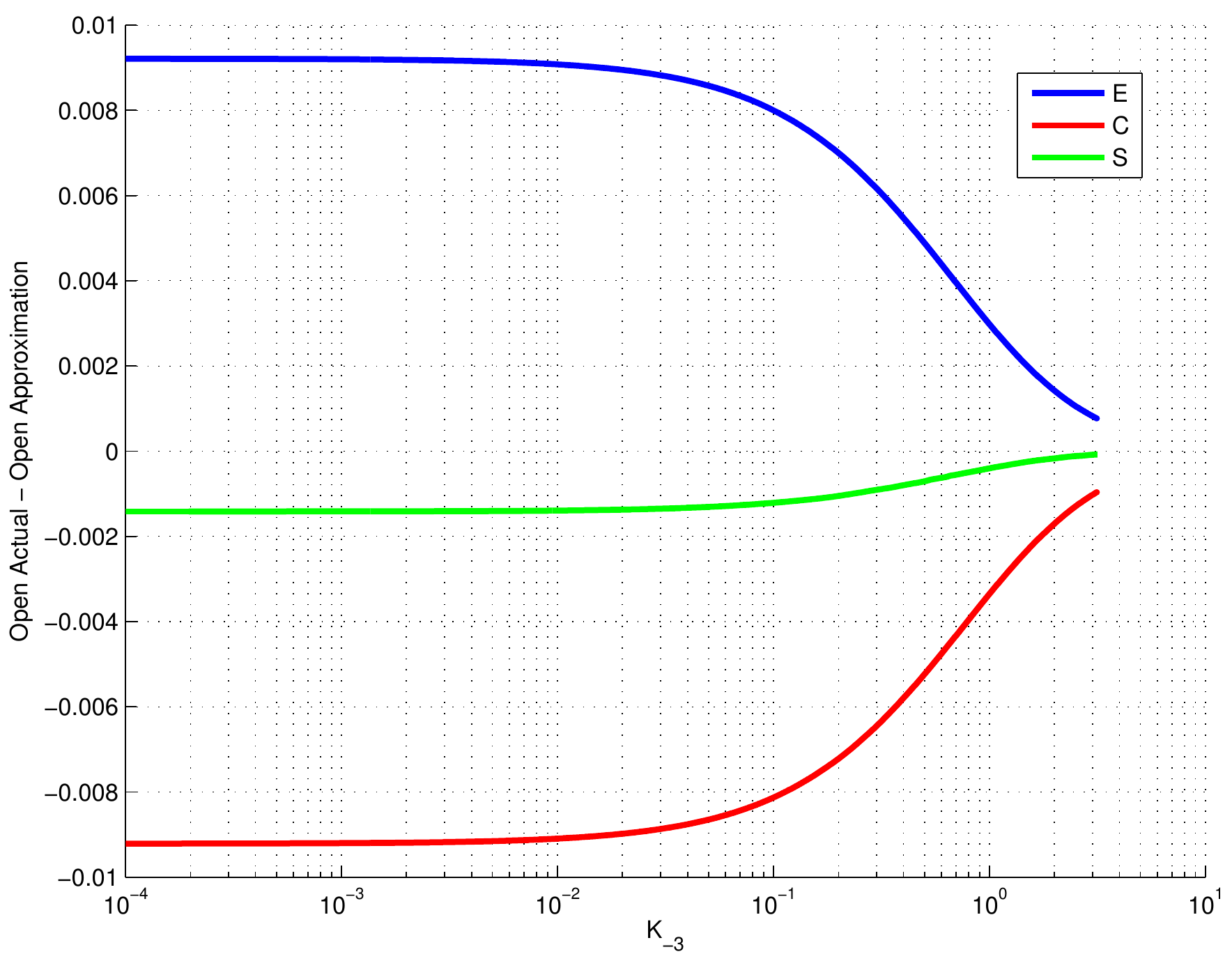}}\hspace{0.5cm}
	\subfloat[][Signed error at $t_{f}$, full open system vs. full closed system]{\label{fig:ES_kn3_a_ErrC}\includegraphics[width= \ewidth]{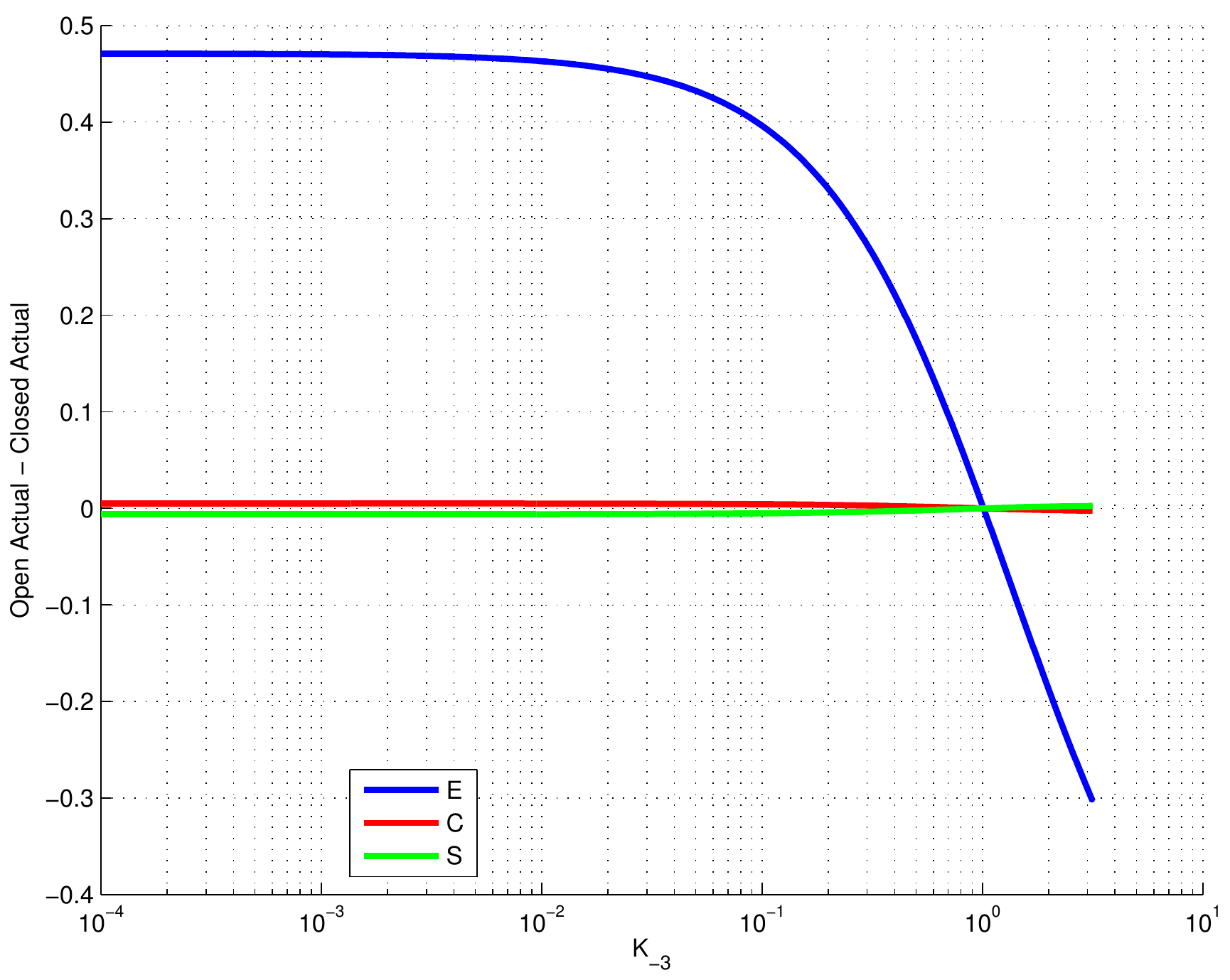}}\\
	\caption[Reverse QSSA regime, varying $\Knk $. $\Kk = 1$]{Phase plane portraits and signed error in the reverse QSSA regime as \Knk\ varies, with $\Kk = 1$.  }
	\label{fig:Q_ES_kn3_a}
\end{figure}



\begin{figure}[!ht]
 	\centering
	\subfloat[][Substrate-Enzyme]{\label{fig:ES_k3_d_SE}\includegraphics[width= \gwidth]{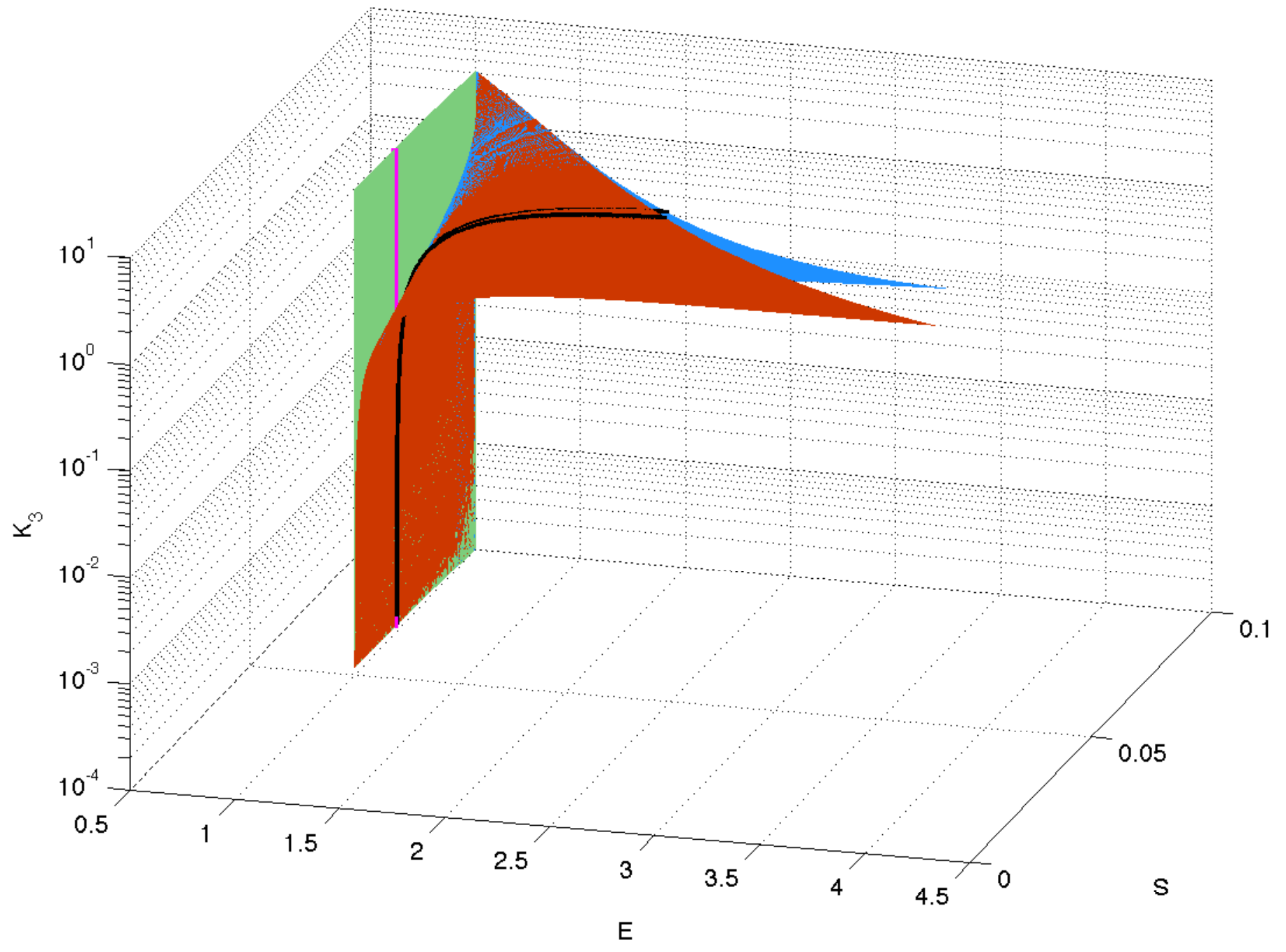}}
	\subfloat[][Complex-Enzyme]{\label{fig:ES_k3_d_CE}\includegraphics[width= \gwidth]{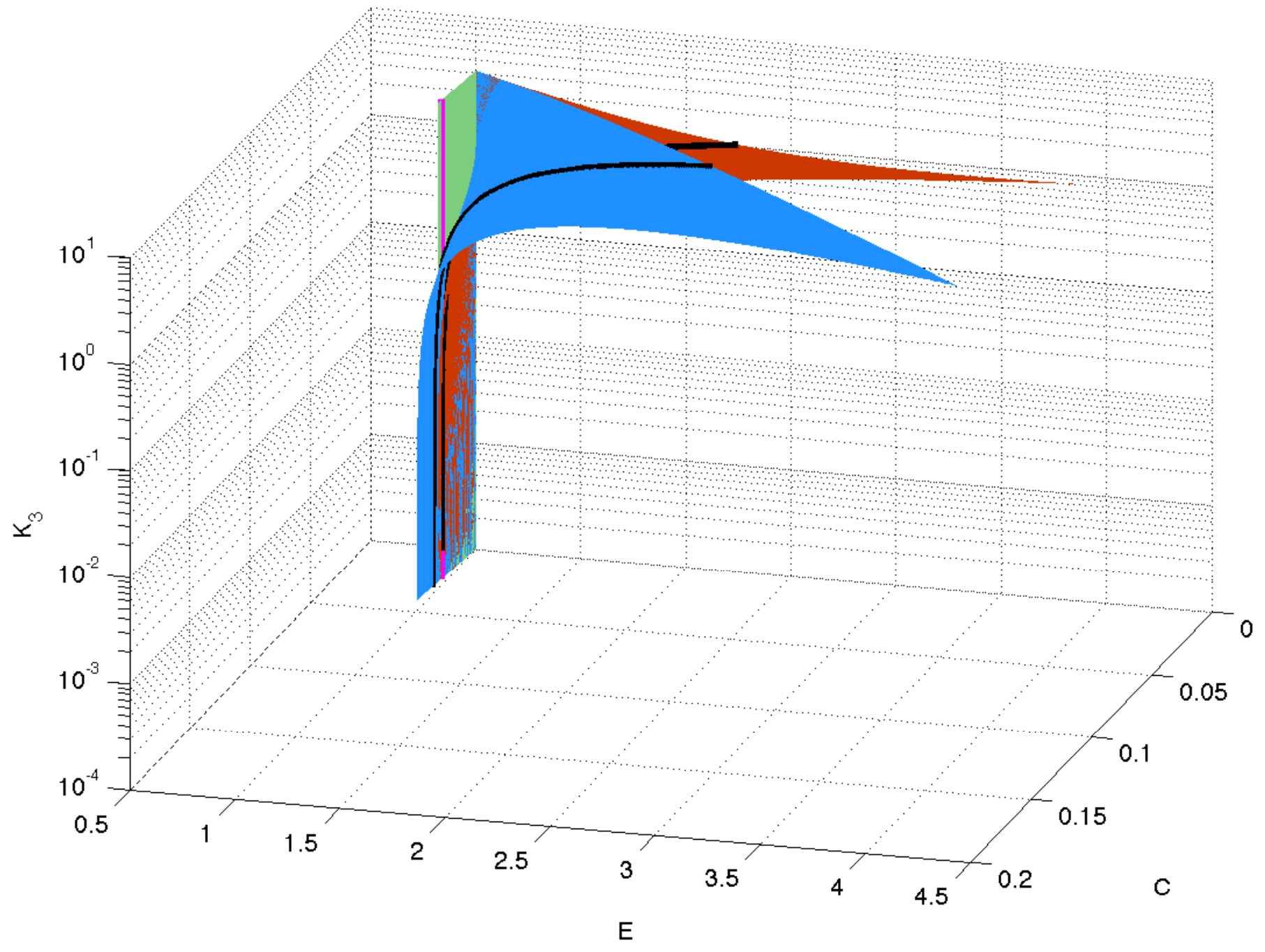}}
	\subfloat[][Complex-Substrate]{\label{fig:ES_k3_d_CS}\includegraphics[width= \gwidth]{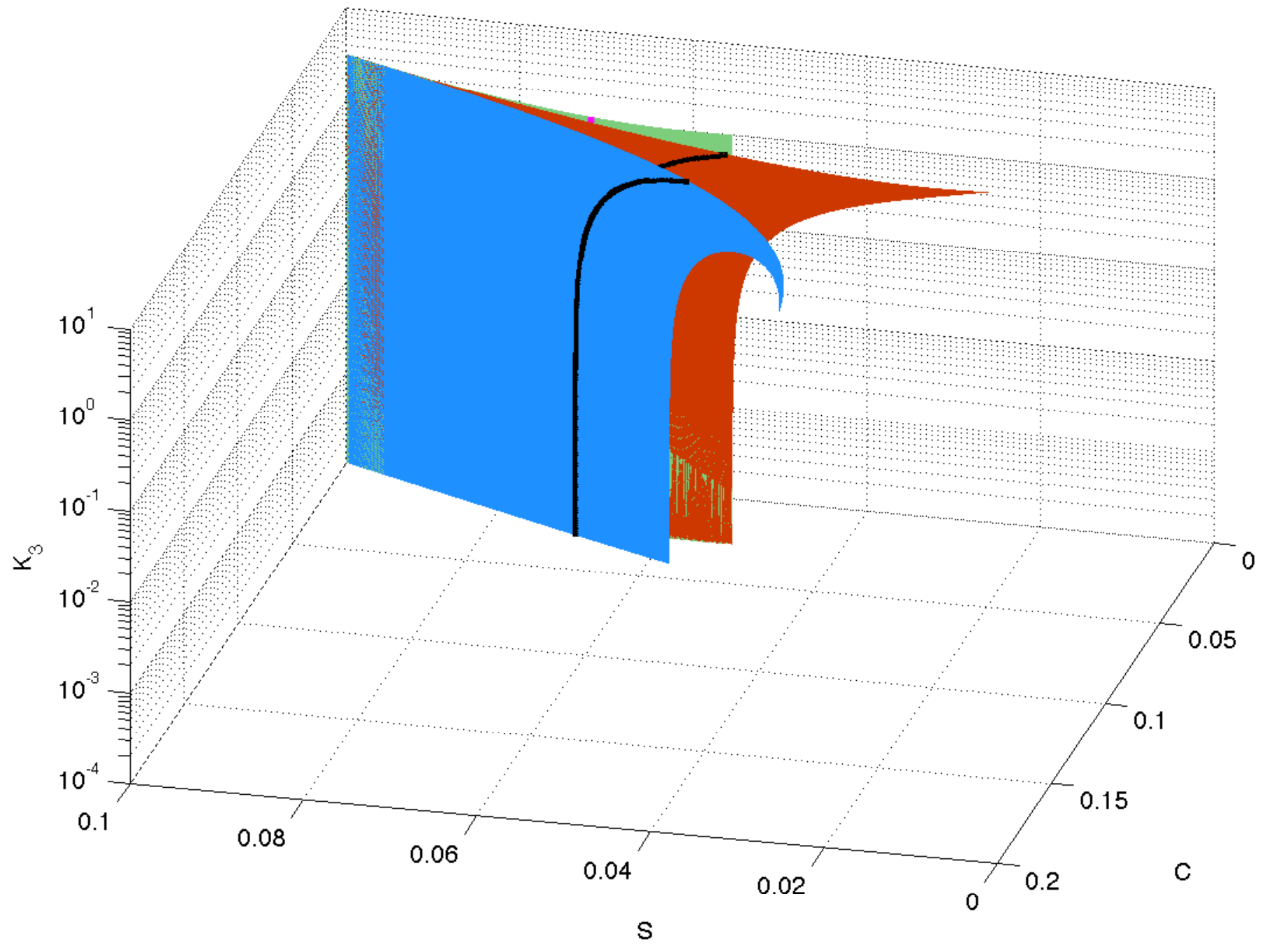}}\\
	\subfloat[][Signed error at $t_{f}$, full open \\system vs. open approximation]{\label{fig:ES_k3_d_Err}\includegraphics[width= \ewidth]{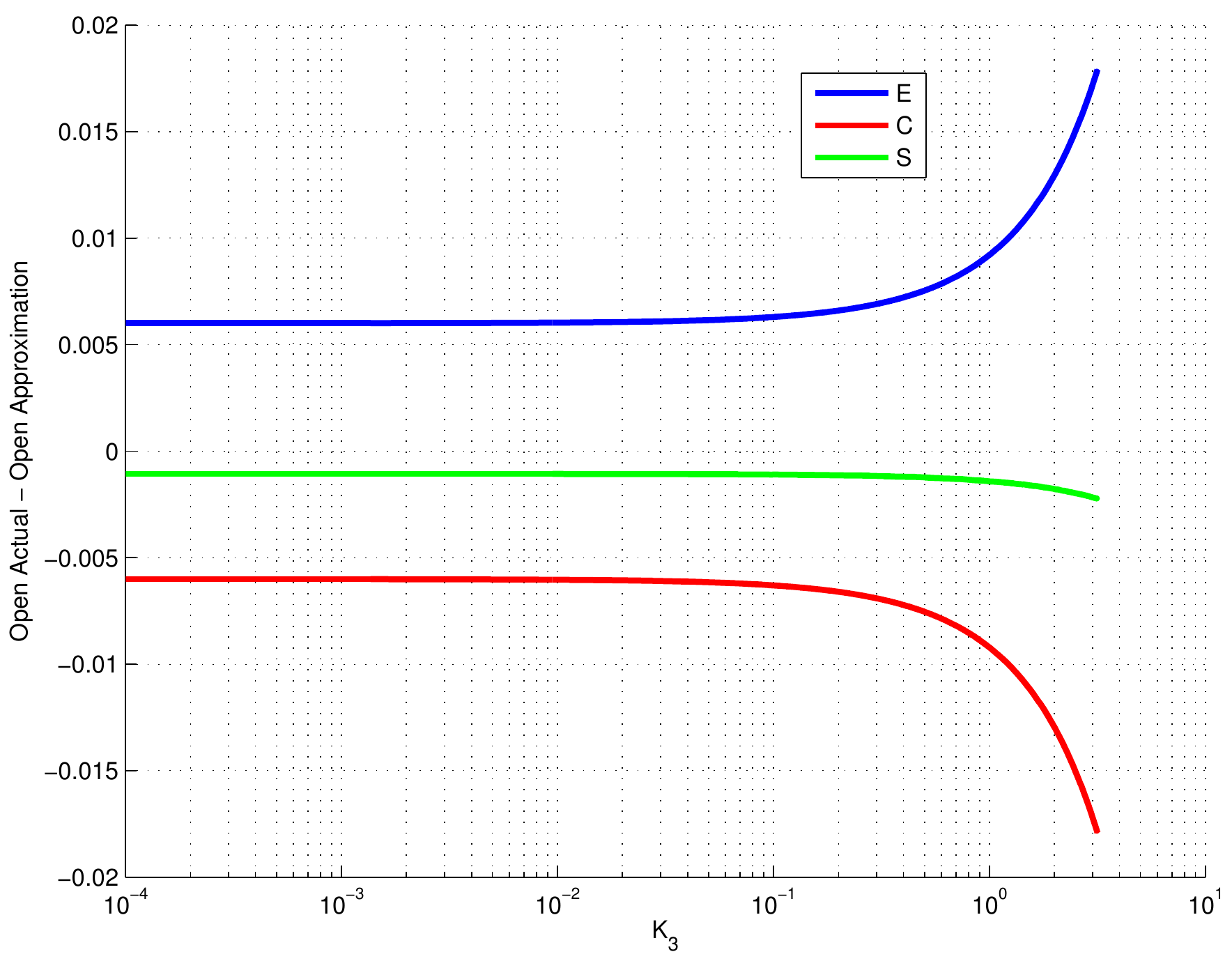}}\hspace{0.5cm}
	\subfloat[][Signed error at $t_{f}$, full open system vs. full closed system]{\label{fig:ES_k3_d_ErrC}\includegraphics[width= \ewidth]{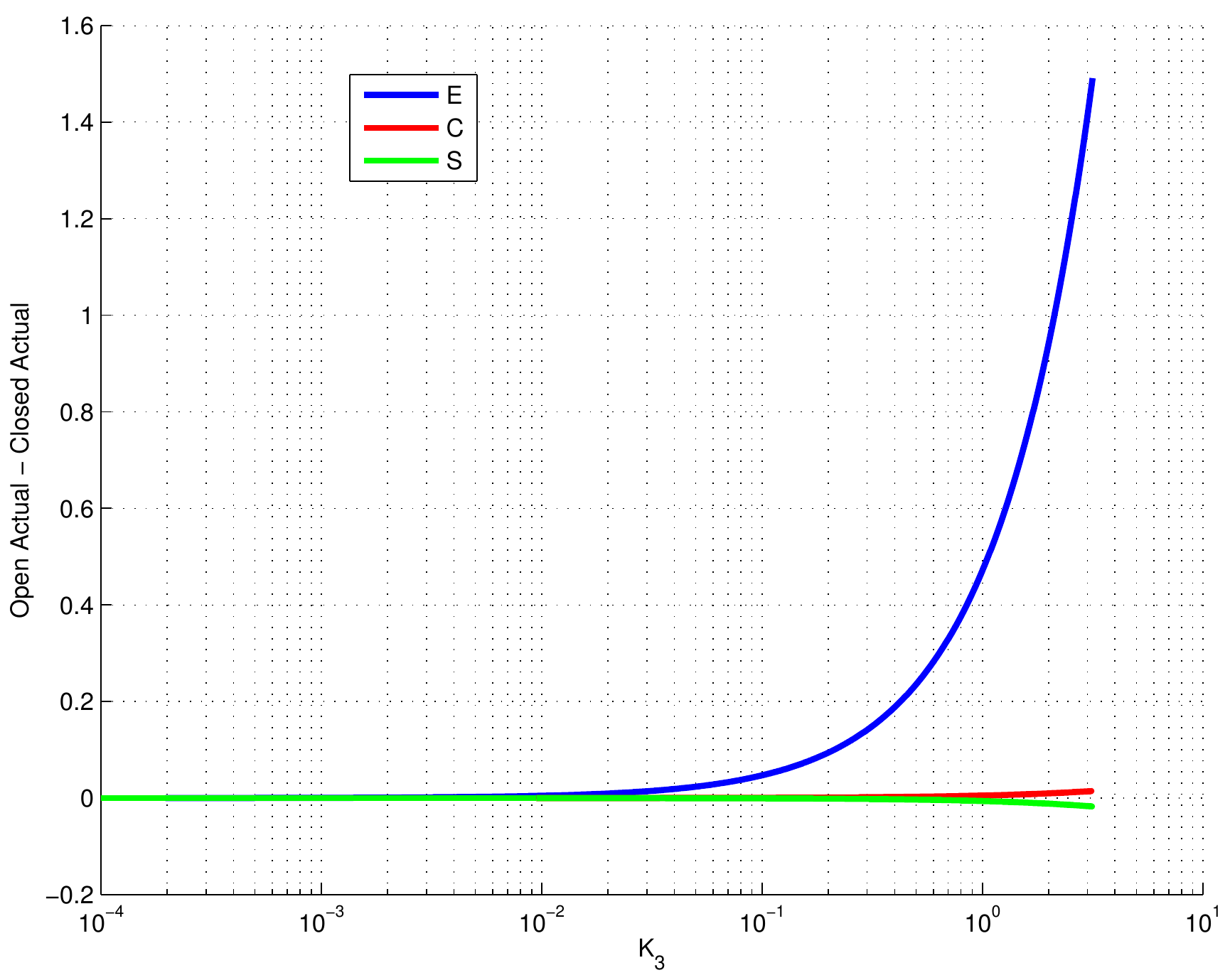}}\\
	\caption[Reverse QSSA regime, varying $\Kk $. $\Knk  = 0.0001$]{Phase plane portraits and signed error in the reverse QSSA regime as \Kk\ varies, with $\Knk = 0.0001$.  }
	\label{fig:Q_ES_k3_d}
\end{figure}

\begin{figure}[!ht]
 	\centering
	\subfloat[][Substrate-Enzyme]{\label{fig:ES_k3_b_SE}\includegraphics[width= \gwidth]{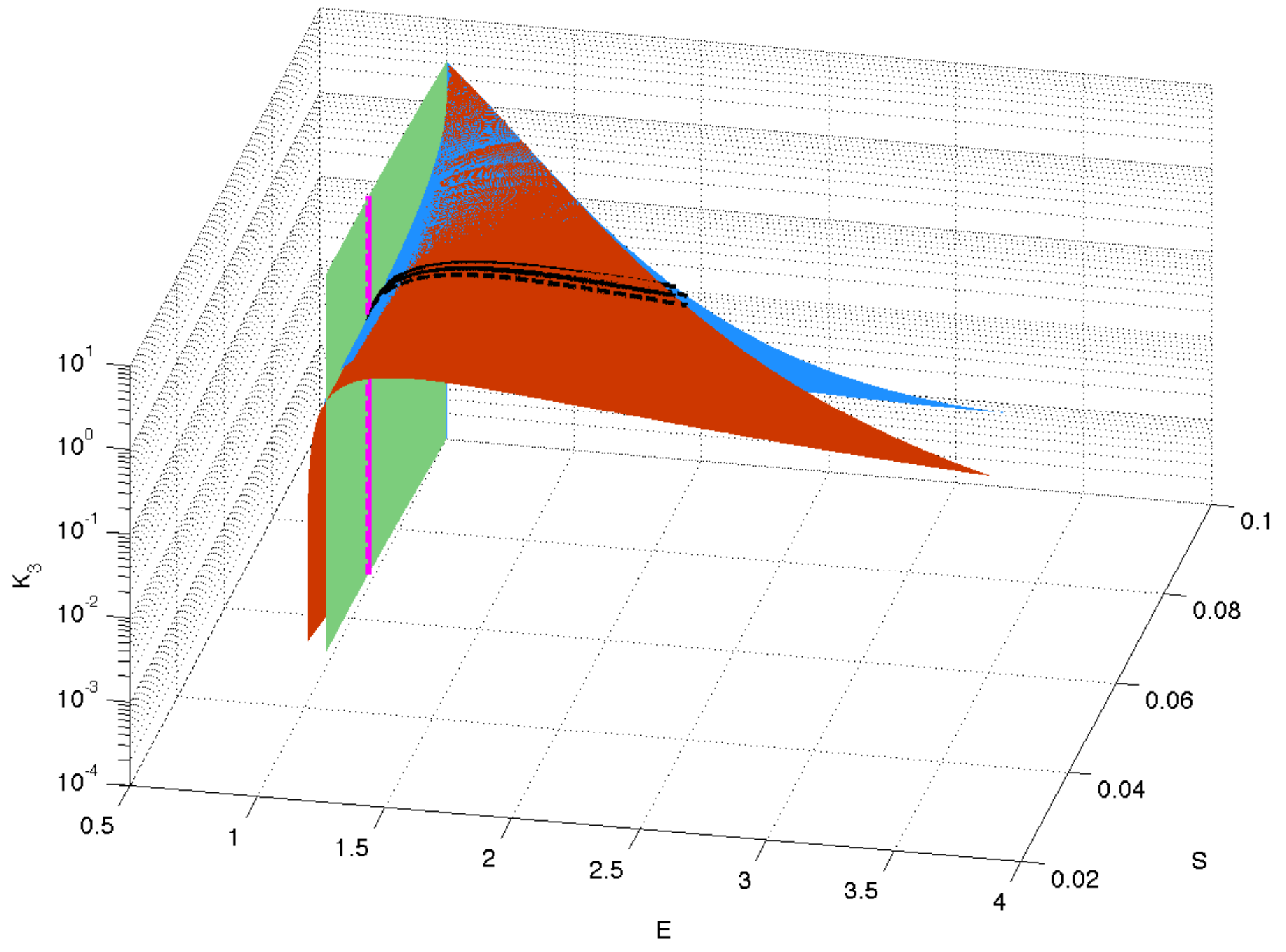}}
	\subfloat[][Complex-Enzyme]{\label{fig:ES_k3_b_CE}\includegraphics[width= \gwidth]{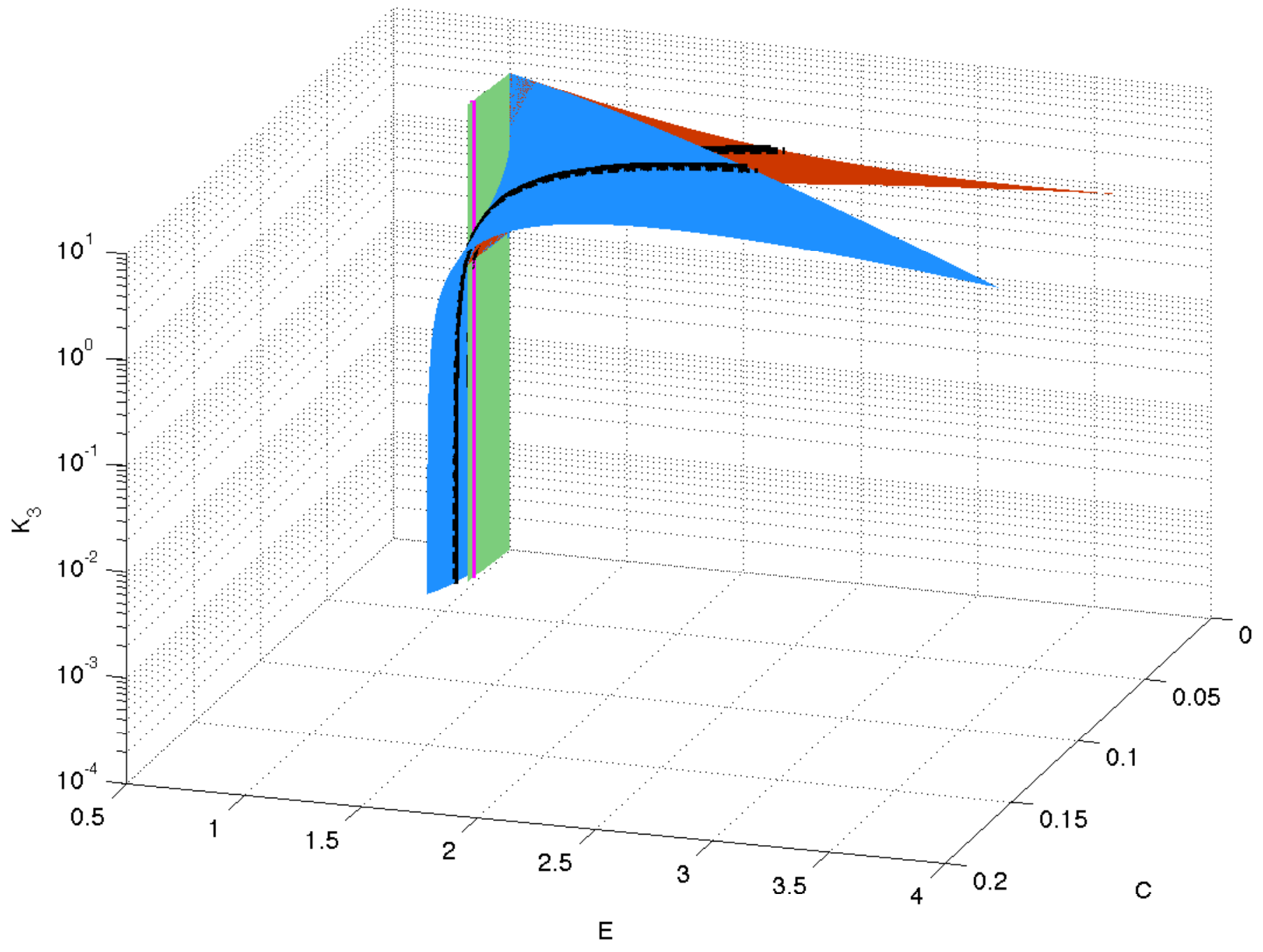}}
	\subfloat[][Complex-Substrate]{\label{fig:ES_k3_b_CS}\includegraphics[width= \gwidth]{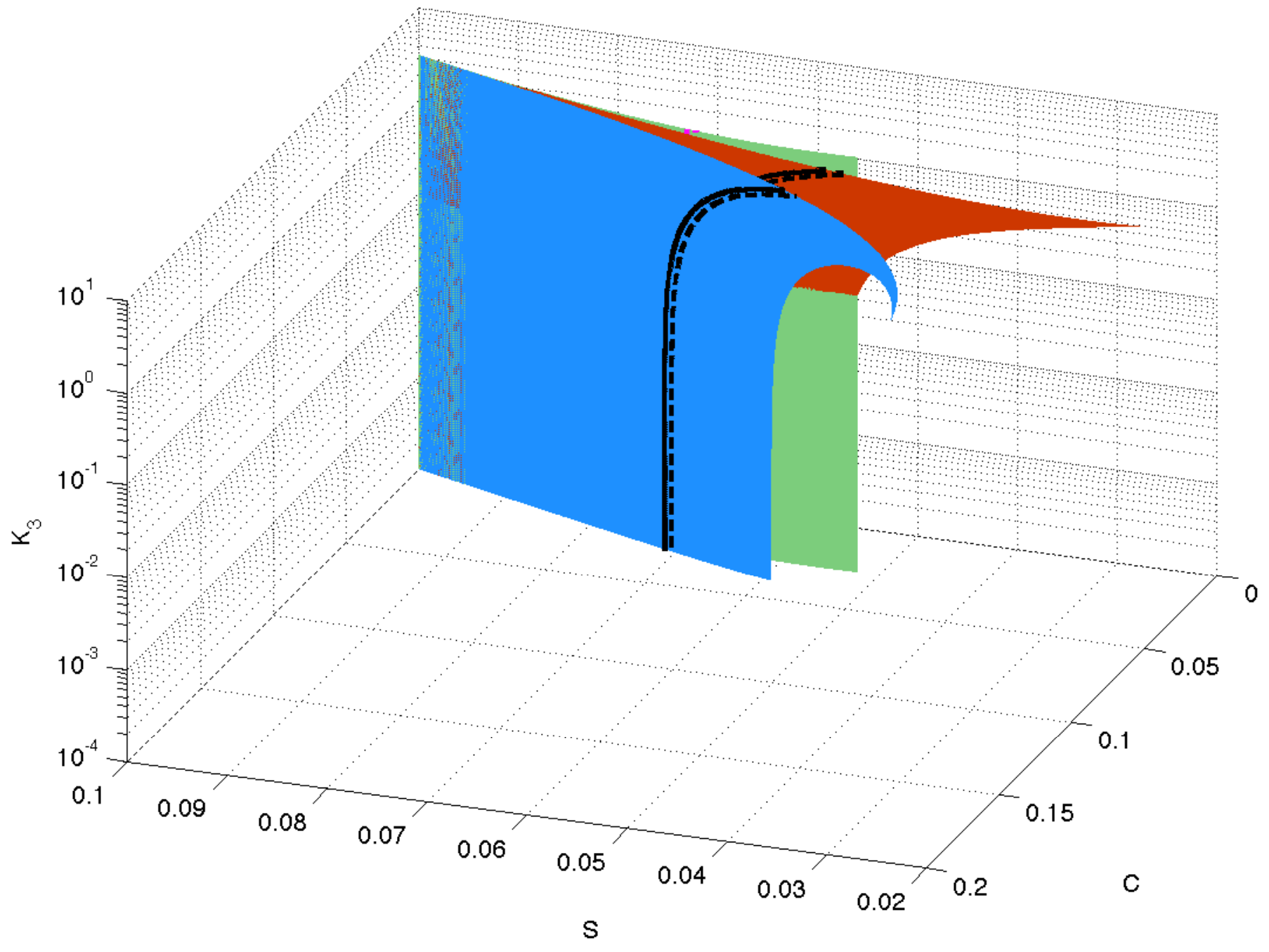}}\\
	\subfloat[][Signed error at $t_{f}$, full open \\system vs. open approximation]{\label{fig:ES_k3_b_Err}\includegraphics[width= \ewidth]{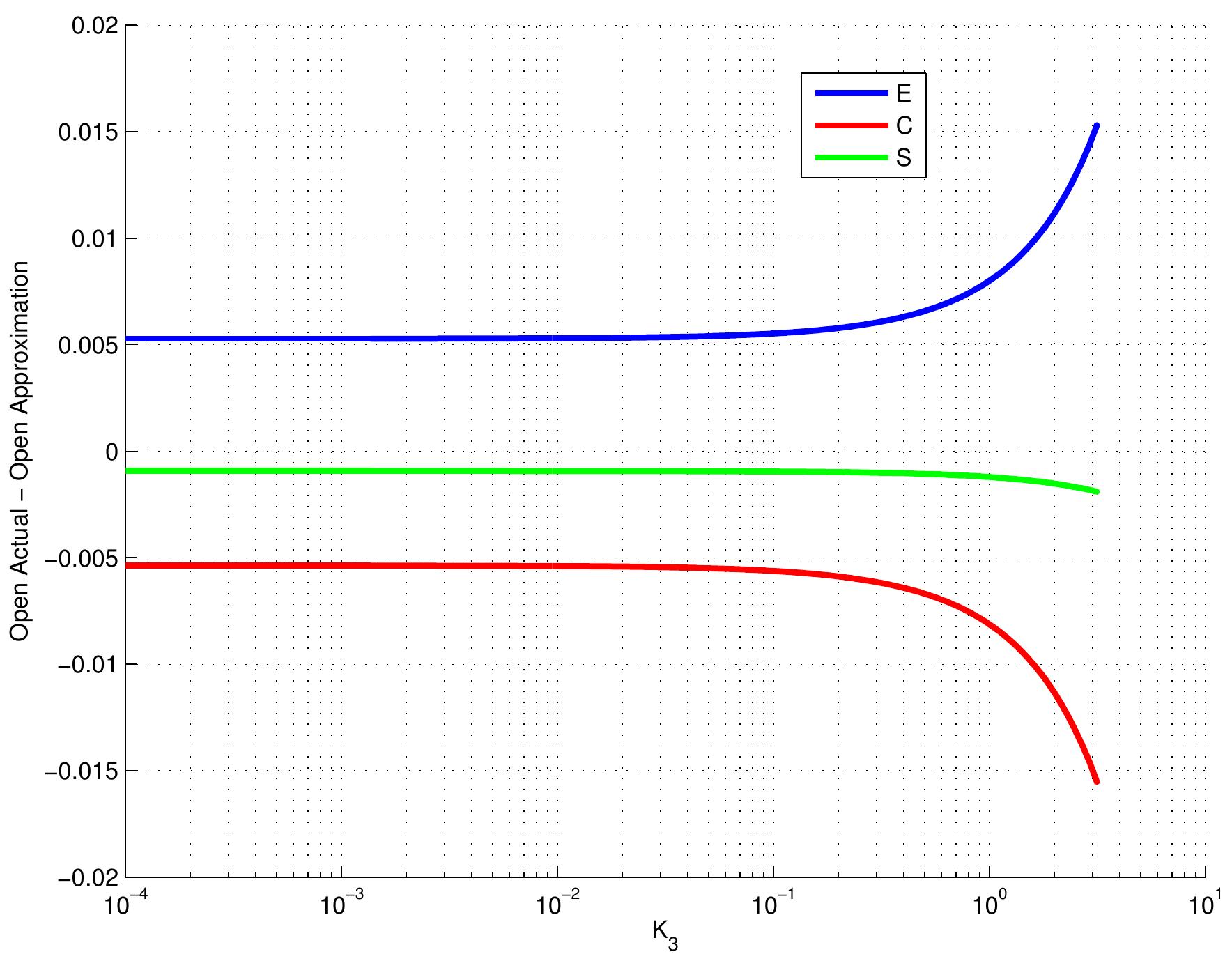}}\hspace{0.5cm}
	\subfloat[][Signed error at $t_{f}$, full open system vs. full closed system]{\label{fig:ES_k3_b_ErrC}\includegraphics[width= \ewidth]{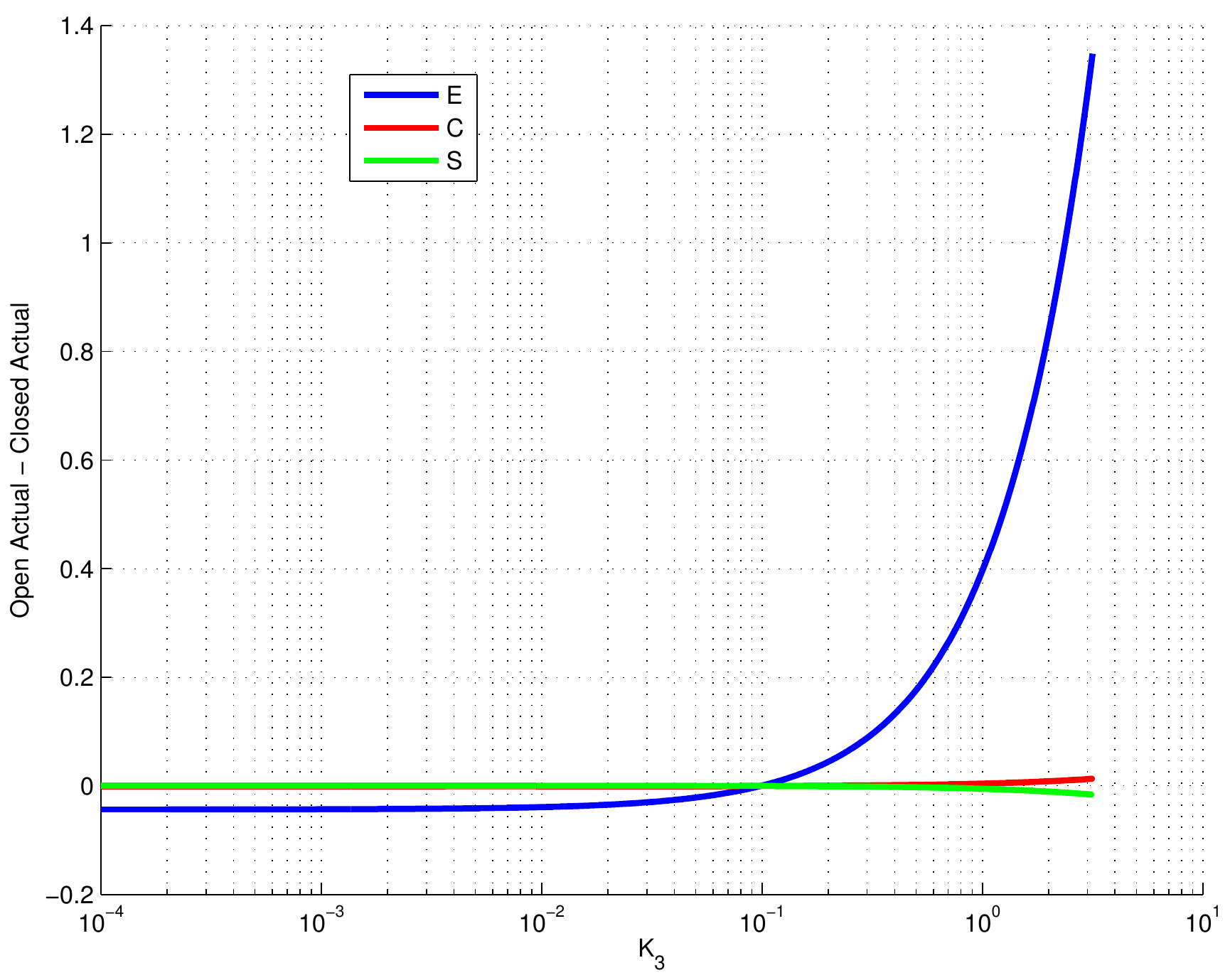}}\\
	\caption[Reverse QSSA regime, varying $\Kk $. $\Knk  = 0.1$]{Phase plane portraits and signed error in the reverse QSSA regime as \Kk\ varies, with $\Knk = 0.1$.  }
	\label{fig:Q_ES_k3_b}
\end{figure}

\begin{figure}[!ht]
 	\centering
	\subfloat[][Substrate-Enzyme]{\label{fig:ES_k3_a_SE}\includegraphics[width= \gwidth]{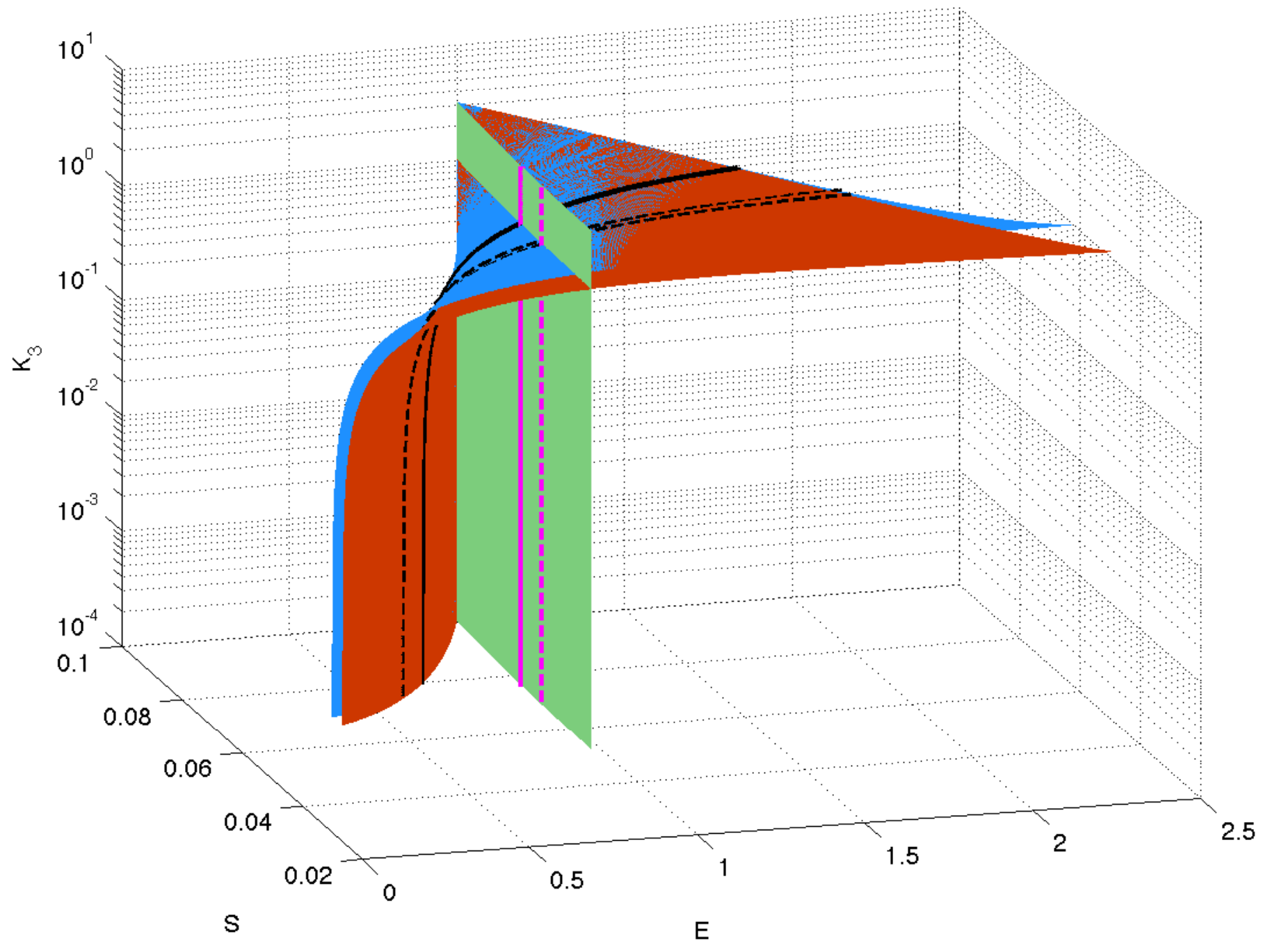}}
	\subfloat[][Complex-Enzyme]{\label{fig:ES_k3_a_CE}\includegraphics[width= \gwidth]{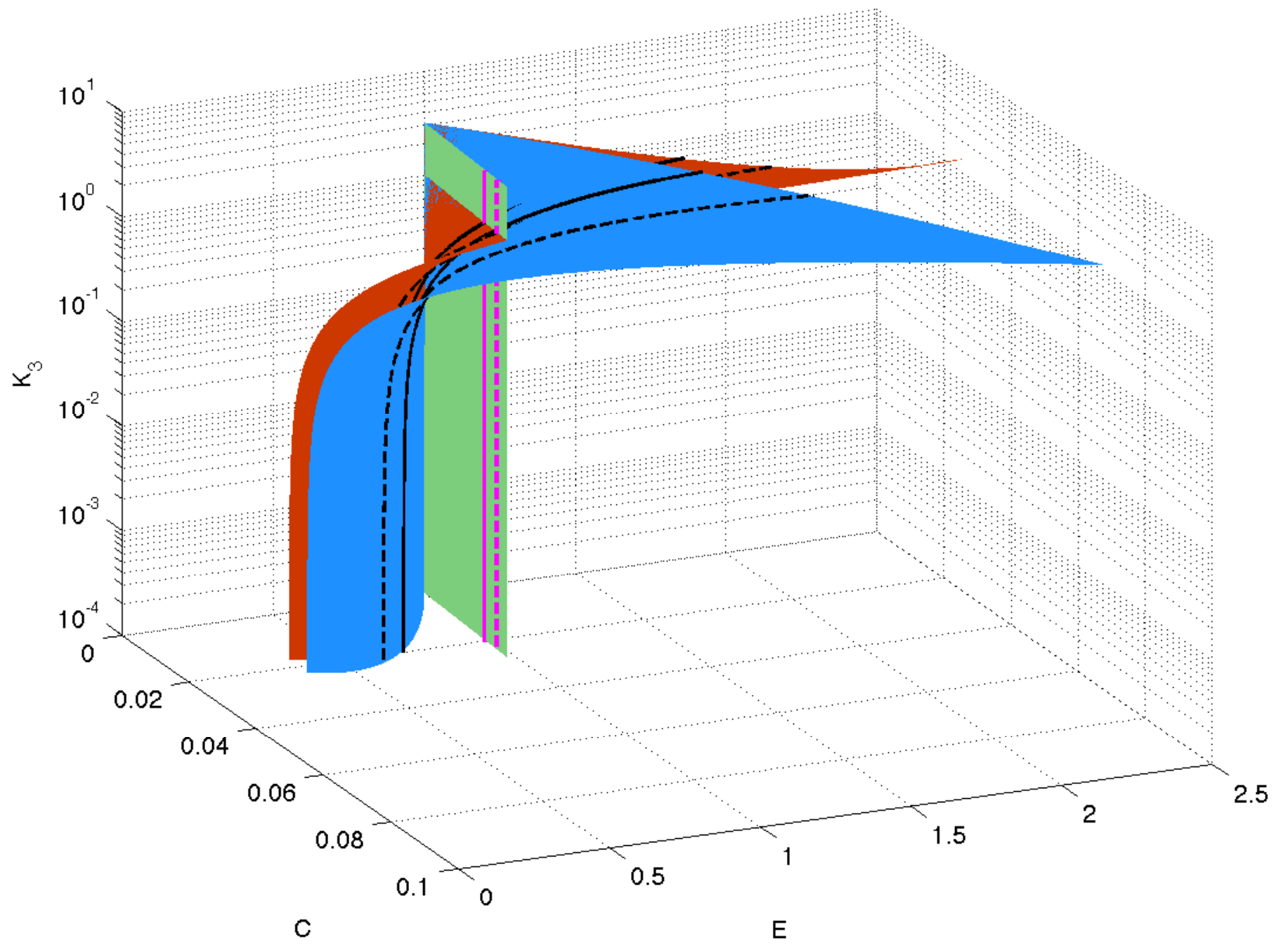}}
	\subfloat[][Complex-Substrate]{\label{fig:ES_k3_a_CS}\includegraphics[width= \gwidth]{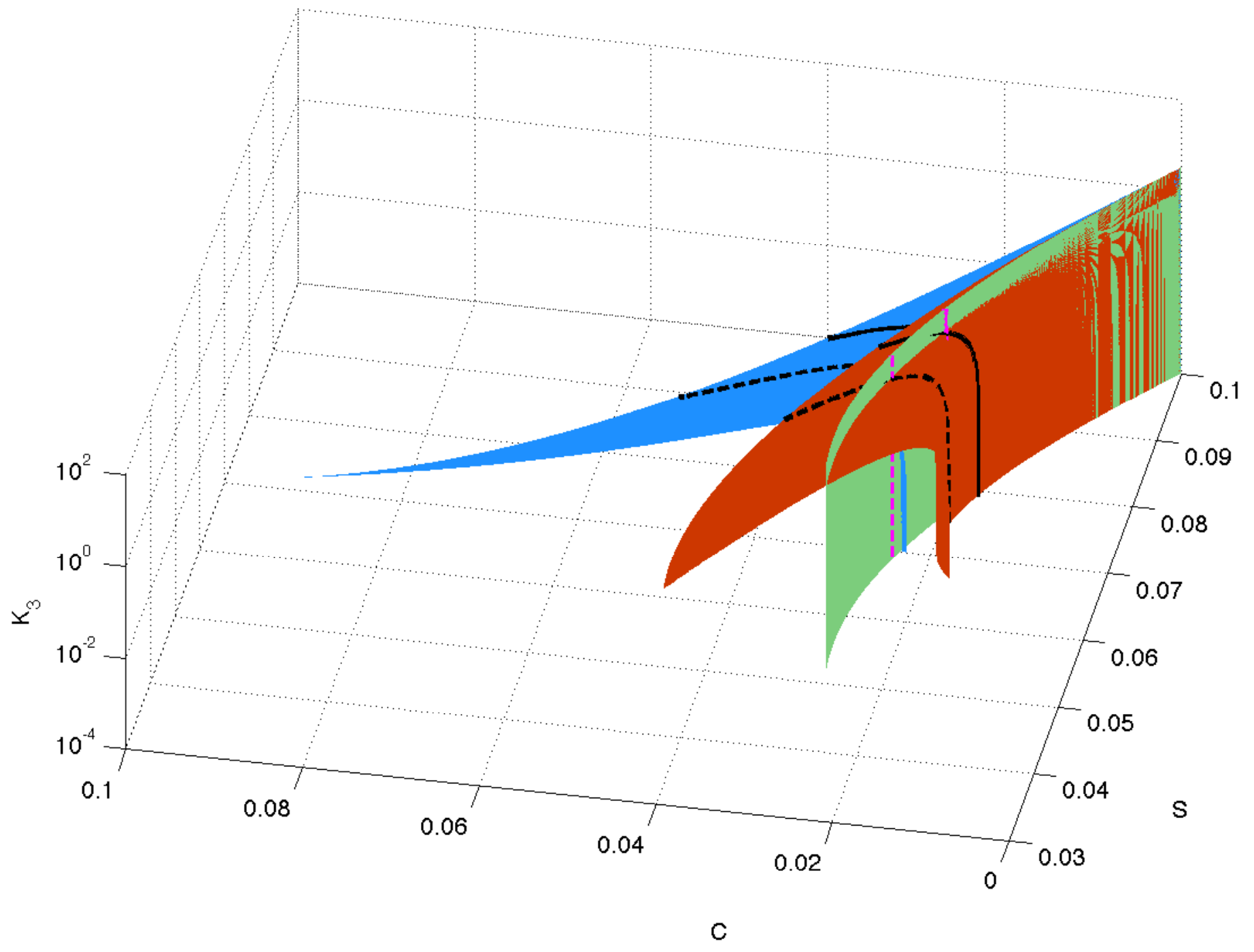}}\\
	\subfloat[][Signed error at $t_{f}$, full open \\system vs. open approximation]{\label{fig:ES_k3_a_Err}\includegraphics[width= \ewidth]{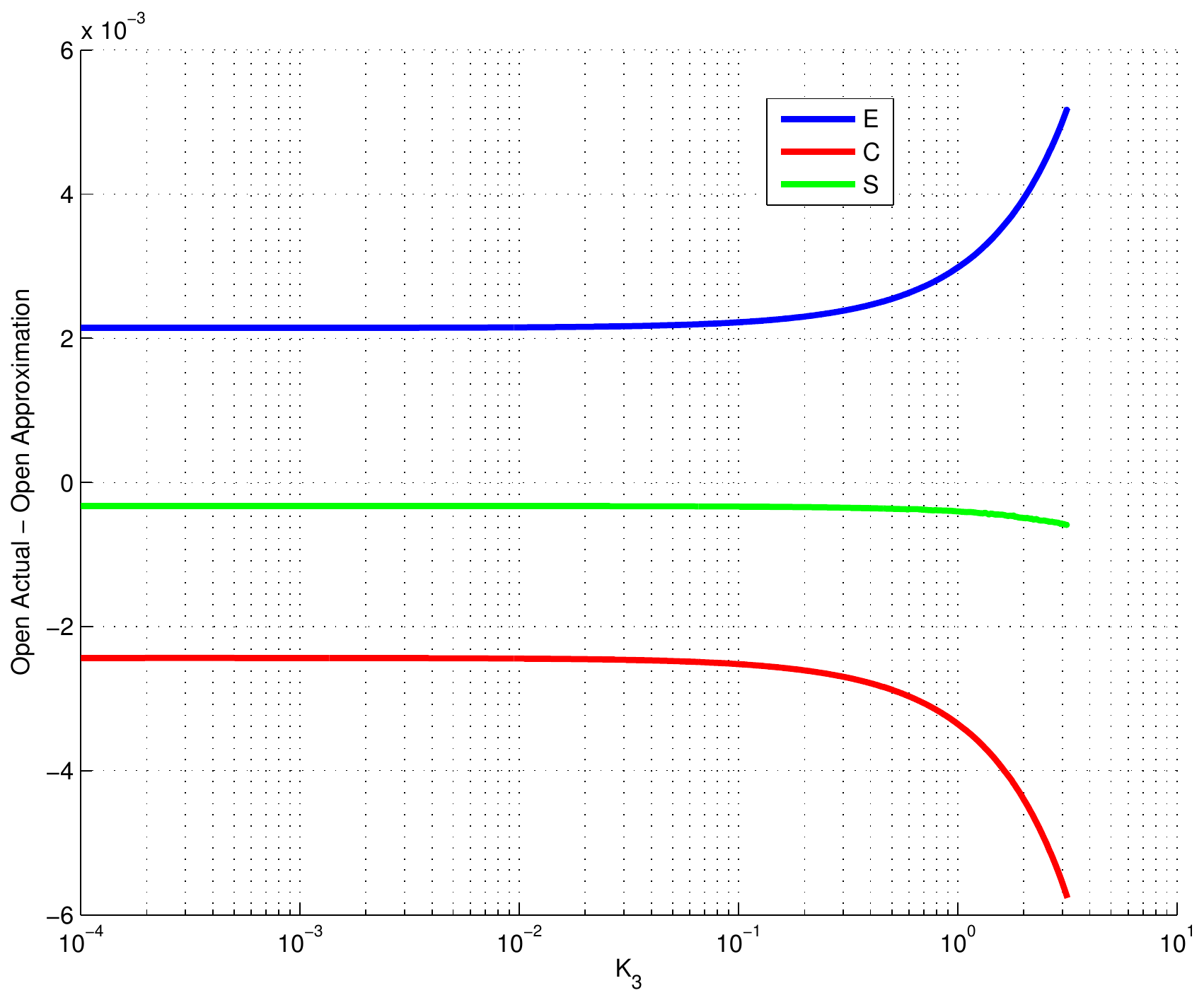}}\hspace{0.5cm}
	\subfloat[][Signed error at $t_{f}$, full open system vs. full closed system]{\label{fig:ES_k3_a_ErrC}\includegraphics[width= \ewidth]{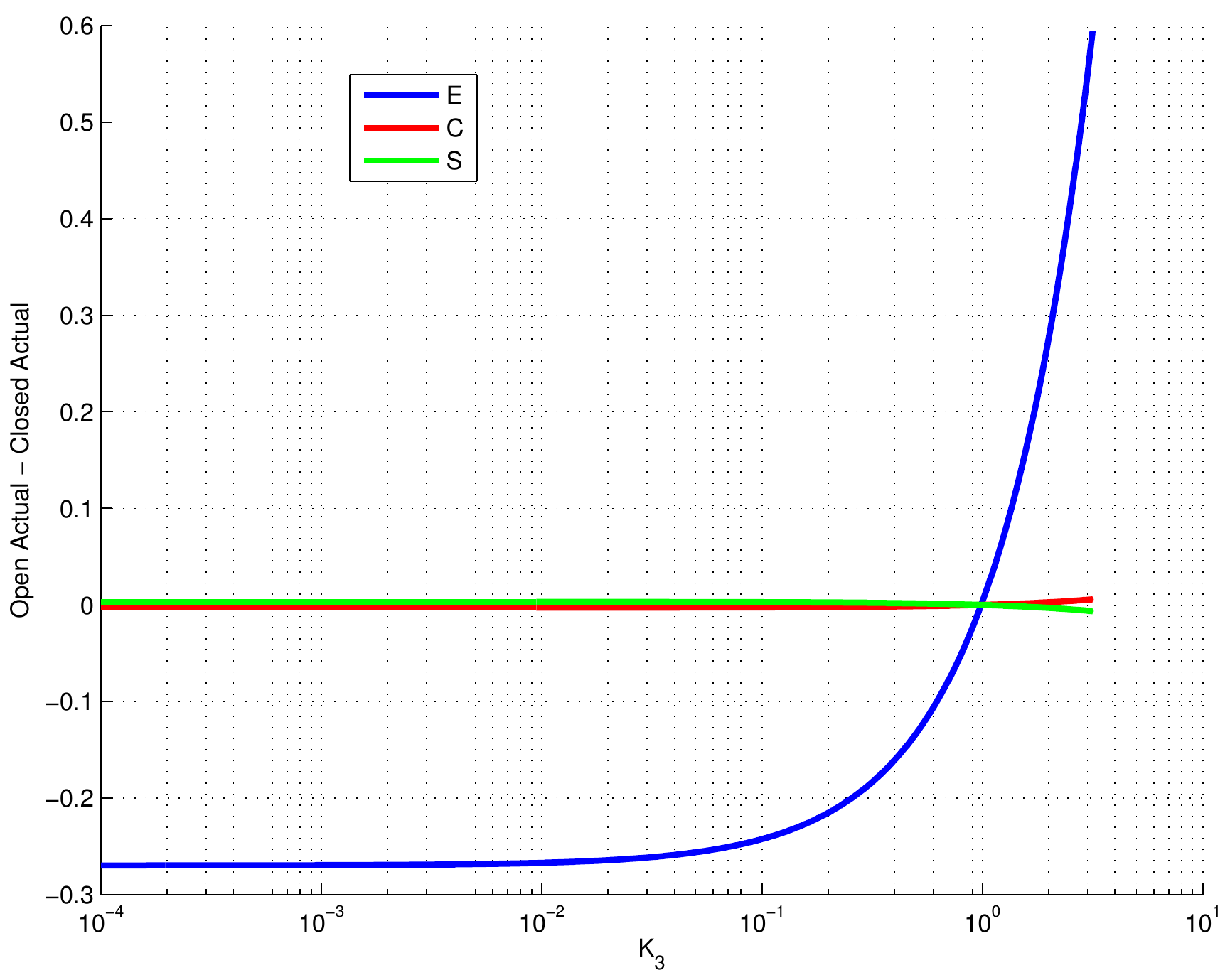}}\\
	\caption[Reverse QSSA regime, varying $\Kk $. $\Knk  = 1$]{Phase plane portraits and signed error in the reverse QSSA regime as \Kk\ varies, with $\Knk = 1$.  }
	\label{fig:Q_ES_k3_a}
\end{figure}

\clearpage

\section{Decaying oscillations?}
\label{sec:oscillation}

Neither the closed system (\ref{eq:iso}) nor the open system (\ref{eq:input}) is capable of sustained oscillations without substrate input. The reaction proceeds until all substrate is irreversibly converted into product. It is conceivable, however, that the approach to the steady state $(E, C, S, P) = (E_0, 0, 0, S_0) \equiv \FP$ could involve decaying oscillations, {\it i.e.} the fixed point \FP\ is a stable spiral. 

It is a simple matter to linearize (\ref{eq:iso}) and (\ref{eq:input}) about \FP\ and calculate the four corresponding eigenvalues $\lambda_{1}, \ldots, \lambda_{4}$. For the closed system (\ref{eq:iso}), $\lambda_{1} = \lambda_{2} = 0$, and for the open system (\ref{eq:input}), $\lambda_{1} = \Knk, \lambda_{2} = 0$. The remaining eigenvalues for both systems are given by

\begin{equation}
\label{eq:eigen}
\lambda_{3,4} = \frac{1}{2}\left[-(E_{0}k_{1} + k_{-1} + k_{2}) \pm \sqrt{(E_{0}k_{1} + k_{-1}+k_{2})^{2} - 4E_{0}k_{1}k_{2}}\right]
\end{equation}

\noindent For decaying oscillations we require a negative discriminant, {\it i.e.}

\begin{subequations}
\label{eq:eigencond1}
\begin{align}
2E_{0}k_{1}k_{2} &> E_{0}^{2}k_{1}^{2} + k_{-1}^{2} + k_{2}^{2} + 2E_{0}k_{1}k_{-1}+2k_{-1}k_{2} \label{eq:eigencond1a}\\
2 &> E_{0}\frac{k_{1}}{k_{2}} + \frac{k_{-1}^{2}}{E_{0}k_{1}k_{2}} + \frac{k_{2}}{E_{0}k_{1}} + 2\frac{k_{-1}}{k_{2}} + 2\frac{k_{-1}}{E_{0}k_{2}}\label{eq:eigencond1b}
\end{align}
\end{subequations}

\noindent Since each of the terms in the right hand side of (\ref{eq:eigencond1b}) is nonnegative, satisfying that inequality entails satisfying
\begin{equation}
\label{eq:eigencond2}
2 > E_{0}\frac{k_{1}}{k_{2}} + \frac{k_{2}}{E_{0}k_{1}}
\end{equation}
\noindent Writing $A = E_{0}\frac{k_{1}}{k_{2}}$, we see that satisfying inequality (\ref{eq:eigencond2}) is equivalent to finding positive $A$ such that $0 > A^{2}-2A+1$, which is impossible. Hence \FP\ cannot be a stable spiral, and the system does not exhibit decaying oscillations. 

It has been shown elsewhere that for some systems, the reduced equations obtained via QSSA analyses may produce dynamical behavior at odds with that of the original system \cite{Flach:2010, Flach:2006}. In particular, the reduced system may be incapable of oscillation though the original system does in fact oscillate \cite{Flach:2006}. Here we briefly investigate whether the QSSA analysis presented above may mislead in the opposite direction, that is, indicating the possibility of (decaying) oscillations when they are in fact impossible in the full system. 
 
Our point of departure is the observation that our method of solution for (\ref{eq:2ndE}) presented in Section (\ref{sec:non-isolatedQSSA}), namely rewriting the first order nonlinear ODE (\ref{eq:inputE}) as a (pseudo-) linear second order ODE (\ref{eq:2ndE}), we obtain the equational form for a forced damped oscillator:

%
%

\begin{equation}
\DD{x}{t} + 2\gamma\D{x}{t} + \omega^{2}x = F(t) \label{eq:osc}
\end{equation}

\noindent In the case of a mechanical oscillator, such as a mass-spring system, $\gamma > 0$ represents the coefficient of friction, $\omega$ represents the natural frequency of oscillation, and $F(t)$ is the forcing term. In the enzyme dynamics equation (\ref{eq:2ndE}) for the open system, these coefficients are 
\begin{subequations}
\label{eq:osc2}
\begin{eqnarray}
\gamma &=& \frac{1}{2}(\Knk + k_{1}S_{0} + k_{-1} + k_{2}) = \frac{1}{2}(\Knk + k_{1}(S + \KM))\\
\omega^{2} &=& \Knk (k_{-1} + k_{2}) = \Knk k_{1} \KM\\
F(t) &=& \Kk(k_{1}+k_{2}) = \Kk k_{1}\KM
\end{eqnarray}
\end{subequations}

\noindent Forcing is constant and proportional to the rate of enzyme input, the putative oscillation frequency is proportional to the rate of enzyme removal, and damping is (in the standard QSSA regime) dominated by the initial substrate concentration or $K_M$. Viewed as a forced damped oscillator, the system would appear to have an natural frequency which is independent of the initial enzyme concentration, a conclusion already at odds with the calculation above. (Obviously, any oscillatory solution may only be valid within the time frame for which Equation (\ref{eq:2ndE})  provides a reasonably accurate approximation to the dynamics of the full system, that is, for time $t < t_{f}$. If $t_{f}$ is less than the expected period of oscillation, then we may not be fully confident of realizing damped oscillations in the system.) 

Considering the approximate solution (\ref{eq:finale}) we obtain for the second order ODE for enzyme, we would anticipate the possibility of damped oscillations if \Rp, \Rm\ are complex valued, that is if
\begin{subequations}
\label{eq:osc-inequality}
\begin{align}
1 - \frac{4\Knk\KM}{k_1(\Knk/k_{1} + S_{0} + \KM)^2} &< 0 \label{eq:osc-inequality1}\\
\intertext{or, equivalently, }
 2\sqrt{\KM\Knk/k_{1}} - (\KM+\Knk/k_{1}) &> S_{0} \label{eq:osc-inequality2}
\end{align}
\end{subequations}

\noindent For any combination of positive parameter values, the maximum value of the lefthand side of inequality (\ref{eq:osc-inequality2}) is zero, so that the inequality cannot be satisfied for positive initial substrate. Thus $r_{1}, r_{2}$ are always real valued, and the QSSA approximation indicates that the system should not oscillate for physically admissible parameter values. 

We may also argue against the possibility of oscillations another way. Turning back to equations (\ref{eq:osc})--(\ref{eq:osc2}), the most propitious situation for oscillation is when there is no forcing, {\it i.e.} $\Kk = 0$, no enzyme input. In this case, decaying oscillations may potentially occur only if the system is underdamped, that is, when $\omega > \gamma$. This requires

\begin{subequations}
\begin{eqnarray}
 \sqrt{\Knk k_{1} \KM} &>& \frac{1}{2}(\Knk + k_{1}(S_{0} + \KM))\\
  4\Knk k_{1} \KM &>&  \Knk^{2} + 2\Knk k_{1}(S_{0} + \KM) + k_1^2(S_{0}+\KM)^{2}\\
 0 &>& (\Knk + k_{1}S_{0})^{2}+ 2k_{1}(k_{1}S_{0}-\Knk)\KM + k_{1}^{2}\KM^{2} \label{eq:condC}
\end{eqnarray}
\end{subequations}

\noindent Condition (\ref{eq:condC}) implies that \KM\ must lie between the roots of the quadratic, $\frac{1}{k_{1}}\left[-(k_{1}S_{0}-\Knk) \pm 2\sqrt{k_{1}S_{0}\Knk}\right]$. The requirement that \KM\ be positive necessitates
\begin{subequations}
\begin{eqnarray}
2\sqrt{k_{1}S_{0}\Knk} &>& k_{1}S_{0}-\Knk\\
2 &>& \frac{k_{1}S_{0}}{\Knk} + \frac{\Knk}{k_{1}S_{0}}\label{eq:condD}
\end{eqnarray}
\end{subequations}

\noindent Applying the same reasoning as we did for inequality (\ref{eq:eigencond2}), we see that inequality (\ref{eq:condD}) cannot be satisfied, and damped oscillations are impossible. 

In the vernacular of harmonic oscillators, thus we find that the enzymatically open system is always overdamped. Decaying oscillations are not observed; in the absence of enzyme input, the free enzyme concentration monotonically decays at an initial exponential rate. This accords with the linear analysis performed above. Though the form of the QSSA-based approximation suggests that the open system might oscillate in certain parameter regimes, multiple complementary arguments demonstrate that this is not in fact the case.

%
\section{Discussion}
\label{sec:discussion}

Biochemical reactions {\it in vivo} take place under conditions in which the concentrations of substrates and active enzymes vary. Yet the dominant framework for modeling biochemical reaction dynamics, the Michaelis-Menten formalism, presumes that such reactions occur in an essentially static environment, isolated from biochemical inputs. This assumption is made explicit, and is indeed crucial, in classical derivations of Michaelis-Menten dynamics using the QSSA \cite{Segel:1989}. Relaxing this assumption in the QSSA would therefore seem to be a natural, if not essential, step in evaluating theoretically the suitability of the Michaelis-Menten formalism for modeling biochemical activity in living organisms. Prior studies have investigated the effects of both constant and time-periodic substrate inputs on the derivation and validity conditions for standard and reverse QSSA \cite{Stoleriu:2004, Stoleriu:2005}. They speak to the fact that in living systems, enzymatic reactions generally occur in networks where the product(s) of one reaction serve as substrate(s) for others, and therefore the amount of substrate available in downstream reactions is hardly constant as assumed in the classical QSSA. 

In the present study, we addressed the fact that in living systems, levels of enzymes in their unbound, activated forms fluctuate for a variety of reasons, among them changing levels of gene activation, allosteric and non-allosteric competition, and enzymatic degradation and denaturation. The effects of such variations are assumed to be negligible in the classical QSSA, and have remained unexamined up to now. In this work, we have performed the first  examination,  to our knowledge, of the effects of enzyme input and removal on the derivation and validity conditions for the QSSA. Our results further delineate the nature and suitability of the Michaelis-Menten formalism for {\it in vivo} biochemical reaction networks. 

We studied a particular modification of the canonical ``closed'' enzymatic reaction which incorporates generic forms of enzyme input and removal and is thus applicable for a variety of biological scenarios. Our approach followed the steps of the classical standard QSSA analysis, with the critical difference that we could not assume the total amount of enzyme (free and bound) remained constant during the reaction. The absence of this conservation relation distinguishes the mathematical problem considered here from all prior work on QSSA, in which the conservation of total enzyme is exploited in the crucial first step of reducing the dimension of the system. We used a different, integral form enzyme conservation relation as the starting point for our analysis of the open system, from which followed a recasting of the system of first order ODEs as second order differential equations. The reformulated system admits straightforward closed form inner solutions for enzyme and complex concentrations (Section \ref{sec:non-isolatedQSSA}). These inner solutions reduce to the standard results from \cite{Segel:1989} as enzyme input and removal rates go to zero. 

Continuing the usual steps of standard QSSA analysis, we also computed expressions for post-transient quasi-equilibria of enzyme and complex in terms of substrate level (Section \ref{subsec:quasi-steady-states}). The expressions we obtained appear as intuitive corrections to the quasi-equilibrium formulas from standard Michaelis-Menten dynamics. They predict, in particular, the elevation of long term enzyme levels by the ratio of enzyme input to the complex formation rate, {\it i.e.} the equilibrium free enzyme accumulation, and the depression of long term enzyme levels by the ratio of enzyme removal to the complex formation rate, {\it i.e.} the equilibrium free enzyme subtraction. We note, however, that the corrected expressions presented here have only limited applicability due to their implicit assumption of static total enzyme levels, an assumption from which the open system deviates increasingly over the course of the reaction. Additional work, likely including novel analytical insight, will be necessary to overcome this limitation in order to obtain accurate outer solutions valid for the entire post-transient period of the open reaction. One potential avenue of investigation would be to modify the form of enzyme input in the open system (\ref{eq:input}) such that it saturates, a scenario perhaps more realistically modeling the influence of varying gene expression on enzyme levels. 

Our examination of the pre-steady-state period was more successful: we calculated the time scale for the initial transient period and derived validity conditions for the inner solutions (Sections \ref{subsec:fast-timescale} and \ref{subsec:validity}). Both sets of expressions closely match standard QSSA results for closed reactions. The time scale for the transient period is shortened slightly from that of the closed system, while the primary inequality for QSSA validity is relaxed. The amount of both corrections depends on the rate of enzyme removal (not enzyme input), which is consonant with the guiding intuition behind the standard QSSA: when substrate significantly exceeds enzyme concentration, free enzyme rapidly binds with substrate and achieves quasi-equilibrium. Enzyme removal increases the amount by which substrate exceeds enzyme, and, by reducing the amount of free enzyme available, decreases the length of the pre-steady-state transient.

With the derivation of inner solutions, time scale estimates, and validity conditions for QSSA for the open system, we thus achieved in large measure our first aim set forth in the Introduction, the extension of standard QSSA to an enzymatically open reaction. We largely relied upon numerical investigations to address our second and third aims. Investigating the suitability of the closed system, and hence classical Michaelis-Menten formulae, in approximating the behavior of open biochemical reactions, we found it to be reasonably accurate so long as free enzyme input and removal occur several orders of magnitude more slowly than other reaction rates. While this may frequently be true {\it in vivo}, it is far from the rule, as described in the Introduction. Furthermore, in approximating the open reaction, the closed system underperforms the inner solutions we derived in (Section \ref{sec:non-isolatedQSSA}) at all rates of enzyme input and removal. Once the enzyme input and/or removal are at one-thousandth to one-hundredth the rate of other reaction steps, the behavior of the open system departs substantially from the Michaelis-Menten formalism. Our inner solutions accurately approximate this behavior for some time beyond the duration of the initial transient period, up to the point that enzyme input and/or removal occur as rapidly as the other components of the reaction. 

Other studies have demonstrated that the reduced systems obtained via QSSA approaches may not faithfully reproduce the full range of behaviors available to the original, unreduced systems \cite{Flach:2010, Flach:2006}. In particular, it has been shown that the canonical enzymatically-catalyzed biochemical reaction (\ref{chem:iso}) admits oscillatory solutions with substrate input, though the QSSA-derived reduced system does not \cite{Flach:2006}. In the course of investigating the range of dynamical behavior in the open system, we noted the apparent possibility of the opposite occurrence (Section \ref{sec:oscillation}): though analysis of the the full open system shows that damped oscillatory solutions are impossible, the reduced equation (\ref{eq:2ndE}), takes the form of a forced damped harmonic oscillator and would at first glance appear to admit decaying oscillatory solutions. The impossibility of finding physically admissible parameter values to realize such solutions in either the full or reduced versions of the approximate open system, however, may be explained by simple arguments. 

The theoretical approach taken in this paper follows the methods and hypotheses of the standard QSSA, with the primary result that a more suitable inner approximation to the dynamics of enzymatically open biochemical reaction was derived. The reverse QSSA scenario was examined numerically; anticipated future work includes closer theoretical examination of reverse QSSA in open systems. An alternative mathematical approach to QSSA involving a different change of variables, {\it total QSSA}, has been successfully used to enlarge the range of validity for QSSA analysis \cite{Borghans:1996,Tzafriri:2004}. We have applied total QSSA techniques to the open system considered here, and we shall report our findings in another paper. 

\bibliographystyle{elsarticle-num}
\bibliography{QSSAwEI}

\begin{thebibliography}{10}
\expandafter\ifx\csname url\endcsname\relax
  \def\url#1{\texttt{#1}}\fi
\expandafter\ifx\csname urlprefix\endcsname\relax\def\urlprefix{URL }\fi
\expandafter\ifx\csname href\endcsname\relax
  \def\href#1#2{#2} \def\path#1{#1}\fi

\bibitem{Michaelis:1913}
L.~Michaelis, M.~Menten, {Die Kinetik der Invertinwirkung}, Biochem Z. 49
  (1913) 333--369.

\bibitem{Briggs:1925}
G.~E. Briggs, J.~B.~S. Haldane, {A Note on the Kinetics of Enzyme Action},
  Biochem. J. (1925) 338--339.

\bibitem{Bowen:1963}
J.~Bowen, A.~Acrivos, A.~Oppenheim, Singular perturbation refinement to
  quasi-steady approximation in chemical kinetics, Chem. Eng. Sci. 18 (1963)
  177--188.

\bibitem{Segel:1988}
L.~A. Segel, On the validity of the steady state assumption of chemical
  kinetics, Bull. Math. Bio. 50 (1988) 579--593.

\bibitem{Segel:1989}
L.~Segel, M.~Slemrod, The quasi-steady state assumption: a case study in
  perturbation, SIAM Review 31 (1989) 446--477.

\bibitem{Borghans:1996}
J.~Borghans, R.~de~Boer, L.~Segel, Extending the quasi-steady state
  approximation by changing variables, Bull. Math. Bio. 58 (1996) 43--63.

\bibitem{Schnell:2000}
S.~Schnell, P.~K. Maini, Enzyme kinetics at high enzyme concentration, Bull.
  Math. Bio. 62~(3) (2000) 483--499.
\newblock \href {http://dx.doi.org/10.1006/bulm.1999.0163}
  {\path{doi:10.1006/bulm.1999.0163}}.

\bibitem{Stoleriu:2004}
I.~Stoleriu, F.~Davidson, J.~Liu, Quasi-steady state assumptions for
  non-isolated enzyme-catalysed reactions, J. Math. Bio. 48 (2004) 82--104.

\bibitem{Tzafriri:2004}
A.~Tzafriri, E.~Edelman, The total quasi-steady-state approximation is valid
  for reversible enzyme kinetics, J. Theor. Bio. 226 (2004) 303--313.

\bibitem{Stoleriu:2005}
I.~Stoleriu, F.~A. Davidson, J.~L. Liu, Effects of periodic input on the
  quasi-steady state assumptions for enzyme-catalysed reactions, J. Math. Bio.
  50 (2005) 115--132.

\bibitem{Flach:2006}
E.~H. Flach, S.~Schnell, Use and abuse of the quasi-steady-state approximation,
  IEE Proc.-Syst. Biol. 153~(4) (2006) 187--191.

\bibitem{Dingee:2008}
J.~W. Dingee, A.~B. Anton, {A New Perturbation Solution to the Michaelis-Menten
  Problem}, AIChE Journal 54~(5) (2008) 1344--1357.

\bibitem{Alon:2006}
U.~Alon, An Introduction to Systems Biology: Design Principles of Biological
  Circuits, Chapman and Hall/CRC, 2006.

\bibitem{Fell:1997}
D.~Fell, Understanding the Control of Metabolism, Portland Press, London, 1996.

\bibitem{Berg:2002}
J.~M. Berg, J.~L. Tymoczko, L.~Stryer, Biochemistry, 5th Edition, W. H.
  Freeman, New York, 2002.

\bibitem{Keener:1998}
J.~Keener, J.~Sneyd, Mathematical Physiology, Vol.~8 of Interdisciplinary
  Applied Mathematics, Springer-Verlag, 1998.

\bibitem{Frenzen:1989}
C.~L. Frenzen, P.~K. Maini, Enzyme kinetics for a two-step enzymatic reaction
  with comparable initial enzyme-substrate ratios, J. Math. Bio. 26 (1989)
  689--703.

\bibitem{Flach:2010}
E.~H. Flach, S.~Schnell, Stability of open pathways, Math. Biosci. 228~(2)
  (2010) 147--152.

\end{thebibliography}


\end{document}